\documentclass[english,11pt,a4paper]{article}
\pdfoutput=1
\usepackage{epsf,amsmath,amssymb,graphicx,dcolumn}
\usepackage{scalefnt,ulem,cite,hyperref}
\usepackage{cleveref,etoolbox,color}
\usepackage{todonotes,xspace}
\usepackage{listings}
\usepackage[utf8]{inputenc}
\usepackage[tocindentauto]{tocstyle}
\apptocmd{\thebibliography}{\setlength{\itemsep}{0.02cm}}{}{}
\Crefname{figure}{Fig.}{Figs.}
\usepackage{placeins}

\newcommand{\SARAH}{{\tt SARAH}\xspace}

\newcommand{\FeynArts}{{\tt FeynArts}\xspace}
\newcommand{\FormCalc}{{\tt FormCalc}\xspace}

\newcommand{\SPheno}{{\tt SPheno}\xspace}

\newcommand{\Mathematica}{{\tt Mathematica}\xspace}

\newcommand{\SFOLD}{{\tt SFOLD}\xspace}
\newcommand{\HFOLD}{{\tt HFOLD}\xspace}
\newcommand{\FVSFOLD}{{\tt FVSFOLD}\xspace}
\newcommand{\CNN}{{\tt CNNDecays}\xspace}
\newcommand{\Fortran}{\texttt{Fortran}\xspace}

\newcommand{\citere}[1]{Ref.\,\cite{#1}}
\newcommand{\citeres}[1]{Refs.\,\cite{#1}}

\newcommand{\abbrev}{\scalefont{.9}}
\newcommand{\eqn}[1]{Eq.\,(\ref{#1})}
\newcommand{\fig}[1]{Fig.\,\ref{#1}}
\newcommand{\sct}[1]{Section~\ref{#1}}
\newcommand{\appref}[1]{Appendix~\ref{#1}}

\newcommand{\lhc}{{\abbrev LHC}}
\newcommand{\qcd}{{\abbrev QCD}}

\newcommand{\sm}{{\abbrev SM}}

\newcommand{\mssm}{{\abbrev MSSM}}
\newcommand{\nmssm}{{\abbrev NMSSM}}
\newcommand{\ckm}{{\abbrev CKM}}
\newcommand{\susy}{{\abbrev SUSY}}
\newcommand{\bsm}{{\abbrev BSM}}

\newcommand{\lo}{{\abbrev LO}}

\newcommand{\lsp}{{\abbrev LSP}}
\newcommand{\nlsp}{{\abbrev NLSP}}

\newcommand{\cp}{{\abbrev $\mathcal{CP}$}}
\newcommand{\nlo}{{\abbrev NLO}}

\newcommand{\drbar}{{\abbrev $\overline{\text{DR}}$}}
\newcommand{\msbar}{{\abbrev $\overline{\text{MS}}$}}

\renewcommand{\Re}{{\rm Re}}

\newcommand{\GeV}{{\rm GeV}}

\newcommand{\nn}{\nonumber}

\newcommand{\sla}[1]{\ooalign{\hfil/\hfil\crcr$#1$}} 
\newcommand{\Ufactors}{$U$-factors}

\numberwithin{equation}{section}

\newcounter{notecount}
\def\ov{\overline}
\def\lagr{\mathcal{L}}
\def\tC{\tilde{C}}

\lstdefinestyle{mathematica}{
        basicstyle=\ttfamily\mdseries,
	language=bash,
	frame=false,
	xleftmargin=.25in}   

\lstdefinestyle{terminal}{
	language=bash,
	frame=lines,
	xleftmargin=.5in,
        numbers=none}
        
\lstdefinestyle{file}{
        basicstyle=\ttfamily\mdseries,
	language=bash,
	frame=shadowbox,
        numbers=left,   
        numberstyle=\tiny} 

\title{\vspace*{-4em}
  \begin{flushright}
    {\sf\small
      DESY-17-042 --- KA-TP-11-2017\\
    }
  \end{flushright}
\vspace*{2em}
Generic calculation of two-body partial decay widths \\ at the full one-loop level\vspace{-0.2cm}}
\author{Mark D. Goodsell${}^{a,b}$, Stefan Liebler${}^c$, Florian Staub${}^{d,e}$\\[1.2em]
{\it ${}^a$ Sorbonne Universit\'es, UPMC Univ Paris 06, UMR 7589}\\[-0.2em]
{\it LPTHE, F-75005, Paris, France}\\[0.2em]
{\it ${}^b$ CNRS, UMR 7589, LPTHE, F-75005, Paris, France}\\[0.2em]
{\it ${}^c$ DESY, Notkestra{\ss}e 85, D-22607 Hamburg, Germany}\\[0.2em]
{\it ${}^d$ Institute for Theoretical Physics (ITP), Karlsruhe Institute of Technology,}\\[-0.2em]
{\it Engesserstra{\ss}e 7, D-76128 Karlsruhe, Germany}\\[0.2em]
{\it ${}^e$ Institute for Nuclear Physics (IKP), Karlsruhe Institute of Technology,}\\[-0.2em]
{\it Hermann-von-Helmholtz-Platz 1, D-76344 Eggenstein-Leopoldshafen, Germany}\\[1.2em]
{\small\tt goodsell@lpthe.jussieu.fr}\\[-0.3em]
{\small\tt stefan.liebler@desy.de}\\[-0.3em]
{\small\tt florian.staub@kit.edu}\\[-0.3em]
}
\date{}

\begin{document}
\maketitle

\begin{abstract}
\noindent
We describe a fully generic implementation of two-body partial decay widths at the full one-loop level
in the \SARAH and \SPheno framework compatible with most supported models.
It incorporates fermionic decays to a fermion
and a scalar or a gauge boson as well as scalar decays into two fermions, two gauge bosons, two scalars
or a scalar and a gauge boson. We present the relevant generic expressions for virtual
and real corrections.
Whereas wavefunction corrections are determined from on-shell conditions,
the parameters of the underlying model are by default renormalised in a \drbar{} (or \msbar{})
scheme. However, the user can also define model-specific counter-terms.
As an example we discuss the renormalisation of the electric charge in the Thomson limit
for top-quark decays in the standard model.
One-loop induced decays are also supported.
The framework additionally allows the addition of mass and mixing corrections induced at higher
orders for the involved external states. We explain our procedure to cancel infra-red
divergences for such cases, which is achieved through an infra-red counter-term taking
into account corrected Goldstone boson vertices.
We compare our results for sfermion, gluino and Higgs decays in the minimal supersymmetric standard model (\mssm{}) against 
the public codes \SFOLD, \FVSFOLD and \HFOLD and explain observed differences. Radiative induced
gluino and neutralino decays are compared against the original implementation in \SPheno in the \mssm{}. We exactly reproduce
the results of the code \CNN for decays of neutralinos and charginos in $R$-parity violating
models. The new version {\tt SARAH 4.11.0} by default includes the calculation
of two-body decay widths at the full one-loop level.
Current limitations for certain model classes are described.
\end{abstract}

\newpage
\tableofcontents
\newpage

\section{Introduction}
While the Large Hadron Collider (\lhc{}) has completed the standard model (\sm{})
of particle physics with the discovery of a scalar which has all expected properties
of the long searched for Higgs boson \cite{Aad:2012tfa,Chatrchyan:2012xdj,Khachatryan:2016vau}, there is no indication for new physics up to now. 
This has lead to impressive exclusion limits for particles predicted by either supersymmetry (\susy{}) or other extensions 
of the \sm{} which were proposed to resolve the open questions of the \sm{}. However, these exclusion limits for beyond the standard model (\bsm{}) particles
depend strongly on the decay properties of these particles. For instance, it is well known that the often cited limits for \susy{} squarks and gluinos
of $1.8$\,TeV and more hold only in vanilla models where these states decay to $100$\,\%
into a given final state \cite{ATLAS-CONF-2016-052,ATLAS-CONF-2016-054,ATLAS-CONF-2016-078,Xie:2223502}. 
Once realistic decay patterns for the particles are used, the limits become much weaker \cite{Han:2013kza,Bharucha:2013epa,Kowalska:2016ent,Buckley:2016kvr}.
Thus, a precise knowledge of the branching ratios of \bsm{} states is necessary to be able
to draw firm conclusions from the null results. On the other hand, once a new particle is discovered,
precise calculations become especially important to extract the underlying parameters and compare against the predictions of many different models.

There has been a lot of effort to improve the predictions of the decay widths for new Higgs-like scalars not only in the minimal supersymmetric standard
model (\mssm{})~\cite{Mendez:1990jr,Li:1990ag,Djouadi:1994gf,Hempfling:1993kv,Hall:1993gn,Carena:1994bv,
Pierce:1996zz,Bartl:1997yd,Arhrib:1997nf,Eberl:1999he,Carena:1998gk,Carena:1999py,Carena:2002bb,Guasch:2003cv,
Heinemeyer:2000fa,Degrassi:2002fi,Curiel:2003uk,Bernreuther:2005gw,Haber:2006kz,Baro:2008bg,Mihaila:2010mp,Hollik:2010ji,Noth:2008tw,
Noth:2010jy,Williams:2011bu,Heinemeyer:2014yya,Heinemeyer:2015pfa,Heinemeyer:2016wey} and the 
next-to-minimal supersymmetric standard (\nmssm{})~\cite{Baglio:2015noa}, but also in several singlet and doublet extensions of the \sm{} 
\cite{Kanemura:2004mg,Kanemura:2015mxa,Bojarski:2015kra,Krause:2016oke,Kanemura:2016lkz,Krause:2016xku,Muhlleitner:2017dkd}. These results are implemented in public tools 
such as {\tt HDECAY}~\cite{Djouadi:1997yw,Djouadi:2006bz}, {\tt FeynHiggs}~\cite{Hahn:2009zz,Heinemeyer:1998yj,Hahn:2013ria} or {\tt NMSSMCALC}~\cite{Baglio:2013iia}.
However, for the plethora of other states,  tree-level results are often used. 
Exceptions are the \mssm{}, where one-loop corrections to all sfermions and gauginos 
were discussed in
\citeres{Baer:1990sc,Beenakker:1996dw,Kraml:1996kz,Bartl:1997pb,Bartl:1998xp,Guasch:2001kz,Hou:2002kv,Li:2002ey,Guasch:2002ez,Arhrib:2004tj,Drees:2006um,Baro:2009gn,
Fritzsche:2011nr,Heinemeyer:2011gk,Heinemeyer:2011ab,Heinemeyer:2012wp,Bharucha:2012re,Hollik:2012rc,Hollik:2013xwa,Gavin:2014yga,Aebischer:2014lfa};
and neutralino and chargino decays in the \nmssm{}~\cite{Liebler:2010bi,Belanger:2016tqb}. For other \susy{} models
with $R$-parity violation and \cp{} violation, only a few selected decay modes were discussed so far in \citeres{Liebler:2011tp,Cheriguene:2014bxa}.
The available codes to study decays at the one-loop level in the \mssm{} are {\tt SDECAY}~\cite{Muhlleitner:2003vg}, {\tt SUSY\_HIT}~\cite{Djouadi:2006bz} and 
\SFOLD~\cite{Hlucha:2011yk} for sfermion decays, \FVSFOLD for flavour violating squark
as well as gluino decays, and {\tt SloopS}~\cite{Belanger:2016tqb} and \CNN~\cite{Liebler:2010bi,Liebler:2011tp} for neutralino
and chargino decays without and with $R$-parity violation.

This limited number of codes and supported models has to be seen in contrast to the increasing number of models which are currently studied. With the increasing limits on 
the \susy{} masses within the \mssm{}, other ideas for new physics are seeing more and more attention. 
In order to be able to also give more accurate predictions for the decays in non-minimal \susy{} models
or also in non-supersymmetric extensions of \sm{}, a high-level of automatisation is needed. 
A very powerful ansatz to obtain robust results for \bsm{} models has been established with the {\tt Mathematica}
package \SARAH~\cite{Staub:2008uz,Staub:2009bi,Staub:2010jh,Staub:2012pb,Staub:2013tta,Staub:2015kfa}: \SARAH derives from a short 
model file all analytical properties of a given model. This information together with generic expressions for various observables is then used to generate {\tt Fortran}
code for \SPheno~\cite{Porod:2003um,Porod:2011nf} 
which can be used to obtain numerical results. Up to now, one- and two-loop masses \cite{Goodsell:2014bna,Goodsell:2015ira,Goodsell:2016udb}, 
one-loop flavour and precision observables \cite{Porod:2014xia}, as well as
two- and three-body tree-level decays could be obtained via this setup. We have now enhanced the decay calculation to the next level by a  generic ansatz to 
calculate two-body decay widths at the full one-loop level. These extensions are now available with {\tt SARAH 4.11.0}. In this paper we give all necessary details
about the calculation, including the renormalisation scheme; the generic expressions for virtual and real corrections; and the handling of ultraviolet and
infra-red divergences.

While wavefunction corrections are determined from on-shell conditions,  the default settings use a \drbar{} (or \msbar{}) renormalisation for the parameters of the underlying
model. However, the user can also define model-specific counter-terms in \SARAH to be used in the numerical evaluation in \SPheno.
For now the self-energies of all particles of the underlying model are available for this purpose.
Since many particle species receive significant higher-order corrections to their masses and mixing beyond tree-level, we also
allow the inclusion of mass and mixing corrections for the involved external states. This needs a careful treatment
of the infra-red divergences, for which we add an infra-red counter-term making use of modified Goldstone boson vertices.
The setup also supports loop-induced decays. An extension to models with \cp{} violation or additional charged
and massive, coloured vector particles is left for future work. A more thorough discussion of Higgs
boson decays, which are very sensitive to corrections of the external states, will be addressed in the future.

The paper is organised as follows: In \sct{sec:technicaldetails} we discuss the technical details of the implementation
employing external tree-level masses. The incorporation of higher-order corrections for the external states is
lined out in \sct{sec:onshellproperties}.
In \sct{sec:usage} we explain how the new features of \SARAH and \SPheno can be used. In \sct{sec:results} we present some 
results obtained with the new machinery: we first show the implementation of counter-terms for two \sm{} examples
and then compare our implementation in \SARAH with other public codes as \SFOLD, \HFOLD, \CNN.
We conclude in \sct{sec:conclusions}. The appendix contains all relevant generic expressions for virtual and
real corrections as well as a derivation of the employed Goldstone boson vertices.

\section{Calculation of decay widths at the full one-loop level}
\label{sec:technicaldetails}

In this section we discuss the technical details of the
calculation of two-body decay widths at next-to-leading order
for decays that are mediated through a tree-level diagram $X\to Y_1Y_2$.
Our implementation can handle the decays 
$S\to SS$, $S\to SV$, $S\to VV$, $S\to FF$, $F\to FS$ and $F\to FV$,
where $S$ denotes a scalar, $F$ a fermion and $V$ a heavy gauge boson.
For loop-induced processes $V$ can also be a photon or gluon.
At next-to-leading order
we include full \qcd{} and electroweak corrections.
Thus, apart from ultraviolet divergences, which we need to address
through the renormalisation of the parameters of the underlying model,
infra-red divergences due to massless photons and gluons have to
be taken care of.
For loop-induced decays the subsequent discussion simplifies substantially,
since neither ultraviolet nor infra-red divergences have to be tamed, i.e.
also the detailed renormalisation of parameters is not of relevance.
We continue as follows: we describe the generic form of unpolarised
squared matrix elements for two-body decays in the subsequent subsection
and thereafter present the various ingredients in terms of tree-level
and one-loop amplitudes. This includes vertex and wavefunction corrections
as well as a discussion of counterterms. Then in \sct{sec:realcorrections} we discuss
the relevant real corrections being $1\to 3$ processes, before we combine
the results in \sct{sec:comresults}. Finally, we list the limitations
of our implementation in \sct{sec:limitations}.

\subsection{Generic unpolarised squared matrix elements}

For any two-body decay we write its partial width in the form
\begin{align}
\label{eq:twobodygamma}
 \Gamma_{X\to Y_1Y_2}=\frac{1}{16\pi m_X^3}\lambda(m_X^2,m_{Y_1}^2,m_{Y_2}^2) C_SC_C \sum_{h,p} |\mathcal{M}|^2\,,
\end{align}
where $m_X, m_{Y_1}$ and $m_{Y_2}$ are the masses of the mother and daughter particles
in the initial and final state, respectively. We denote their momenta with $p_0,p_1$ and $p_2$, respectively.
The sum runs over all helicities ($h$) and polarisations ($p$) in the initial and final state.
A symmetry factor $C_S$ and colour factor $C_C$ have to
be employed. The symmetry factor is $C_S=1$ by default. For $X=F$ we have $C_S=\frac12$, if $\overline{Y}_1=Y_1$
and $\overline{Y}_2=Y_2$. For $X=S$ it is $C_S=\frac12$, if $\overline{Y}_1=Y_1$
and $\overline{Y}_2=Y_2$ and $Y_1=Y_2$. Therein $\overline{Y}$ denotes the antiparticle of $Y$.
The colour factor $C_C$ for a decaying colour singlet is equal to the dimension of the final states under SU$(3)_C$.
For example, for a colour octet decaying into triplets, it yields $C_C=\frac12$, while for more complicated colour configurations $C_C$ can be easily extracted
from the colour-dependent part of the vertex triggering the decay: the colour of the initial state is fixed and a sum 
over all possible colour combinations in the final state is performed. The K\"all\'en function $\lambda$ is given by
\begin{align}
\label{eq:kaellen}
\lambda(p_0^2,p_1^2,p_2^2)=\sqrt{p_0^4+p_1^4+p_2^4-2p_0^2p_1^2-2p_1^2p_2^2-2p_0^2p_2^2}\,.
\end{align}

For decay modes with fermions and gauge bosons in the initial and/or final state the matrix elements are a sum over Lorentz structures; we label these with a lower index as $M_i$ and 
therefore split the total squared amplitude in sums of contributions $M_iM_j^*$, which are multiplied with different kinematic dependences obtained from helicity and polarisation sums.
The structures and their sums are given by:
\begin{align}\nn
F \to FS:&\\\label{eq:matrixelementsplit1}
\mathcal{M} &\equiv M_1 \bar{v} (p_0) P_L v (p_1) + M_2 \bar{v} (p_0) P_R v (p_1) \nn\\
\sum_{h,p} |\mathcal{M}|^2 &= \frac{1}{2}(m_X^2 + m_{Y_1}^2 - m_{Y_2}^2)(M_{1}M^{*}_{1} + M_{2}M^{*}_{2})
+m_{Y_1}m_{Y_2}(M_{1}M^{*}_{2} + M_{2}M^{*}_{1})\\\nn
S \to FF:&\\
\mathcal{M} &\equiv M_1 \bar{u} (p_1) P_L v (p_2) + M_2 \bar{u} (p_1) P_R v (p_2)  \nn\\
\sum_{h,p} |\mathcal{M}|^2 &= (m_X^2 - m_{Y_1}^2 - m_{Y_2}^2)(M_{1}M^{*}_{1} + M_{2}M^{*}_{2})
-2m_{Y_1}m_{Y_2}(M_{1}M^{*}_{2} + M_{2}M^{*}_{1})\\\nn
S \to SV:&\\
\mathcal{M} &\equiv \epsilon_\mu^*(p_2) (p_0^\mu + p_1^\mu) M \nn\\ 
\sum_{h,p} |\mathcal{M}|^2 &= \frac{1}{4m_{Y_2}^2}\left[m_X^4 + (m_{Y_1}^2 - m_{Y_2}^2)^2 - 2m_X^2(m_{Y_1}^2+m_{Y_2}^2)\right]MM^{*}\\\nn
S \to SS:&\\
\mathcal{M} &\equiv M \nn\\
\sum_{h,p} |\mathcal{M}|^2 &= MM^*
\end{align}

For $S\to VV$ we split the squared amplitude as follows:
\begin{align}\nn
\mathcal{M}\equiv& \epsilon_\mu^*(p_1) \epsilon_\nu^* (p_2) \bigg( M_1 \eta^{\mu\nu} + M_2 p_0^\mu p_0^\nu \bigg)\\\nn
\sum_{h,p} |\mathcal{M}|^2 = &\frac{1}{2m_{Y_2}^2m_{Y_3}^2}\left[m_X^4 + m_{Y_1}^4 + 10 m_{Y_1}^2m_{Y_2}^2 + m_{Y_2}^4 - 2 m_X^2(m_{Y_2}^2+m_{Y_3}^2)\right]M_1M^*_1\\\nn
&+ \frac{1}{8m_{Y_2}^2m_{Y_3}^2}\left[m_X^4 + (m_{Y_1}^2-m_{Y_2}^2)^2 - 2 m_X^2(m_{Y_2}^2+m_{Y_3}^2)\right]^2M_2M^*_2\\\nn
&+ \frac{1}{4m_{Y_2}^2m_{Y_3}^2}\left[m_X^6 - 3m_X^4(m_{Y_1}^2 + m_{Y_2}^2) - (m_{Y_1}^2 - m_{Y_2}^2)^2(m_{Y_1}^2 + m_{Y_2}^2)\right.\\
&\qquad \left.+ m_{Y_1}^2(3m_{Y_1}^4 + 2m_{Y_1}^2m_{Y_2}^2 + 3m_{Y_2}^4)\right](M_{1}M^{*}_{2} + M_{2}M^{*}_{1})
\end{align}

Lastly the squared amplitude for $F\to FV$ is given by:
\begin{align}\nn
&\mathcal{M}\equiv  \epsilon_\mu^* (p_2) \bigg( M_1 \bar{v} (p_0) \gamma^\mu P_L v (p_1) + M_2 \bar{v} (p_0)\gamma^\mu 
P_R v (p_1) \\ &\qquad\qquad+ M_3  p_0^\mu  \bar{v} (p_0) P_L v (p_1) + M_4 p_0^\mu  \bar{v} (p_0) P_R v (p_1)\bigg) \nn\\\nn
&\sum_{h,p} |\mathcal{M}|^2 = \frac{1}{2m_{Y_2}^2}\left[m_X^4 + m_{Y_1}^4 + m_{Y_1}^2m_{Y_2}^2 - 2m_{Y_2}^4 + m_X^2(-2m_{Y_1}^2 + m_{Y_2}^2)\right](M_1M^*_1+M_2M^*_2)\\\nn
& + \frac{1}{8m_{Y_2}^2}\left[(m_X^2 + m_{Y_1}^2 - m_{Y_2}^2)(m_X^4 + (m_{Y_1}^2 - m_{Y_2}^2)^2 - 2m_X^2(m_{Y_1}^2 + m_{Y_2}^2))\right](M_3M^*_3+M_4M^*_4)\\\nn
& - 3m_Xm_{Y_1}(M_1M^*_2+M_2M^*_1)\\\nn
& - \frac{1}{4m_{Y_2}^2}\left[m_{Y_1}(m_X^4 + (m_{Y_1}^2 - m_{Y_2}^2)^2 - 2m_X^2(m_{Y_1}^2 + m_{Y_2}^2))\right](M_1M^*_3+M_3M^*_1 + M_2M^*_4+M_4M^*_2)\\\nn
& - \frac{1}{4m_{Y_2}^2}\left[m_X(m_X^4 + (m_{Y_1}^2 - m_{Y_2}^2)^2 - 2m_X^2(m_{Y_1}^2 + m_{Y_2}^2))\right](M_1M^*_4+M_4M^*_1 + M_2M^*_3+M_3M^*_2)\\
& + \frac{1}{4m_{Y_2}^2}\left[m_Xm_{Y_1}(m_X^4 + (m_{Y_1}^2 - m_{Y_2}^2)^2 - 2m_X^2(m_{Y_1}^2 + m_{Y_2}^2))\right](M_3M^*_4+M_4M^*_3)
\label{eq:matrixelementsplit2}
\end{align}

We implemented special cases for final states with vanishing masses, which
are not given here.\footnote{In the calculation of one-loop decays we 
introduced a minimal allowed mass for fermions and scalars of $10^{-15}$\,GeV. Smaller masses
are set to zero to stabilise numerics, 
see \sct{sec:cnnvsspheno} for a discussion in the context of $R$-parity violation.}

\subsection{Tree-level amplitudes}

For the two-body decays at tree-level the contributions to the matrix elements~$M^T_i$
can be directly identified with the (left- and right-handed) couplings
as follows:
\begin{align}
\label{eq:treelevelamplitudes}
& F\to FV:\quad M^T_1=ic_R,M^T_2=ic_L\,,   &\qquad F\to FS:\quad M^T_1=-ic_R,M^T_2=-ic_L\\
& S\to FF:\quad M^T_1=-ic_R,M^T_2=-ic_L\,, &\qquad S\to SS:\quad M^T=ic\\
& S\to SV:\quad M^T=-2ic\,,              &\qquad S\to VV:\quad M^T_1=ic
\end{align}
The conventions for the parametrisation of the vertices are summarised in \appref{app:genericexp}.
For $F\to FV$ $M^T_3$ and $M^T_4$ and
for $S\to VV$ $M^T_2$ vanish at tree-level, but contributions are generated at the one-loop level.

\subsection{One-loop amplitudes}

Before discussing the detailed form of vertex and wavefunction corrections, we show
their combination with the previously presented results. Once the amplitude
due to vertex corrections $M^V$ and due to wavefunction corrections $M^W$ are split
into $M^V_i$ and $M^W_i$, which encode
the various contributions to different combinations of helicities and polarisations,
they can be added to the tree-level amplitudes as follows
\begin{align}
\label{eq:loopamplitudes}
 M_i = M^T_i + 2M^V_i + 2M^W_i\,.
\end{align}
For an exact next-to-leading order calculation the complex-conjugated part of
this amplitude~$M_i^*$ is inserted in the complex-conjugated matrix elements
of \eqn{eq:matrixelementsplit1} to \eqn{eq:matrixelementsplit2}, whereas $M^T_i$ is used for the non-conjugated ones.
The total partial width is obtained from the real part of the full expressions in \eqn{eq:matrixelementsplit1}
to \eqn{eq:matrixelementsplit2}.
When squaring the amplitude one needs to be careful if external coloured particles are involved. For these cases, \SARAH calculates
individual colour factors for tree- and loop-level contributions and sums them up congruently. 
For loop-induced decays the sum of the amplitudes of the vertex corrections~$M^V_i$ and wavefunction corrections $M^W_i$
is inserted into all occurrences of matrix elements
in \eqn{eq:matrixelementsplit1} to \eqn{eq:matrixelementsplit2}.

\subsection{Vertex corrections}
\label{sec:vertexcorrections}
\begin{figure}[h] 
\centering 
\includegraphics[width=0.5\linewidth]{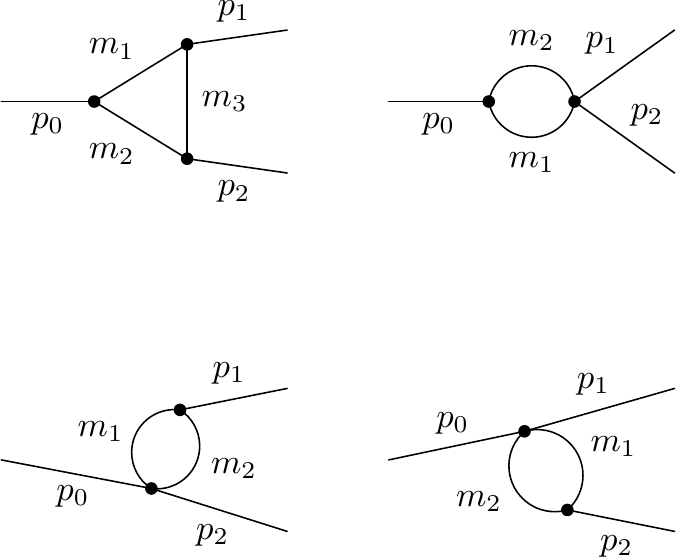} 
\caption{Possible topologies contributing to the vertex corrections. }
\label{fig:topologies}
\end{figure} 
In general, there are four different topologies contributing to the vertex corrections~$M^V$ which are 
shown in \fig{fig:topologies}. For decays involving fermions only the first topology is of relevance. 
Depending on the considered decay, different generic diagrams are associated with these topologies.
They are depicted for the different decays in \appref{app:genericexp} in Figs.~\ref{fig:FFS}--\ref{fig:SVV}.
The results are a function of internal
masses $m_1, m_2$ and $m_3$ entering the loop diagrams, but also of the external momenta $p_0, p_1$ and $p_2$.
In this \sct{sec:technicaldetails} their squared values correspond to the
squared external masses of the particles, i.e. $m_X^2$, $m_{Y_1}^2$ and $m_{Y_2}^2$, respectively.
The calculation of the generic amplitudes~$M^{(k)}_i$ for each diagram is straightforward and all results 
are given in \appref{app:genericexp}. They
are obtained with \FeynArts~\cite{Hahn:2000kx} and \FormCalc~\cite{Hahn:1998yk} in Feynman-'t~Hooft gauge,
i.e. charged and neutral Goldstone bosons are included in the calculation and cancel the
unphysical contributions from heavy gauge bosons.

Our results are expressed in terms of Passarino-Veltman integrals obtained
through dimensional reduction (\drbar{}).\footnote{\FormCalc works with
constrained differential renormalisation~\cite{delAguila:1998nd},
which equals dimensional reduction at the one-loop level~\cite{Hahn:1998yk}.} Thus,
the ultraviolet divergences can be
split off in terms of $\Delta=\tfrac{1}{\epsilon}-\gamma_E+\ln(4\pi)$, where $\epsilon$ regularises
the divergence and equals the difference to four dimensions $d=4-2\epsilon$ and $\gamma_E$ is
the Euler-Mascheroni constant.
Terms denoted by $r$ need to be set to zero in dimensional reduction and correspond to the difference with respect
to dimensional regularisation, i.e. it yields
$r=1$ for calculations performed in the minimal subtraction scheme (\msbar{}). By default \SARAH sets
$r=0$ in \susy{} models and $r=1$ in non-\susy{} models. 
In order to match mass dimensions correctly in less than $4$ dimensions,
dimensional reduction also introduces a new scale~$Q$, the renormalisation scale.
The generic particle~$U$ denotes a Faddeev-Popov ghost. In Feynman-'t~Hooft gauge
their masses are identical masses to the gauge bosons masses, i.e. also ghosts obtain
the subsequently discussed regulator mass.
Infra-red divergences due to massless photons and gluons are regularised through
a finite regulator mass. There are no diagrams that contain both photons and gluons.
The cancellation of infra-red divergences will be addressed
in \sct{sec:realcorrections}, whereas the cancellation of ultraviolet
divergences is obtained by adding the subsequently discussed
corrections~$M^W$.

The combinatoric part to populate the generic diagrams with all possible field insertions 
in a given model is done by  \SARAH. \SARAH also checks for possible symmetry factors which appear if 
in the topologies $2-4$ in \fig{fig:topologies} two real and identical particles are in the loop. In addition,
it calculates relevant colour factors to be multiplied with the interference terms $M^T(M^V)^*$. 

\subsection{Wavefunction corrections}
\label{sec:wvcorrections}

The amplitude~$M^W$ contains the corrections due to wavefunction normalisation as well as
the counter-term for the tree-level coupling. They cancel the ultraviolet
divergences of the vertex corrections~$M^V$ and
are mostly determined through renormalisation prescriptions,
in contrast to $M^V$.
Omitting the complication of fermions and gauge bosons for a moment
the amplitude $M^W_{ijk}$ for a vertex of the form $c_{ijk}X_iY_{1j}Y_{2k}$ for the process $X_i\to Y_{1j}Y_{2k}$ yields
\begin{align}
 M^W_{ijk}=i\left(\delta c_{ijk} +  \frac12 c_{ljk}\delta Z_{X_l X_i} + \frac12 c_{ilk}\delta Z_{Y_{1l} Y_{1j}} + \frac12 c_{ijl}\delta Z_{Y_{2l} Y_{2k}}\right)
\end{align}
with the counter-term $\delta c$ of the tree-level coupling $c$ and the wavefunction corrections $\delta Z$
for the three particles involved. In the last three terms a sum over $l$ has to be performed.
In the following we will first describe the derivation
of the wavefunction corrections and then comment on the counter-term for the tree-level coupling.

For the wavefunction corrections we employ an on-shell scheme for the three fields $S$, $V$ and~$F$. 
For the fermions we distinguish left- and right handed components $F^L$ and $F^R$.
In all cases we allow for mixing among particles induced through loop effects,
such that the wavefunction corrections are generally matrices.
\begin{align}
V^{\mu,0}_{i}&\rightarrow Z_{V_i V_j} V_{\mu,j} = (\delta_{ij}+\frac12 \delta Z_{V_i V_j})V^{\mu}_{j}\\
S^0_i&\rightarrow Z_{S_i S_j}S_j = (\delta_{ij} + \frac12\delta Z_{S_i S_j})S_j\\
F^{L0}_i&\rightarrow Z_{F_i F_j}^LF_j = (\delta_{ij} + \frac12\delta Z^L_{F_i F_j})F^L_j\\
F^{R0}_i&\rightarrow Z_{F_i F_j}^RF_j = (\delta_{ij} + \frac12\delta Z^R_{F_i F_j})F^R_j
\end{align}
In order to determine the wavefunction corrections $\delta Z$ from on-shell conditions,
we need the self-energies for our three particle species. Their notation can be read
off from the inverse propagators at the one-loop level, which we write as follows:
\begin{align}
\label{eq:loopmasses}
\Gamma_{S_i S_j}(p^2) &= i(p^2 - m_S^2)\delta_{ij} + i\hat\Pi_{S_i S_j}(p^2)\\
\Gamma^{\mu \nu}_{V_i V_j}(p^2) &=  -i g^{\mu \nu}(p^2-m_V^2)\delta_{ij} - i\left(g^{\mu\nu} - \frac{p^\mu p^\nu}{p^2}\right)\hat\Pi_{V_i V_j}(p^2) - i\frac{p^\mu p^\nu}{p^2} \hat\Pi^L_{V_i V_j}(p^2)\\
\Gamma_{F_i F_j}(p) &=i(\sla{p} - m_{F})\delta_{ij}+i\left[\sla{p}(P_L\hat\Sigma_{ij}^{L}(p^2)+ P_R\hat\Sigma_{ij}^{R}(p^2)) + P_L\hat\Sigma_{ij}^{SL}(p^2) + P_R\hat\Sigma_{ij}^{SR}(p^2)\right]
\end{align}
The renormalised self-energies are   indicated through $\hat\Pi$ and $\hat\Sigma$ compared to the unrenormalised
ones $\Pi$ and $\Sigma$, which are of relevance for the subsequent discussion.
$\Pi_{VV}$ and $\Pi_{SS}$ are the self-energies of the gauge bosons and scalars, respectively.
For the gauge bosons we are only interested in the transverse part $\Pi_{VV}$.
The only mixing induced between the gauge bosons of the \sm{} is among the photon
and the $Z$ boson.
$P_L$ and $P_R$ are the left- and right-handed projection operators, which split the
self-energies of the fermions in $\Sigma^L$, $\Sigma^R$, $\Sigma^{SL}$ and $\Sigma^{SR}$.
\begin{figure}[t] 
\centering 
\includegraphics[width=0.5\linewidth]{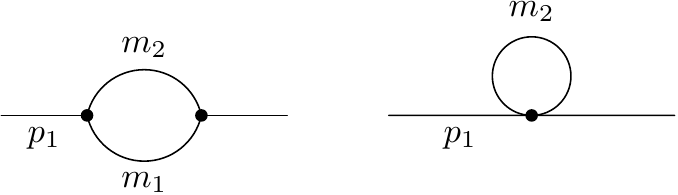} 
\caption{Possible topologies contributing to the wavefunction renormalisation.}
\label{fig:wavetop}
\end{figure} 
The topologies which can contribute are shown in \fig{fig:wavetop}. Moreover, all 
possible generic diagrams contributing to the fermion, scalar and vector bosons self-energies 
are shown in \appref{app:genericexp} in Figs.~\ref{fig:WV_F}--\ref{fig:WV_V}. We give also in \appref{app:genericexp}
the expressions for the generic amplitudes for the self-energies and their derivatives. 
Note that the above structure for the gauge bosons implies the usage of Feynman-'t~Hooft gauge.
The derivatives of the wavefunction corrections,
denoted with $\dot{\Pi}$ and $\dot{\Sigma}$, are defined as follows:
\begin{align}
\dot{\Pi}(k^2) = \left.\frac{\partial}{\partial p^2}\Pi(p^2)\right|_{p^2=k^2}\qquad\text{and}\qquad
\dot{\Sigma}(k^2) = \left.\frac{\partial}{\partial p^2}\Sigma(p^2)\right|_{p^2=k^2}
\end{align}
Demanding on-shell conditions for the external states fixes the wavefunction corrections.
Their derivation can for example be found in \citeres{Denner:1991kt,Liebler:2010bi} and results in similar
expressions for scalars and gauge bosons:
\begin{align}
\delta Z_{S_i S_i} = & - \widetilde{\Re} \dot\Pi_{S_i S_i}(m_{S_i}^2)\,, &\delta Z_{V_i V_i} =  - \widetilde{\Re} \dot\Pi_{V_i V_i}(m_{V_i}^2)\\
\delta Z_{S_i S_j} = & \frac{2}{m_{S_i}^2 - m_{S_j}^2} \widetilde{\Re}\Pi_{S_i S_j}(m_{S_j}^2)\,, & \delta Z_{V_i V_j} =  \frac{2}{m_{V_i}^2 - m_{V_j}^2} \widetilde{\Re}\Pi_{V_i V_j}(m_{V_j}^2)
\end{align}
For the fermions we need to distinguish four cases:
%{\allowdisplaybreaks
\begin{align}\nn
\delta Z^L_{F_i F_i} = & -\widetilde{\Re}\left[ \Sigma^L_{ii}(m_{F_i}^2) +  m_{F_i}^2 \left(\dot \Sigma^L_{ii}(m_{F_1}^2) + \dot \Sigma^R_{ii}(m_{F_i}^2)\right)
 + m_{F_i} \left(\dot \Sigma^{SL}_{ii}(m_{F_i}^2) + \dot \Sigma^{SR}_{ii}(m_{F_i}^2)\right)\right] \\\nn
\delta Z^R_{F_i F_i} = & - \widetilde{\Re}\left[  \Sigma^R_{ii}(m_{F_i}^2) +  m_{F_i}^2 \left(\dot \Sigma^L_{ii}(m_{F_1}^2) + \dot \Sigma^R_{ii}(m_{F_i}^2)\right)
 + m_{F_i} \left(\dot \Sigma^{SL}_{ii}(m_{F_i}^2) + \dot \Sigma^{SR}_{ii}(m_{F_i}^2)\right)\right] \\\nn
 \delta Z^L_{F_i F_j} = & \frac{2m_{F_j}}{m_{F_i}^2 - m_{F_j}^2} \widetilde{\Re}\left[ m_{F_j} \Sigma^L_{ij}(m_{F_j}^2) +  m_{F_i}  \Sigma^R_{ij}(m_{F_j}^2)
+ \frac{m_{F_i}}{m_{F_j}} \Sigma^{SL}_{ij}(m_{F_j}^2) + \Sigma^{SR}_{ij}(m_{F_j}^2)  \right]\\
\delta Z^R_{F_i F_j} = & \frac{2m_{F_j}}{m_{F_i}^2 - m_{F_j}^2} \widetilde{\Re}\left[ m_{F_i} \Sigma^L_{ij}(m_{F_j}^2) +  m_{F_j}  \Sigma^R_{ij}(m_{F_j}^2) 
 + \Sigma^{SL}_{ij}(m_{F_j}^2) + \frac{m_{F_i}}{m_{F_j}} \Sigma^{SR}_{ij}(m_{F_j}^2)  \right]
\end{align}%}
By $\widetilde{\Re}$ we indicate that $\Pi$ and $\Sigma$ entering $\delta Z$ include only the real parts of the loop functions,
whereas couplings enter with real and imaginary components. In case of \cp{} violation the definition of wavefunction
corrections usually distinguishes between in- and outgoing particles in order to correctly multiply absorbative parts of
self energies with complex couplings, see the Appendix of \citere{Fritzsche:2011nr}. This is beyond our implementation.
We note that despite the fact that we
employ on-shell conditions to determine the wavefunction corrections 
our external particles are not necessarily on-shell particles, see the discussion at the end of this section.

With this setup at hand we can also define counter-terms to be used for tree-level rotation matrices,
which at lowest order transform gauge into mass eigenstates. Those counter-terms enter 
the counter-term of the tree-level coupling. For any particle species $\Phi$
in mass eigenstates, which is obtained from gauge eigenstates $\Phi'$ through $\Phi_i=R^\Phi_{ij}\Phi'_j$,
the counter-term is given by 
\begin{equation}
\label{eq:CTrot}
\delta R^\Phi_{ij} = \frac{1}{4} \sum_k \left(\delta Z_{\Phi_i\Phi_k} - (\delta Z_{\Phi_k\Phi_i})^*\right) R^\Phi_{kj}\,.
\end{equation}
For fermions left- and right-handed states are rotated with two matrices, such that
two counter-terms employing left- and right-handed wavefunction corrections also  need to be defined.
For Majorana fermions we refer to \citere{Liebler:2010bi}.
It is well-known that the definition of such counter-terms for mixing matrices based on the wavefunction
corrections needs a proper treatment of Goldstone boson tadpole contributions in order to achieve gauge invariance,
see \citere{Liebler:2010bi} for a more detailed discussion.
Since we work in Feynman-'t Hooft gauge we can completely omit these Goldstone boson tadpole contributions, since
they ultimately cancel between the wavefunction corrections and the counter-term of the 
mixing matrices. As for the vertex corrections,  \SARAH inserts all combination of particle species in the generated code,  and includes colour as well as
symmetry  factors.

The non-trivial and mostly non automatisable part of the calculation of two-body partial decay width
is the renormalisation prescription used for the bare parameters of the
underlying theory, that enter the tree-level coupling counter-term of the two-body decay under consideration.
The counter-terms are usually chosen depending on the model and process.
However, a simple \drbar{} (or \msbar{}) prescription 
for the renormalisation of the parameters of the underlying theory is always easily applicable:
From the $\beta$ functions and anomalous dimensions used for the renormalisation group equations 
implemented in \SARAH~\cite{Martin:1993zk,Yamada:1994id,Jack:1997eh,Fonseca:2011vn,Fonseca:2013bua,Goodsell:2012fm,Sperling:2013eva,Sperling:2013xqa}
we can define all counter-terms of the parameters of the underlying theory
to be just proportional to the pure ultraviolet divergence only.
We will refer to this scheme as \drbar{} (or \msbar{}) scheme in the following.
It is well-known that this scheme will not perform well in various cases.
Therefore, the user of \SARAH can define their own counter-terms, see
\sct{sec:setuploopdecays} for a more detailed discussion. We also
add an example of a proper renormalisation of the electric charge in \sct{sec:results}.

A consequence of the application of the \drbar{} (or \msbar{}) scheme is that our partial
decay widths are left with a dependence on the renormalisation scale~$Q$ introduced through the regularisation
of ultraviolet divergences. This is most prominent in the running of the parameters
that enter the tree-level coupling obtained from the renormalisation group equations,
which is not cancelled at the one-loop level. In the generated code
the scale $Q$ is by default set to the average stop mass $\sqrt{m_{\tilde{t}_1}m_{\tilde{t}_2}}$
in supersymmetric models and the top-quark mass $m_t$ in non-supersymmetric models.
However, the user can control the scale~$Q$ in the input file, either throughout \SPheno or only
for the calculation of the decays at one-loop level, see \sct{sec:setuploopdecays}.
A common choice for the renormalisation scale is also $Q\sim m_X$ close to the mass of the decaying particle~$X$.
We refrain from making it the default option, since $Q\sim m_X$ slows down the numerical evaluation
substantially. In this case loop contributions need to be evaluated multiple times.
We recommend to vary the scale to check the stability of the partial decay width calculation, as we demonstrate
in \sct{sec:results} for the decay of the \sm{} Higgs boson into bottom quarks.
If the scale is changed throughout \SPheno keep in mind that also masses and thus kinematics can change.
For a full on-shell calculation the scale dependence also completely vanishes. We demonstrate
this for the decay of the top-quark in \sct{sec:results}. In order to achieve a renormalisation-scale independent result,
external states have to have fixed masses and mixing, which for gauge bosons and
fermions can be achieved through the settings explained in \sct{sec:setuploopdecays}.

Until now we ignored the fact that particles receive higher-order mass corrections.\footnote{Also on-shell schemes can lead to finite
shifts at higher orders, see e.g. \citeres{Eberl:2001eu,Fritzsche:2002bi,Baro:2009gn} for examples in the neutralino and chargino sector.}
By construction we have to employ the mass values at lowest order
throughout the calculation. We will discuss in \sct{sec:onshellproperties}
how for the external states mass corrections and mixing beyond tree-level can be incorporated
into our calculation. If we allow for mass corrections we limit the discussion to the inclusion of \drbar{} (\msbar{}) corrections
to the masses, whereas full on-shell prescriptions for \bsm{} particles (as it would
  be appropriate) are left for future work.
With the outlined procedure in the previous subsections, we obtain
a gauge-independent and ultraviolet finite result for the partial
width $X\to Y_1Y_2$, which in the most general case however is scale dependent.
As mentioned, the cancellation of infra-red
divergences is addressed in the next section.

\subsection{Real corrections}
\label{sec:realcorrections}

In the previous calculation of vertex and wavefunction corrections
we regularised infra-red divergences through the introduction of
a finite, but small regulator mass for the photon and/or gluon.
The artificial dependence of the cross section calculation on that
mass is cancelled by adding the real emission of a photon and/or gluon
to the two-body decay, i.e. by adding three-body decays.
For a soft photon and/or gluon a divergence is induced, which
can again be regularised through a mass and cancels the mass dependence
from the vertex and wavefunction corrections. %This is the result
%of the Kinoshita-Lee-Nauenberg theorem~\cite{Kinoshita:1962ur,Lee:1964is}.
With the help of {\tt FeynArts}~\cite{Hahn:2000kx} and {\tt FormCalc}~\cite{Hahn:1998yk}
we generated generic results for the emission of one additional
photon~$\gamma$ or gluon~$g$ for the previously discussed two-body
processes $S\to SS$, $S\to SV$, $S\to VV$, $S\to FF$, $F\to FS$ and $F\to FV$.
We denote the real corrections for $X\to Y_1Y_2+\gamma/g$ with
\begin{align}
\label{eq:threebodygamma}
\Gamma_{X\to Y_1Y_2+\gamma/g} = \frac{1}{(4\pi)^3m_X}\frac{1}{\pi^2}\int
\frac{d^3p_1}{2p_1^0}\frac{d^3p_2}{2p_2^0}\frac{d^3k}{2k^{0}}\delta^{4}\left(p_0-p_1-p_2-k\right)C'_S\sum_{h,p,c}|\mathcal{M}|^2 \,,
\end{align}
where $k$ denotes the momentum of the photon or gluon and momenta with upper index $0$ equal the zeroth component of the corresponding four vector.
External momenta are set to $p_0^2=m_X^2$, $p_1^2=m_{Y_1}^2$, $p_2^2=m_{Y_2}^2$ and $k^2=0$.
We then have $C'_S=C_S$ for $S\to VV$ and $S\to SV$, otherwise $C'_S=\tfrac{1}{2}C_S$.
The charge and colour structure is encoded in matrices $C_{ij}$ explained in \appref{app:realcorr}, which
is why \eqn{eq:threebodygamma} only contains an additional $c$ for colour to be summed over.
It is clear that the real corrections due to photon emission and gluon
emission can be calculated individually and summed up afterwards.
By rewriting denominators in terms of eikonal factors, the above integrals can be mapped onto
\begin{align}
I^{j_1j_2}_{i_1i_2}(m_X,m_{Y_1},m_{Y_2}) =\frac{1}{\pi^2}\int
\frac{d^3p_1}{2p_1^0}\frac{d^3p_2}{2p_2^0}\frac{d^3k}{2k^{0}}\delta^4\left(p_0-p_1-p_2-k\right)
\frac{(\pm 2p_{j_1}\cdot k)(\pm 2p_{j_2}\cdot k)}{(\pm 2p_{i_1}\cdot k)(\pm 2p_{i_2}\cdot k)}\,,
\end{align}
where $p_{i,j}\in\lbrace p_0,p_1,p_2\rbrace$ and the minus signs refer to cases where $p_{i,j}$ equals the momentum~$p_0$ of the initial particle~$X$.
The notation follows \citere{Denner:1991kt}, where also results for the relevant integrals are shown.
Only integrals with double lower indices are infra-red divergent and thus dependent on the regulator mass in addition.
We present our results in \appref{app:realcorr}.
Through our procedure we calculate the full soft- and hard emission of such photons and gluons
and thus for the three-body decay $S\to SV+\gamma/g$ also include the
four-point interaction, which does not diverge as the regulator mass approaches zero.
The correct charge and colour factor assignments in the real corrections
are done by \SARAH as explained in \appref{app:realcorr}.
Where possible we compared to the analytic results for real corrections
implemented in {\tt SFOLD}~\cite{Hlucha:2011yk} and {\tt HFOLD}~\cite{Frisch:2010gw}
as well as the result presented in \citere{Liebler:2010bi}.
Apart from finite contributions in $S\to SV$ we found complete
agreement.

We avoid additional collinear divergences by keeping finite masses for
all three particles in the initial and final state of our two-body decay
calculation, if they interact with photons or gluons.
Thus, this problem does not arise for e.g. final-state neutrinos.
For fully massless charge- and colour neutral particles in the final state
we implemented dedicated routines for $F\to F'S\gamma$ and $S\to F'F\gamma$, where
one final state fermion $F'$ can be massless.
Keep in mind that if final-state charged or coloured particles are very light,
large collinear logarithms can induce a bad numerical behaviour of our routines.
This shouldn't cause problems in practical applications unless charged or coloured states
with very smalles masses ($\ll$keV) are present.  

Lastly note that since the real correction decay widths are gauge independent,
we performed the calculation in unitary gauge for simplicity. This ensures that the results
depend only on the gauge couplings and the original tree-level vertex, and we are not obliged
to include would-be Goldstone boson vertices as we do for the corresponding loop corrections. The 
exception is the decay $S\to SV+\gamma/g$, where gauge invariance fixes the form of the four-point vertex
in terms of the three-point one, and we implicitly assume this relation. 

\subsection{Combination of results}
\label{sec:comresults}

The partial width at next-to-leading order is thus obtained as follows
\begin{align}
 \Gamma^{\text{\nlo{}}}_{X\to Y_1Y_2} = \Gamma_{X\to Y_1Y_2} + \Gamma_{X\to Y_1Y_2+\gamma/g}\,,
\end{align}
where $\Gamma_{X\to Y_1Y_2}$ is obtained from \eqn{eq:twobodygamma}
with the squared amplitudes given in \eqn{eq:matrixelementsplit1} to \eqn{eq:matrixelementsplit2}.
The individual parts $M_i$ are taken from \eqn{eq:loopamplitudes} and \eqn{eq:treelevelamplitudes}
for the complex conjugated and non-complex conjugated squared amplitudes, respectively.
$\Gamma_{X\to Y_1Y_2+\gamma/g}$ are the real corrections from \eqn{eq:threebodygamma},
which are calculated individually for photons and gluons and summed up.
For loop-induced decays the virtual contributions in $M^V$ are
by definition ultraviolet finite, still we also include the described wavefunction corrections.
We note that ultraviolet finiteness can be checked through a variation of $\Delta$ defined
in \sct{sec:vertexcorrections} and infra-red finiteness through a variation of the
regulator mass for the photon and/or gluon, see \sct{sec:usage} for a description how to access them.

\subsection{Current limitations}
\label{sec:limitations}
In the approach described so far, we made some assumptions which make the results not applicable to all models 
which are currently supported by \SARAH.
\begin{itemize}
 \item While complex parameters in all calculation can be handled in principle, the setup is not yet supposed to be 
 used for \cp{} violation. The reason is that for decays of real particles into complex final states, only the decay mode $Y_1 Y_2$ is calculated, while 
 $\overline{Y}_1 \overline{Y}_2$ is assumed to have the same partial width. Also note that in case of \cp{} violation a common approach is to define
 extended wavefunction corrections as discussed in the Appendix of \citere{Fritzsche:2011nr}.
 \item The calculation of the real divergences has neglected the possibility of massive, coloured vector bosons as for instance in Pati-Salam, deconstructed, or trinification models. 
 \item When using loop-corrected external masses as described in the next section, we need to cure all infra-red divergences through a proper treatment of Goldstone boson vertices. Currently we assume
 that the $W$ boson is the only massive, charged vector boson, such that models with a $W'$ cannot be used with loop-corrected external masses.
 \item Gauge boson decays are not implemented yet. This is partially due to the previously two mentioned limitations. On the other hand for decays of
 the gauge bosons of the \sm{} our framework can be easily extended, which we leave for future work.
\end{itemize}

\section{Higher-order corrections to the external states}
\label{sec:onshellproperties}

\begin{figure}[!h]
\centering 
\includegraphics[width=0.45\linewidth]{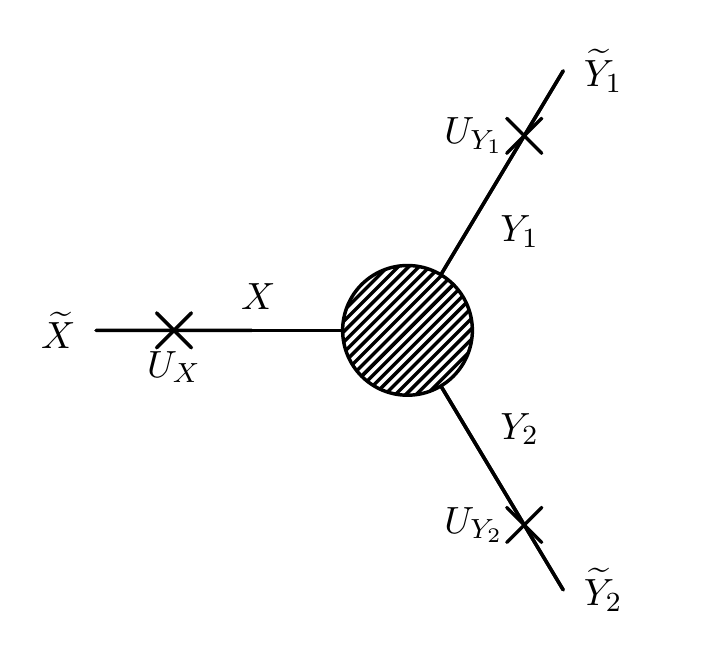} 
\caption{Schematic picture of our method to include higher-order mass and mixing corrections. The calculation 
presented in \sct{sec:technicaldetails} is combined with external normalisation factors and external loop-corrected
masses.}
\label{fig:scheme}
\end{figure}

Our previous discussion was based on the usage of tree-level masses
for internal as well as external particles. However the masses of various particle species
receive significant higher-order contributions. 
One way to address this is to adopt an on-shell scheme throughout the calculation,
but pure on-shell schemes are not always the best choice for such calculations, as is well-known
from the Higgs sector of the \mssm{}. 
Also even if the calculation is performed in terms of on-shell states
in particular in supersymmetric models the limited number
of renormalisable parameters in the Lagrangian does not allow for a renormalisation procedure where
all on-shell masses correspond to their tree-level values.

Instead, if we do not want to change our renormalisation prescription, we should use the LSZ reduction formula
to connect S-matrix elements of on-shell states with Feynman diagrams in our scheme. 
This results in external normalisation factors
and external loop-corrected masses (which need to be distinguished from the
previously discussed tree-level masses and mixing matrices). A schematic picture 
is shown in Fig.~\ref{fig:scheme}. For Higgs bosons the approach is discussed
in detail in \citere{Frank:2006yh}, where such wave function normalisation
factors are denoted $Z$-factors. In particular, as noted there, since we are working
at only one loop, there are different truncations of the perturbative series that we can make and
different approximations can also be made for expediency.
Here we outline the choice(s) that we have made.

Firstly, in the the gauge boson sector we
recommend to use on-shell values for gauge bosons, see the discussion
in \sct{sec:usage}. For scalars and fermions, however, we introduce matrices that we denote $U$. 
Let us introduce our notation for the example of $n$ scalars~$S_i$, which
are mass eigenstates obtained from gauge eigenstates $S_i=R^S_{ij}S'_j$ at tree-level
and mix at higher orders: following \eqn{eq:loopmasses} the mass matrices
beyond tree-level take the form
\begin{align}
\mathcal{M}_{ij}(p^2)=m_i^2\delta_{ij}-\hat\Pi_{ij}(p^2) 
\end{align}
with the tree-level masses $m_i$. Let's suppose that for the calculation of the external masses
a \msbar{} or \drbar{} scheme is preferred, i.e.
the self-energies $\hat\Pi$ and $\hat\Sigma$ are renormalised such that only the corresponding
ultraviolet divergent part is omitted.
In a first approximation we set $p^2=0$
and diagonalize the obtained mass matrix $\mathcal{M}(0)$ through a unitary $(n\times n)$ matrix $U^0$.
The tree-level mass eigenstates $S_i$ are thus rotated into states $\tilde{S}_i=U^0_{ij}S_j$
with masses $\widetilde{m}_i$.
This matrix $U^0$ incorporates the additional mixing induced at higher orders and in principle
corresponds to the $Z$-factors in the $p^2=0$ approximation of \citere{Frank:2006yh}.\footnote{It differs by the
prescription how the self-energies $\Pi$ are renormalised.}
It is used to rotate the tree-level, vertex and counter-term corrections
uniformly by applying it at the amplitude level. For the decay $\tilde{S}_i\to \tilde{S}_j\tilde{S}_k$
we for example shift the amplitudes by
\begin{align}
\label{eq:Urotation}
\widetilde{M}_{ijk} = \sum_{s,t,u} U^0_{is}U^0_{jt}U^0_{ku}M_{stu}
\end{align}
for $M=M^T,M^V$ and $M^W$. Also we define rotated tree-level couplings $\widetilde{c}_{ijk}$
in the same manner to be used in the calculation of tree-level ampltiude and 
real corrections as discussed subsequently.
This concept can be very similarly employed for fermions, where again left- and right-handed
mixing matrices $U$ have to be introduced.
Before we discuss the cancellation of ultraviolet and infra-red divergences
let us note that we also implemented two more methods to obtain the mixing matrix~$U$:
instead of setting $p^2=0$ an alternative choice is to use $p^2=m_i^2$.
This results in the mass eigenvalue $\widetilde{m}_i$, which is used to repeat the procedure iteratively
with $p^2=\widetilde{m}_i$ until the mass determination stabilizes. 
Our default choice is that the relative change between the masses of two iterations should
be below $10^{-6}$.
This procedure needs to be performed for each mass eigenstate $\tilde{S}_i$ separately
and the matrix $U^p$ is determined row by row and is thus not unitary any more,
as it is also well-known from the general form of $Z$-factors.
Lastly a possible choice is $p^2=m_1^2$, i.e. the external momenta is chosen to
be equal to the lightest mass eigenstate. In this case $U^{m_1}$ is again a unitary matrix.
We note that the outlined procedure to determine $\widetilde{m}$ and $U$ can be 
performed beyond one-loop level, i.e. for supersymmetric Higgs boson masses corrections
at the two-loop level can be incorporated.

\subsection{Ultraviolet and infra-red divergences}
The application of external masses~$\widetilde{m}$ different from the tree-level values~$m$
and mixing matrices~$U$ in addition to tree-level mixing
induces a problem with the cancellation
of ultraviolet and infra-red divergences. The first problem can be solved easily. 
We employ tree-level masses~$m$ for all propagators of loop functions
as well as external momenta entering loop functions. This applies to
vertex and counter-term corrections and guarantees the cancellation
of ultraviolet divergences.

The infra-red problem is more demanding. In order to achieve the cancellation of infra-red divergences
we define infra-red counter-terms. These counter-terms
encode the mismatch between the 
 masses and mixings of internal and external states
and are formally of higher order. These counter-terms are
used to shift the the wavefunction and vertex corrections:
\begin{align}
M^V \to M^V + \delta M^V  \\
M^{W} \to M^{W} + \delta M^{W}
\end{align}
The aim is to cancel the infra-red divergences stemming from $2 M^{T} (M^V + \delta M^V  + M^W + \delta M^W)^*$ against the ones from 
the real emission calculated with loop-corrected masses. The counter-terms $\delta M^V$, $\delta M^W$ are defined to be the difference
of the infra-red divergences of our default scheme and the one with loop-corrected masses $\widetilde{m}$
\begin{equation}
\delta M^{V,W} =   \text{IR}\left(\widetilde{M}^{V,W}\right) - \text{IR}\left(M^{V,W}\right)
\end{equation}
where $\text{IR}(M)$ takes only the infra-red divergent part of the amplitude~$M$. 
The definition of such counter-terms in
supersymmetric models is a common strategy:
\citere{Fritzsche:2011nr} introduces an infra-red counter-term
for the decays~$\tilde{t}_1\rightarrow \tilde{b}_iW^+$, see Eq. (191) in \citere{Fritzsche:2011nr},
which exactly encodes
the difference between on-shell, i.e. loop-corrected, and tree-level masses due to
the limited number of renormalisable parameters in the stop and sbottom sectors.
\citere{Gonzalez:2012mq} discusses
the introduction of such counter-terms for the heavy Higgs boson decay~$H\to W^+W^-$.
\citere{Fuchs:2014ola} discusses these aspects in detail in the context of a generalised narrow-width
approximation, where also infra-red divergent parts in the loop contributions are
sorted out and evaluated at a common mass scale.

It is clear that the subtraction and re-addition of infra-red divergent logarithms
as discussed before induces a spurious dependence on other masses, namely on the masses being the
counterpart of the regulator mass(es) in the logarithms. This is unavoidable,
however numerically of minor relevance.

In practice, the following procedure is applied:
\begin{enumerate}
\item We calculate the virtual corrections using tree-level masses.
\item We extract the infra-red divergences of all two and three-point function using the results given in 
\appref{sec:appendix_ir}. These result are used to obtain $\text{IR}\left(M^{V,W}\right)$. 
\item We use loop-corrected masses~$\widetilde{m}$ throughout all infra-red divergent diagrams 
for the external legs and the particles in the loop. We take again the infra-red divergent parts of 
these amplitudes to obtain $\text{IR}\left(\widetilde{M}^{V,W}\right)$.
\item The calculation of the kinematics as well as of the helicity and polarisations sums for both, the virtual and 
real corrections, is done with loop-corrected masses. 
\item Lastly, the usage of an additional external mixing applied
through the mixing matrices~$U$, named \Ufactors{}, works as follows: we rotate the amplitudes of the tree-level,
wavefunction and virtual corrections according to \eqn{eq:Urotation}. Instead for the contribution
of the infra-red counter-term we use rotated tree-level couplings~$\widetilde{c}$ rather than rotated
amplitudes~$\widetilde{M}$. Those rotated couplings also enter the calculation of the real corrections.
In this context we note that by construction the infra-red counter-term always
contains exactly one occurrence of the coupling~$c$ of the tree-level two-body decay.
\end{enumerate}
These steps give for most cases infra-red finite results. However, 
there is one complication: 
if the infra-red counter-term contains loops with massive gauge bosons, then there
will necessarily also be related diagrams with charged Goldstone bosons, and the  gauge symmetries require several relationships between 
the couplings -- and masses -- of the internal and external particles in order for the infra-red divergences to cancel. If we were to 
apply the above procedure then the infra-red counter-term would not be gauge invariant; for these diagrams
we therefore use loop-corrected masses and couplings.
Note that if we used unitary gauge we could avoid a discussion of corrected couplings in Goldstone boson vertices.
Denoting a would-be Goldstone boson by $G$, massive gauge bosons by $V^G$ with
masses $m_V^G$ and massless ones by $\gamma^a$, real scalars as $S_i$ with masses $m_i$ and Dirac fermions as $F_I$ with masses $m_I$, the relevant couplings are
\begin{align}
\lagr \supset&  \frac{1}{2} c_{ij}^G V^{G\, \mu} (S_j \partial_\mu  S_i  -  S_i \partial_\mu S_j) + \frac{1}{2} c_{ij}^a \gamma^{a\, \mu} (S_j \partial_\mu  S_i  -  S_i \partial_\mu S_j) \nn\\
& + \frac{1}{2} c_{GG'}^{G''} V^{G''\, \mu} (G' \partial_\mu  G  -  G \partial_\mu G') + \frac{1}{2} c_{GG'}^{a} \gamma^{a\, \mu} (G' \partial_\mu  G  -  G \partial_\mu G') + c_{G}^{aG'} G \gamma_\mu^a V^{G'\,\mu} \nn\\
& + \frac{1}{2} c_{ijG} S_i S_j G +\frac{1}{2} c_{iGG'} S_i G G' + \frac{1}{2} c_{iG}^{G'} V^{G'\, \mu} (G \partial_\mu S_i - S_i \partial_\mu  G) + c_i^{GG'} S_i V_{\mu}^G V^{G'\mu} \nn\\
& + c^{aGG'} \bigg(\partial^\mu \gamma_\nu^a V^{G}_\mu V^{G'\,\nu} + \gamma^{a\,\mu} \partial_\nu V_\mu^G V^{G'\,\nu}  + \gamma^{a}_\mu V^{G}_\nu\partial^\mu V^{G'\,\nu} \bigg)\nn\\
& + (c_{IJ}^{a,L}  \gamma^a_\mu + c_{IJ}^{G,L}  V^G_\mu) \ov{F}_I \gamma^\mu P_L F_J + (c_{IJ}^{a,R}  \gamma^a_\mu + c_{IJ}^{G,R}  V^G_\mu) \ov{F}_I \gamma^\mu P_R F_J \nn\\
& + c_{IJG}^L G \ov{F}_I P_L F_J + c_{IJG}^R G \ov{F}_I P_R F_J .
\end{align}
The couplings $c^a_{ij}, c_{GG'}^a, c_{IJ}^{a,L/R} $ are just generators of the unbroken gauge group in the appropriate representation multiplied by the unbroken gauge coupling. 
We find that we must enforce the following relations, which we derive in \appref{app:goldstones}:  
\begin{align}
c^{aGG'} =& c^a_{GG'} = \frac{1}{m_V^{G'}} c_{G}^{aG'}  \\
c_{ijG} =& \frac{1}{m_V^G} (m_i^2 - m_j^2) c_{ij}^G \\ 
c_{iGG'} =& \frac{m_i^2}{m_V^G m_V^{G'}} c_{i}^{GG'}, \qquad c_{iG}^{G'} = \frac{1}{m_V^G} c_i^{GG'}  \\
c_{IJG}^{L} =& \frac{1}{m_V^G} \big[  m_I c_{IJ}^{G, L} - m_J c_{IJ}^{G, R} \big], \qquad c_{IJG}^{R} = -\frac{1}{m_V^G} \big[  m_J^* c_{IJ}^{G, L} - m_I^* c_{IJ}^{G, R} \big]\,.
\end{align}
The implementation in \SARAH currently assumes that the gauge sector is that of the \sm{}; so there are no infra-red divergent diagrams
with neutral Goldstone bosons, and we do not shift their couplings to loop-corrected masses. 
In practice, a new set of Goldstone vertices is derived by the following relations which is then used in the calculation 
of the IR shifts.
\begin{align}
&c^L_{F_1 F_2 G^+} = \frac{m_{F_1} c^L_{F_1 F_2 W} - m_{F_2} c^R_{F_1 F_2 W}  }{m_W}\,,\qquad &
c^R_{F_1 F_2 G^+} = \frac{m_{F_1} c^R_{F_1 F_2 W} - m_{F_2} c^L_{F_1 F_2 W}  }{m_W} \label{EQ:practicalIRFF}\\
&c_{S_1 S_2 G^+} =  \frac{m_{S_1}^2-m_{S_2}^2}{m_W} c_{S_1 S_2 W}\,, \qquad
&c_{S G^+ W} = \frac{1}{2 m_W} c_{S W W}\\
&c_{G^+ W \gamma} = - m_W c_{W W \gamma}\,,
\end{align}
Note, that in \eqn{EQ:practicalIRFF} we explicitly assume no \cp{} violation.

Employing the outlined procedure we obtain partial decay widths at next-to-leading order
with full cancellation of ultraviolet and infra-red divergences.
Though the application of loop-corrected masses in the infra-red counter-term
can induce a spurious higher-order gauge dependence, for
phenomenological purposes this is however small, see e.g. \citeres{Liebler:2010bi,Liebler:2011tp}.
Note that for external heavy gauge bosons of the \sm{} we give the option to put the heavy gauge bosons on-shell, 
such that the cancellation of a gauge dependence in the real corrections among internal
gauge bosons and Goldstone bosons is always guaranteed.

\subsection{Mixing of species}
Particular attention is needed in the calculation of processes where self-energy diagrams allow for the
mixing between different particle species beyond tree-level. As an example (\cp{}-odd) Higgs bosons including
the neutral Goldstone boson can mix with the $Z$ boson and even the photon.
Then wave function corrections to the two-body decay come
with an internal propagator with a state different from the
external state. Such diagrams potentially need to sum up correctly to ensure 
a gauge-independent partial width.
For this purpose, in order to avoid unphysical poles the momenta flowing through the propagators have to match.
As an example,  \citere{Williams:2011bu} keeps tree-level masses
in diagrams mixing Higgs bosons and the $Z$ boson/Goldstone boson
in the calculation of Higgs decays to Higgs bosons. 
Slavnov-Taylor identities then ensure that the sum of the $Z$ and Goldstone contributions give zero, see also \citeres{Dabelstein:1995js,Krause:2016xku}.
Generally such
diagrams are beyond our generic implementation described here and require a
process-dependent treatment, i.e. they are not included. Still, in our setup they can be easily added.

\subsection{Loop-induced decays}
We finish with a remark about loop-induced decays like $F_i \to F_j \gamma$. 
Since infra-red divergences do not appear for these processes at the one-loop level,
there are fewer restrictions on which masses should be used to calculate the vertex one-loop diagrams.
As a default setting we therefore use loop-corrected masses everywhere.
The reason is that these decays are of particular importance in regions of kinematical
thresholds. Thus, the mass splitting between the two massive states should be
taken properly into account in the one-loop calculation. 
\section{Implementation in \SARAH}
\label{sec:usage}

\subsection{\SARAH--\SPheno interface}
The possibility to calculate one-loop decay widths is available from {\tt SARAH 4.11.0}. This is a new feature of the \SARAH interface to \SPheno which 
was established with {\tt SARAH 3.0.0}: \SARAH generates \Fortran code which can be compiled together with the standard \SPheno package 
to obtain a spectrum generator for a given model. The main features of a spectrum generator obtained in that way are a precise mass spectrum calculation
including two-loop corrections to real scalars \cite{Goodsell:2014bna,Goodsell:2015ira,Goodsell:2016udb}, 
a prediction for many precision and flavour observables \cite{Porod:2014xia} and up to now the 
calculation of two- and three-body decays mainly at tree-level. \\
The general procedure to obtain the \SPheno code for a given model starts with the download of the most recent \SARAH version from {\tt HepForge}:
\begin{lstlisting}
http://sarah.hepforge.org/ 
\end{lstlisting}
Then the user should copy the tar-file into a directory called {\tt \$PATH} in the following and extract it through:
\begin{lstlisting}[style=terminal]
tar -xf SARAH-4.11.0.tar.gz 
\end{lstlisting}
Afterwards, start \Mathematica, load \SARAH, run a  model {\tt \$MODEL} and generate a \SPheno version through
\begin{lstlisting}[style=mathematica]
<< $PATH/SARAH-4.11.0/SARAH.m;
Start["$MODEL"];
MakeSPheno[];
\end{lstlisting}
The last command initialises all necessary calculations and writes all \Fortran files into the output directory
of the considered model. These files can be compiled together with \SPheno version 3.3.8 or later. \SPheno is also available at {\tt HepForge}:
\begin{lstlisting}
http://spheno.hepforge.org/ 
\end{lstlisting}
The necessary steps to compile the new files are:
\begin{lstlisting}[style=terminal]
tar -xf SPheno-4.0.2.tar.gz
cd SPheno-4.0.2
mkdir $MODEL
cp -r $PATH/SARAH-4.11.0/Output/MODEL/EWSB/SPheno/* MODEL
make Model=$MODEL
\end{lstlisting}
This creates a new binary {\tt bin/SPheno\$MODEL} which reads all input parameters from an external file. 
\SARAH writes a template for this input file which can be used after filling it with numbers by typing:
\begin{lstlisting}[style=terminal]
./bin/SPheno$MODEL $MODEL/LesHouches.in.$MODEL 
\end{lstlisting}
The output is written to
\begin{lstlisting}
 SPheno.spc.$MODEL
\end{lstlisting}
and contains all running parameters at the renormalisation scale, the loop-corrected mass spectrum,
as well as all other observables calculated for the given model and parameter point. 

The time for generating the \Fortran code for the one-loop two-body decays as well as the compilation time of \SPheno are extended by these new 
routines. Therefore, in case that the user in not interested in the loop-corrected two-body decays, they can be turned off via:
\begin{lstlisting}[style=mathematica]
MakeSPheno[IncludeLoopDecays->False]; 
\end{lstlisting}
They can be permanently turned off for a given model by adding
\begin{lstlisting}[style=file]
SA`AddOneLoopDecay = False;
\end{lstlisting}
to {\tt SPheno.m}.
Usually, the calculation of the one-loop decays triggers also the calculation of the RGEs even when using the 
option ``{\tt OnlyLowEnergySPheno = True;}'' to generate the \SPheno code. 
The reason is that the $\beta$-functions are used to check the cancellation of ultraviolet divergences. 
However, for non-supersymmetric models, in particular in the presence of many quartic couplings in the scalar potential, the RGE calculation
can be very time-consuming. In this case the option
\begin{lstlisting}[style=file]
SA`NoRGEsforDecays=True; 
\end{lstlisting}
skips the RGE calculation. Of course, the verification of the cancellation of ultraviolet divergences will not be performed with this setting. 

\subsection{Definition of counter-terms}
\label{sec:setuploopdecays}
We included the possibility to define counter-terms
to be used in the calculation of the one-loop decays. This is done in {\tt SPheno.m} via the new array {\tt RenConditionsDecays}.
For instance, the standard renormalisation conditions for the electroweak gauge couplings are set via:
\begin{lstlisting}[style=file]
RenConditionsDecays={
{dCosTW, 1/2*Cos[ThetaW] * (PiVWm/(MVWM^2) - PiVZ/(mVZ^2))},
{dSinTW, -dCosTW/Tan[ThetaW]},
{dg2, 1/2*g2*(derPiVPheavy0 + PiVPlightMZ/MVZ^2
 - 2*dSinTW/Sin[ThetaW] + (2*PiVZVP*Tan[ThetaW])/MVZ^2)},
{dg1, dg2*Tan[ThetaW]+g2*dSinTW/Cos[ThetaW]
 - dCosTW*g2*Tan[ThetaW]/Cos[ThetaW]}
};
\end{lstlisting}
We give an example for the application of the above electroweak counter-terms and their derivation in \sct{sec:renormalpha}.
If {\tt RenConditionsDecays} is not defined, a pure \msbar{}/\drbar{} renormalisation for the bare parameters
of the underlying model is performed. 
The counter-terms can also be turned on/off in the numerical session via new flags in the \SPheno 
input file as explained in the next subsection.
The conventions are:
\begin{itemize}
 \item The names for the counter-terms are the names of the corresponding parameter starting with {\tt d}.
 \item For a rotation angle {\tt X}, no counter-term for the angle itself is introduced, but for the trigonometric functions involving that angle. Those
 are called  {\tt dCosX}, {\tt dSinX} and {\tt dTanX}. 
\end{itemize}
The following objects can be used to define the counter-terms:
\begin{itemize}
 \item {\bf Parameters of the model}: the internal \SARAH names must be used.
 \item {\bf Masses of particles in the model}: those are called {\tt MX} where {\tt X} is the name of the particles in \SARAH.
 \item {\bf Self-energies for scalars and vector bosons}: those are called {\tt PiX} where {\tt X} is the name of the particles in \SARAH.
 \item {\bf The derivatives of self energies of scalars and vector bosons}: those are called {\tt derPiX} where {\tt X} is the name of the particles in \SARAH.
 \item {\bf Self-energies and their derivatives mixing vector bosons}: those are called {\tt PiXY} respectively {\tt derPiXY} where {\tt X} and {\tt Y} are the names 
 of the particles in \SARAH and $p^2=m_Y^2$.
 \item {\bf Special self-energies for vector bosons containing only light/heavy states}:
 \begin{itemize}
  \item {\tt PiVPlight0}/{\tt derPiVPlight0}: only light degrees of freedom are included in the loops; external momentum $p^2=0$.
  \item {\tt PiVPlightMZ}/{\tt derPiVPlightMZ}: only light degrees of freedom are included in the loops; external momentum $p^2=m_Z^2$.
  \item {\tt PiVPheavy0}/{\tt derPiVPheavy0}: only heavy degrees of freedom are included in the loops; external momentum $p^2=0$.
  \item {\tt PiVPheavyMZ}/ {\tt derPiVPheavyMZ}: only heavy degrees of freedom are included in the loops; external momentum $p^2=m_Z^2$.
 \end{itemize}
 \item {\bf The different parts of the fermion self-energies and their derivatives}: those are called {\tt SigmaLX}, {\tt SigmaRX}, {\tt SigmaSLX}, {\tt SigmaSR} respectively 
 {\tt DerSigmaLX}, {\tt DerSigmaRX}, {\tt DerSigmaSLX}, {\tt DerSigmaSR}, where {\tt X} is the name of the particles in \SARAH.
\end{itemize}
When \SARAH is finished generating the \SPheno output, a list of all self-energies and their derivatives which are available in \SPheno
is stored in {\tt SA`SelfEnergieNames}, and the names for all counter-terms are saved in {\tt SA`ListCounterTerms}. \\
One needs to be careful when using self-energies or their derivatives for particles which come with several generations.
In this case, the objects defined above are arrays with three indices. The last two indices give the involved generations, the first one the external momentum, e.g.
\begin{align}
\Pi_{ij}(m^2_{S_k}) & \to \quad {\tt PiS(k,i,j)} \\ 
\Sigma^L_{ij}(m^2_{F_k}) & \to \quad {\tt SigmaLF(k,i,j)}\,.
\end{align}
When defining the counter-terms, commands for matrix or tensor operators should already be evaluated in {\tt Mathematica}. Although we offer the possibility to the user to define 
counter-terms in that way, we want to stress that it has not been tested in practice beyond the examples given in this paper.
Thus, this option should be used carefully and the results should be tested throughout, e.g. the ultraviolet finiteness of
the partial decay widths is a first test to be performed. Again we emphasize that such counter-terms are for now
only applied in the calculation of decay widths. Thus, on-shell prescriptions for the calculation of masses
as e.g. known from the neutralino and chargino sector, see \citeres{Eberl:2001eu,Fritzsche:2002bi,Baro:2009gn,Liebler:2010bi}, can not yet be incorporated.\\

\subsection{Options for the evaluation with \SPheno}
There are several options to steer the performed one-loop calculations which can be controlled via the block {\tt DECAYOPTIONS} in the Les Houches input file for \SPheno. 
In practice the most important options are:
\begin{lstlisting}[style=file]
Block DECAYOPTIONS #
...
1001 ... # One-loop decays of particle X
1002 ... # One-loop decays of particle Y
...
1114 ... # U-factors (0: off, 1:p2_i=m2_i, 2:p2=0, p3:p2_i=m2_1) 
1115 ... # Use loop-corrected masses for external states
1116 ... # OS values for W,Z and fermions
           (0: off, 1:g1,g2,v 2:g1,g2,v,Y_i) 
1117 ... # Use defined counter-terms 
1118 ... # Use loop-corrected masses for loop-induced decays 
\end{lstlisting}
\newpage
The following settings are possible:
\begin{itemize}
 \item {\tt DECAYOPTIONS[10XY]}: the one-loop decays for each particle can individually be turned 
 on ({\tt 1}) or off  ({\tt 0}) via these flags. The particle to which a given flag corresponds to is written as comment by \SARAH. The default value is {\tt 1}.
 \item {\tt DECAYOPTIONS[1114]}: this defines the choice for the external \Ufactors{}:
  \begin{itemize}
   \item {\tt 0}: no \Ufactors{} are applied.
   \item {\tt 1}: the \Ufactors{} including the full $p^2$ dependence are used ($U^p$).
   \item {\tt 2}: the \Ufactors{} calculated for $p^2=0$ are used ($U^0$).
   \item {\tt 3}: the \Ufactors{} are calculated from the loop-corrected rotation matrix for the lightest mass eigenstate ($U^{m_1}$).
  \end{itemize}
  The default value is {\tt 1}.
 \item {\tt DECAYOPTIONS[1115]}: 
  \begin{itemize}
   \item {\tt 0}: the kinematics is done with tree-level masses.
   \item {\tt 1}: the kinematics is done with loop-corrected masses.
  \end{itemize}
  The default value is {\tt 1}.
 \item {\tt DECAYOPTIONS[1116]}: 
  \begin{itemize}
   \item {\tt 0}: \msbar{}/\drbar{} parameters are used for gauge couplings, $v$ and Yukawa couplings.
   \item {\tt 1}: $g_1$, $g_2$ and $v$ are set to reproduce the measured values of $M_Z$, $\alpha_{ew}(M_Z)$ and $\sin\Theta_W$.
   \item {\tt 2}: same as {\tt 1}, but in addition the Yukawa couplings are set to reproduce the measured values of SM fermions.
  \end{itemize}
  The default value is {\tt 0}.
 \item {\tt DECAYOPTIONS[1117]}: 
  \begin{itemize}
   \item {\tt 0}: the counter-terms defined in {\tt RenConditionsDecays} are not used.
   \item {\tt 1}: the counter-terms defined in {\tt RenConditionsDecays} are used.
  \end{itemize}
  The default value is {\tt 0}.
 \item {\tt DECAYOPTIONS[1118]}: 
  \begin{itemize}
   \item {\tt 0}: for loop-induced decays tree-level masses are used.
   \item {\tt 1}: for loop-induced decays loop-corrected masses are used.
  \end{itemize}
  The default value is {\tt 1}.
\end{itemize}
In addition, the following options exist which are mainly supposed to be used for testing and validation of the virtual and real corrections:
\begin{lstlisting}[style=file]
Block DECAYOPTIONS #
...
1101 ... # Only ultraviolet divergent parts of integrals 
1102 ... # Ultraviolet divergence 
1103 ... # Debug information 
1104 ... # Only tree-level decay widths
...
1201 ... # Photon/Gluon regulator mass 
1205 ... # Renormalisation scale
\end{lstlisting}
The following settings are possible:
\begin{itemize}
 \item {\tt DECAYOPTIONS[1101]}: this option can be used to check the cancellation of ultraviolet divergences.
 \begin{itemize}
   \item {\tt 0}: One-loop functions employed in the calculation of one-loop decay widths return the finite part and the ultraviolet divergence defined in {\tt DECAYOPTIONS[1102]}.
   \item {\tt 1}: Only the ultraviolet divergence defined in {\tt DECAYOPTIONS[1102]} is returned.
  \end{itemize}
  The default value is {\tt 0} .
 \item {\tt DECAYOPTIONS[1102]}: this option can be used to check the cancellation of ultraviolet divergences.
 {\tt X} sets the value used for the ultraviolet divergence $\Delta$ defined in \sct{sec:vertexcorrections}. The default value is {\tt 0}.
 \item {\tt DECAYOPTIONS[1103]}:
 \begin{itemize}
   \item {\tt 0}: No debug information is shown.
   \item {\tt 1}: Additional information is shown on the screen. This includes the individual contributions from vertex, wavefunction and real corrections,
   which are useful to check the cancellation of ultraviolet and infra-red divergences.
  \end{itemize}
  The default value is {\tt 0}.
  \item {\tt DECAYOPTIONS[1104]}: this option can be used to check the consistency between the tree-level and one-loop calculation of decay widths.
 \begin{itemize}
   \item {\tt 0}: The one-loop routines return decay widths at \nlo{}.
   \item {\tt 1}: The one-loop routines only return the tree-level decay widths, which can be compared to the tree-level results contained in Block {\tt DECAY}.
  \end{itemize}
    The default value is {\tt 0}.
 \item {\tt DECAYOPTIONS[1201]}: this option can be used to check the cancellation of infra-red divergences.
 {\tt X} defines the value (in GeV) used for the photon/gluon mass. The default value is {\tt 1.0E-5}.
  Note that this option does not work with loop-corrected masses. The user should
  ensure that the regulator mass dependence of vertex and wavefunction corrections cancels against the one of the
  real corrections and yields a regulator-mass independent decay width. In order to show the individual contributions {\tt DECAYOPTIONS[1103]}
  should be set to {\tt 1}.
 \item {\tt DECAYOPTIONS[1205]}: this option can be used to check the renormalisation scale dependence.
  If defined, {\tt X} sets the value (in GeV) used for the renormalisation scale~$Q$ in all one-loop functions employed in the calculation of decay widths.
  The default option is to use the same renormalisation scale as used in the calculation of masses, see \sct{sec:wvcorrections}.
\end{itemize}

\subsection{Output of \SPheno}
The results of the one-loop calculation of decay widths are written in the \SPheno output. For this purpose,
we introduced the keyword {\tt DECAY1L} beside the standard Block {\tt DECAY} which lists the 
results of the `old', i.e. leading order, calculation. Thus, for an arbitrary \mssm{} point, the output file contains:
\begin{lstlisting}[style=file]
DECAY   1000001     5.03001929E+01   # Sd_1
#  BR              NDA     ID1      ID2
   2.91393772E-01  2       6  -1000024   # BR(Sd_1 -> Fu_3 Cha_1)
   1.70527978E-01  2       6  -1000037   # BR(Sd_1 -> Fu_3 Cha_2)
   2.29399038E-04  2       3   1000023   # BR(Sd_1 -> Fd_2 Chi_2)
   7.62216405E-03  2       5   1000022   # BR(Sd_1 -> Fd_3 Chi_1)
   1.47930395E-01  2       5   1000023   # BR(Sd_1 -> Fd_3 Chi_2)
   1.87674294E-03  2       5   1000025   # BR(Sd_1 -> Fd_3 Chi_3)
   2.08779423E-03  2       5   1000035   # BR(Sd_1 -> Fd_3 Chi_4)
   3.78314020E-01  2 1000002       -24   # BR(Sd_1 -> Su_1 VWm )
...
DECAY1L   1000001     5.07518318E+01   # Sd_1
#  BR              NDA     ID1      ID2
   2.86487000E-01  2       6  -1000024   # BR(Sd_1 -> Fu_3 Cha_1)
   1.63886304E-01  2       6  -1000037   # BR(Sd_1 -> Fu_3 Cha_2)
   2.35164668E-04  2       3   1000023   # BR(Sd_1 -> Fd_2 Chi_2)
   6.82861774E-03  2       5   1000022   # BR(Sd_1 -> Fd_3 Chi_1)
   1.50793873E-01  2       5   1000023   # BR(Sd_1 -> Fd_3 Chi_2)
   1.91193810E-03  2       5   1000025   # BR(Sd_1 -> Fd_3 Chi_3)
   2.14376064E-03  2       5   1000035   # BR(Sd_1 -> Fd_3 Chi_4)
   3.87696880E-01  2 1000002       -24   # BR(Sd_1 -> Su_1 VWm )
\end{lstlisting}
Although this block {\tt DECAY1L} is not officially supported by the Les Houches conventions, there are the following
reasons not to overwrite the results of the `old' calculation:
\begin{itemize}
 \item The sizes of the one-loop corrections are immediately apparent.
 \item The results given in {\tt DECAY} are not only pure tree-level decay widths, but include in particular for the
 Higgs decays crucial higher order corrections adapted from literature. Those are beyond the one-loop corrections
 which we can provide in the new automatised framework at the moment. 
 \item The `old' calculations also include  tree-level three-body decays. We leave the choice to the user
 how to combine them with the two-body decay widths obtained at the one-loop level.
\end{itemize}

\section{Numerical results}
\label{sec:results}

We start this section with two examples for the calculation of two-body decay width
in the \sm{}, where we demonstrate the relevance of model- and process-dependent counter-terms.
Our default implementation makes use of an \msbar{} or \drbar{} renormalisation
of all parameters of the underlying theory. However, for many processes different schemes
are actually better suited. This is particularly true for the calculation of electroweak corrections.
For this purpose the user of \SARAH can define their own counter-terms, as outlined in \sct{sec:setuploopdecays}.
We show two simple examples in the \sm{}, namely
the calculation of the partial decay width $t\to Wb$ and $H\to b\bar b$.
In the first example we discuss different schemes for the renormalisation of the electric charge,
in the second example we show that our \msbar{} renormalisation for the bottom-quark
Yukawa coupling is actually sufficient.
After these examples we continue with a detailed comparison of our implementation
with existing codes, among them \SFOLD, \HFOLD and \CNN. Whereas \SFOLD and \HFOLD
are also based on a \drbar{} renormalisation of the parameters of the \mssm{}, the code
\CNN calculates neutralino and chargino decays in the \mssm{}, \nmssm{} and in models
with $R$-parity violation again renormalising the electric charge in the Thomson limit.
Thereafter, we compare loop-induced decays with the original implementation of \SPheno
and lastly show the effect of \Ufactors{} in the calculation of two-body decay widths.
A more thorough comparison for Higgs boson decays is left for future work.

\subsection{Renormalisation of $\alpha$ and the top-quark width}
\label{sec:renormalpha}

First we perform a calculation of the top-quark partial width
in the decay $t\to Wb$ including electroweak and \qcd{} corrections
using a \sm{} version of \SPheno.
Since this process is mediated through the gauge coupling $g_2$ of SU$(2)_L$
at tree-level, we will discuss the renormalisation of $g_2$ in this context.
We choose the following input parameters
\begin{align}
m_t=173.3\,\GeV,\quad m_b=4.75\,\GeV,\quad m_W=80.350\,\GeV\\
\alpha_s(m_Z)=0.1187,\quad \alpha(m_Z)=1/127.9,\quad V_{tb}=1\,.
\end{align}
We neglect quark mixing (i.e. the \ckm{} matrix is approximated by the identity matrix). Note that in a more general approach
the renormalisation prescription introduced in \eqn{eq:CTrot} can be applied to quark mixing.
Subsequently we employ
external tree-level masses without running, i.e. we effectively calculate
with on-shell masses for all three involved particles (setting flag {\tt SPHENOINPUT[61]}$=$0 to disable 
the RGE running for the parameters and and flag {\tt DECAYOPTIONS[1116]}$=$2 to use on-shell mass
values). This also fixes $g_1, g_2$ and $v, m_W$ from $G_F, m_Z$ and $\alpha(m_Z)$.
Our simple \msbar{} scheme for the renormalisation of $g_2$ (named scheme $(1)$) yields
\begin{align}
 (1) \quad\delta g_2 = -\frac{1}{16\pi^2}\frac{19}{12}g_2^3\Delta\quad.
\end{align}
Next, we provide the decay width for the renormalisation of the electric charge
in the Thomson limit of the $ff\gamma$-vertex, i.e. at zero momentum transfer~\cite{Denner:1991kt}.
The counter-terms for the electric charge are given by (see \citere{Liebler:2011qsa} for an overview)
\begin{align}
&(2) \quad\alpha(0) \quad\text{and}\quad \delta Z_e(0)=\frac12 \dot \Pi_{\gamma\gamma}(0)-\frac{\tan \theta_W}{m_Z^2}\Pi_{Z\gamma}(0)\\
&(3) \quad\alpha(m_Z) \quad\text{and}\quad \delta Z_e(m_Z)=\delta Z_e(0)-\frac12 \dot \Pi_{\gamma\gamma,\text{light}}(0)+\frac{1}{2m_Z^2}\widetilde{\Re}\Pi_{\gamma\gamma,\text{light}}(m_Z^2)\,,
\end{align}
where we distinguish two schemes: At \nlo{} we can make use of the very precise value of $\alpha(0)$
together with the corresponding counter-term $ \delta Z_e(0)$ or we employ $\alpha(m_Z)$ and compensate 
for the shift through the additional terms in $\delta Z_e(m_Z)$. The relevant self-energies include only contributions from
light fermions.
Ultimately we also need the renormalisation of the weak mixing angle, which is given by
\begin{align}
\delta\cos\theta_W=\frac12 \cos\theta_W\left(\frac{1}{m_W^2}\Pi_{WW}(m_W^2)-\frac{1}{m_Z^2}\Pi_{ZZ}(m_Z^2)\right)=-\tan\theta_W\delta\sin\theta_W\,,
\end{align}
such that in schemes $(2)$ and $(3)$ we obtain
\begin{align}
\delta g_2 = \left(\delta Z_e-\frac{\delta\sin\theta_W}{\sin\theta_W}\right)g_2\,.
\end{align}
In this section we keep $\alpha$ and $\alpha_s$ fixed as a function of the renormalisation scale~$Q$
and therefore schemes $(2)$ and $(3)$ lead to a scale-independent partial decay width, whereas
scheme $(1)$ develops a scale dependence. Also, in scheme $(1)$ it is a priori not clear
whether $\alpha(0)$ or $\alpha(m_Z)$ is the better choice, so we provide both values.
The results are shown in Table~\ref{tab:tWb}. We include the \lo{} partial width as well
as the \nlo{} partial width only including electroweak and including electroweak and
\qcd{} corrections. We also provide (absolute and) relative corrections in brackets. For 
$\Gamma^{\text{\nlo,EW+\qcd}}_{t\to Wb}$ they are with respect to the partial width~$\Gamma^{\text{\nlo,EW}}_{t\to Wb}$.

\begin{table}
\begin{center}
\begin{tabular}{l|ccc}
Scheme & $\Gamma^{\text{\lo}}_{t\to Wb}$ [GeV] & $\Gamma^{\text{\nlo,EW}}_{t\to Wb}$ [GeV] & $\Gamma^{\text{\nlo,EW+\qcd}}_{t\to Wb}$ [GeV] \\
\hline
$(1)$, $\alpha(0)$,   $Q=173$\,GeV & $1.443$ & $1.487$ $[+3.0\%]$ & $1.352$ $[-0.135][-9.1\%]$\\
$(1)$, $\alpha(m_Z)$, $Q=173$\,GeV & $1.546$ & $1.596$ $[+3.2\%]$ & $1.452$ $[-0.144][-9.0\%]$\\
$(2)$, $\alpha(0)$                   & $1.443$ & $1.519$ $[+5.3\%]$ & $1.384$ $[-0.135][-8.9\%]$\\
$(3)$, $\alpha(m_Z)$                 & $1.546$ & $1.522$ $[-1.6\%]$ & $1.378$ $[-0.144][-9.5\%]$\\
\end{tabular}
\caption[]{\label{tab:tWb} Partial decay width $t\to Wb$ in different schemes, see text for details.}
\end{center}
\end{table}

For scheme~$(1)$ evaluated with $\alpha(m_Z)$ we obtain $\Gamma_{t\to Wb}^{\text{\nlo{},EW+\qcd{}}} = 1.434~\GeV$ at $Q=90$\,GeV,
which demonstrates that the renormalisation scheme dependence is not very pronounced.
The absolute \qcd{} correction remains constant (for fixed values of $\alpha$ and $\alpha_s$)
and yields $\sim -9$\% as expected~\cite{Jezabek:1988iv,Jezabek:1993wk}.
It is apparent that schemes $(2)$ and $(3)$ yield very comparable results
at \nlo{} despite the different input values for $\alpha$. This is due to the compensation
through the shift in the counter-term, which guarantees that the electric charge is renormalised in the Thomson limit.
In contrast scheme~$(1)$ shows a significant dependence on the input value, where not surprisingly
the choice $\alpha(0)$ comes closer to the results in schemes $(2)$ and $(3)$.

We conclude that for precision predictions the proper renormalisation
of certain parameters is rather important.
The relevant counter-terms for scheme $(3)$ can be defined by the user in the \SARAH framework
as discussed in \sct{sec:setuploopdecays} through {\tt RenConditionsDecays}.
Note that such counter-terms will only apply at the moment to the calculation of decay widths,
not to the calculation of masses.

\subsection{Renormalisation of Yukawa couplings and fermionic Higgs decays}

We also shortly discuss the calculation of $H\to b\bar b$ in the \sm{}, which is mediated
through the bottom-quark Yukawa coupling~$Y_b$, and it turns out that the \msbar{} renormalisation
of $Y_b$ is the preferred choice. A priori, we would expect that the calculation of \nlo{}
electroweak corrections~\cite{Fleischer:1980ub,Bardin:1990zj,Dabelstein:1991ky}
would again be optimally performed using an on-shell renormalisation of all parameters involved. The counter-term of
the bottom-quark Yukawa coupling in the on-shell case is given by
\begin{align}
 \delta Y_b^{\rm os}=\frac{1}{v}\left(\sqrt{2}\delta m_b - Y_b\delta v\right)\,,
\end{align}
such that renormalisation prescriptions for $\delta m_b$ and $\delta v$ are needed.
Whereas $\delta m_b$ can be obtained from the self-energies of the down-type quarks,
the on-shell renormalisation of the vacuum expectation value depends on other parameters:
one requests that renormalised tadpoles vanish as well as the on-shell renormalisation of the Higgs
mass and the Higgs self coupling, see e.g. \citere{Dabelstein:1991ky}. Also, such
counter-terms can be implemented in principle through {\tt RenConditionsDecays}.
However, it turns out that electroweak corrections are small ($\sim 1\%$) for a Higgs mass of $m_H=125$\,GeV.
In contrast \qcd{} corrections are much larger and for them the renormalisation of
the Yukawa coupling in the \msbar{} scheme is more convenient, since it resums large logarithms~\cite{Braaten:1980yq,Drees:1990dq}.
We demonstrate this effect in Table~\ref{tab:Hbb}, which is obtained with the \sm{} version of \SPheno
setting flag {\tt DECAYOPTIONS[1116]}$=$1 and flag {\tt SPHENOINPUT[61]}$=$1. Through these settings $Y_b$ as well as the gauge couplings, in particular $\alpha_s$,
are evaluated at the renormalisation scale~$Q$ and for $Y_b$ the \msbar{} scheme is employed.
We again depict the \lo{} as well as the \nlo{} partial width
with only electroweak as well as electroweak and \qcd{} corrections including relative corrections.
The most relevant parameters are $m_H=125$\,GeV, $m_b(m_b)=4.18$\,GeV and $\alpha_s(m_Z)=0.1187$.
The running to $Q=m_H$ yields $Y_b\propto m_b(m_H)=2.781$\,GeV and $\alpha_s(m_Z)=0.1133$.

\begin{table}
\begin{center}
\begin{tabular}{l|ccc}
Scale & $\Gamma^{\text{\lo}}_{H\to b\bar b}$ [MeV] & $\Gamma^{\text{\nlo,EW}}_{H\to b\bar b}$ [MeV] & $\Gamma^{\text{\nlo,EW+\qcd}}_{H\to b\bar b}$ [MeV] \\
\hline
$Q=125$\,GeV  & $1.959$ & $1.972$ $[+0.6\%]$ & $2.376$ $[+20.5\%]$\\
$Q=62.5$\,GeV & $2.188$ & $2.215$ $[+1.5\%]$ & $2.473$ $[+11.6\%]$\\
$Q=250$\,GeV  & $1.778$ & $1.783$ $[+0.3\%]$ & $2.280$ $[+27.8\%]$\\
\end{tabular}
\caption[]{\label{tab:Hbb} Partial decay width $H\to b\bar b$ for different values of $Q$, see text for details.}
\end{center}
\end{table}

We see from Table~\ref{tab:Hbb} that the electroweak corrections are indeed small. The \qcd{} corrections coincide with the term found
in the literature, being $5.667\alpha_s(m_H)/\pi\sim 20.4\%$~\cite{Chetyrkin:1996sr}. The depicted
scale dependence can be used to estimate the remaining uncertainties, which can be
significantly reduced by including higher order \qcd{} corrections beyond one-loop level.

\subsection{Comparison with other codes}

In order to further validate our calculations and implementations, we compared the obtained results for the \mssm{}
and in $R$-parity violating models against other public tools. 

Our comparison is two-fold: we compared neutralino and chargino decays into neutralinos and charginos and heavy gauge
bosons with \CNN, where we employ a full on-shell scheme for the gauge couplings, but work with tree-level neutralino
and chargino masses. We use the counter-terms for the electroweak sector as
outlined in \sct{sec:setuploopdecays}. Given that we adjust all input parameters to be identical, we can therefore
exactly reproduce the results of \CNN.

Secondly, we made use of the three codes \SFOLD, \HFOLD and \FVSFOLD which also use a \drbar{} scheme for the renormalisation
of the parameters of the \mssm{}
to calculate one-loop corrections to two-body decays.
Since these codes do not make use of external \Ufactors, we have turned them 
off in our evaluation with \SPheno. In addition, we forced all codes to use tree-level (\drbar{}) masses in all loops and for the kinematics. 
Thus, the partial widths and the size of the loop corrections presented in the following might be of limited physical interest since the
inclusion of \Ufactors{} and external loop-corrected masses can change the results substantially, see also \sct{sec:externalufactors}.
Thus in this section our aim is to only demonstrate the agreement (and disagreement) between the codes.
For the comparison with \SFOLD, \HFOLD and \FVSFOLD
we have chosen a parametrisation for the general \mssm{} which depends only on one dimensionful parameter $m$ as follows:
\begin{align}\nonumber
& M_1 =  0.3 m\,, \quad M_2 = 0.75 m\,,\quad M_3 = 2.5 m & \\\nonumber
& \mu = 0.5 m\,, \quad M_A^2 = 3 m^2 & \\\nonumber
& m_{\tilde d,11}^2 = m^2\,,\quad  m_{\tilde d,22}^2 = m^2\,,\quad m_{\tilde d,33}^2 = 0.5 m^2 & \\\nonumber
& m_{\tilde u,11}^2 = m^2\,,\quad  m_{\tilde u,22}^2 = m^2\,,\quad m_{\tilde u,33}^2 = 0.5 m^2 & \\\nonumber
& m_{\tilde q,11}^2 = m_{\tilde q,22}^2 = m^2\,,\quad m_{\tilde q,33}^2 = 2 m^2 & \\\nonumber
& m_{\tilde e,11}^2 =  m_{\tilde e,22}^2 = m_{\tilde e,33}^2 = 0.25 m^2 & \\\nonumber
& m_{\tilde l,11}^2 =  m_{\tilde l,22}^2 = 0.25 m^2\,,\quad m_{\tilde l,33}^2 = m^2 & \\
& T_{u,33} = m\,,\quad T_{e,33} = 0.5 m& 
\end{align}
All other soft-terms are set to zero. This parametrisation has no physical motivation but was chosen in a way to open many different decay 
channels to be compared among the codes. In addition, we fixed $\tan\beta=10$. We show results for the predicted partial widths when varying
$m$ from $300$\,GeV to $2500$\,GeV.

We also performed a comparison for the loop-induced neutralino and gluino decays which were already implemented in \SPheno. 
The details of this comparison and the outcome are summarised in \sct{sec:ComparisonSPheno}.

\subsubsection{Neutralino and Chargino decays in the \mssm{} and in bilinear $R$-parity violation: \CNN vs. \SARAH}
\label{sec:cnnvsspheno}

We compared the decay modes $\tilde{\chi}_i^\pm\to \tilde{\chi}_j^0 W^\pm$ and $\tilde{\chi}_i^0\to \tilde{\chi}_j^\mp W^\pm$ as well as $\tilde{\chi}_i^0\to \tilde{\chi}_j^0 Z$ and
$\tilde{\chi}_i^\pm \to \tilde{\chi}_j^\pm Z$ in the $R$-parity conserving \mssm{} and adding bilinear $R$-parity violation.
Bilinear $R$-parity violation allows an explanation of neutrino masses, but makes the lightest supersymmetric
particle (\lsp{}) unstable. Since its decay modes are related to the $R$-parity violating parameters, which are small in order to explain the size of 
neutrino masses, the decay width of the \lsp{} is also small.
We remain with real parameters, both in the \mssm{} as well as for the $R$-parity breaking parameters.
We adjust the tree-level masses and mixing as well as the gauge couplings $g_1$ and $g_2$ to be exactly identical in both codes
and also ensure to choose the same renormalisation scale (through {\tt DECAYOPTIONS[1205]}), namely $Q=m_Z$.
Since we employ the renormalisation of the electric charge in the Thomson limit as outlined in \sct{sec:setuploopdecays},
the partial decay widths are in principle all renormalisation-scale independent, however we also want to compare
wavefunction and vertex corrections individually.
We find full agreement between both codes, i.e. numerically identical results beyond $8$ digits in the \mssm{}. In particular this is also true
for the vertex and wavefunction corrections individually as well as the individual pieces to the counter-terms.
Also for the $R$-parity violating decays $\tilde{\chi}_4^0\to l^\mp W^\pm$ in bilinear $R$-parity violation we find agreement 
at the per mille level; the smallness of couplings and masses makes those decay modes more sensitive to numerical errors  (factors too small or large for the precision of the code).
Decay modes into light neutrinos and a gauge boson or a scalar like e.g. $\tilde{\chi}_4^0\to \nu Z$ or $\nu S$, which are of relevance for $R$-parity violating scenarios,
suffer from bad numerical errors.
Therefore, neutrino masses, which are analytically zero at tree-level, are set to zero in the calculation of one-loop decays.
In contrast, the mixing matrix of neutrinos
(and neutralinos) remains exact, such that the associated error is small.
As we already explained a detailed check of \cp{}-violating scenarios as discussed in \citeres{Heinemeyer:2011gk,Bharucha:2012re,Cheriguene:2014bxa} is left for future work
for the reasons explained in \sct{sec:limitations}.

\subsubsection{Sfermion decays in the \mssm{}: \SARAH vs. \SFOLD}
\begin{figure}[t]
\includegraphics[width=0.45\linewidth]{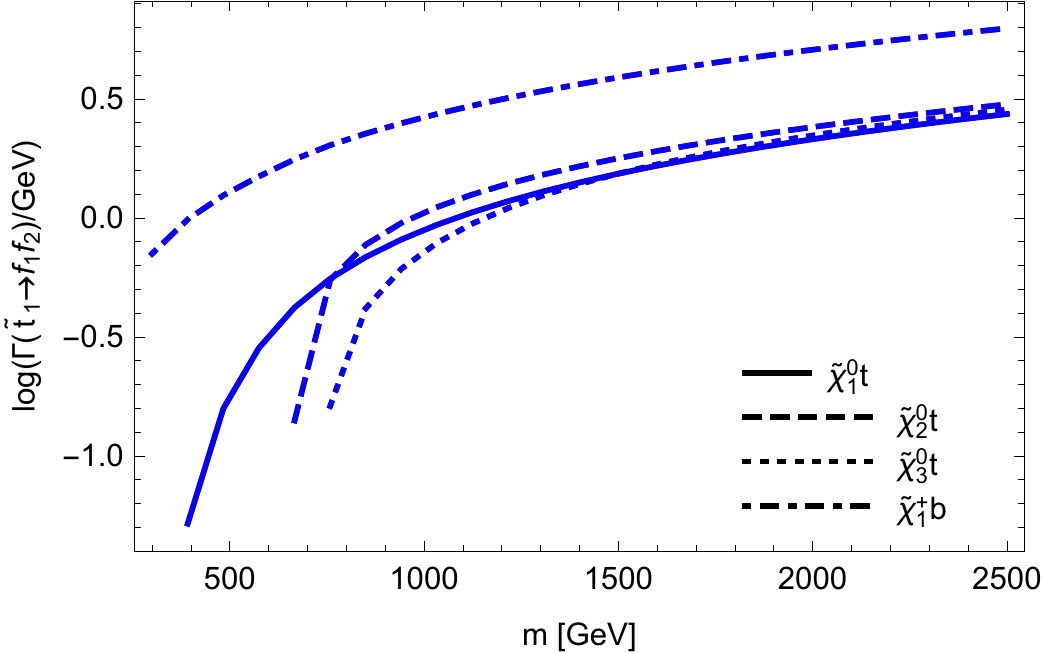} \hfill 
\includegraphics[width=0.45\linewidth]{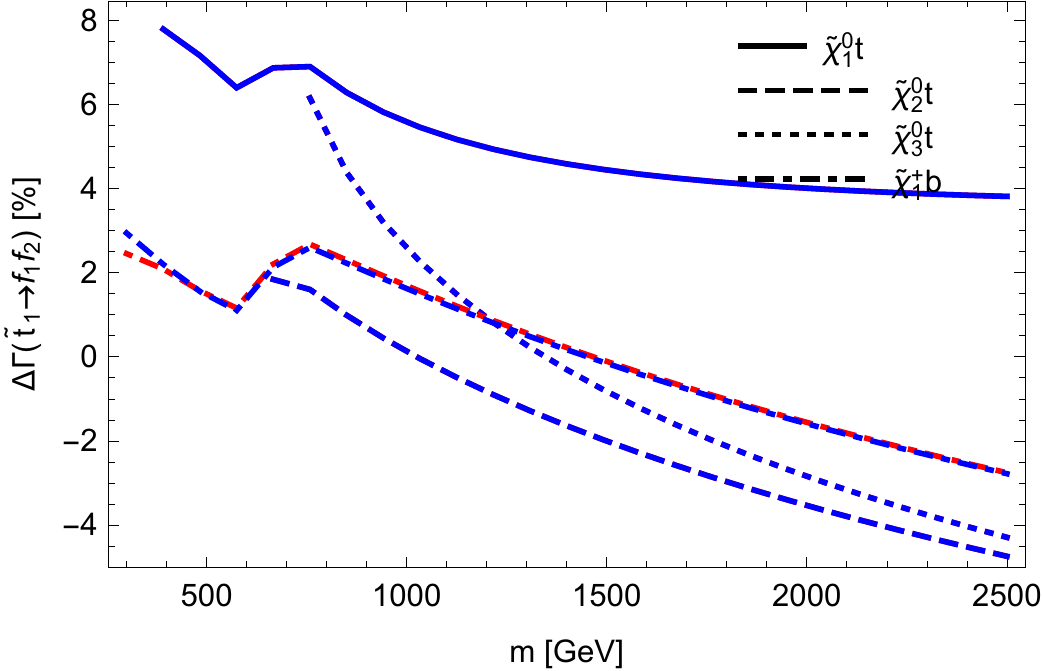}  \\
\includegraphics[width=0.45\linewidth]{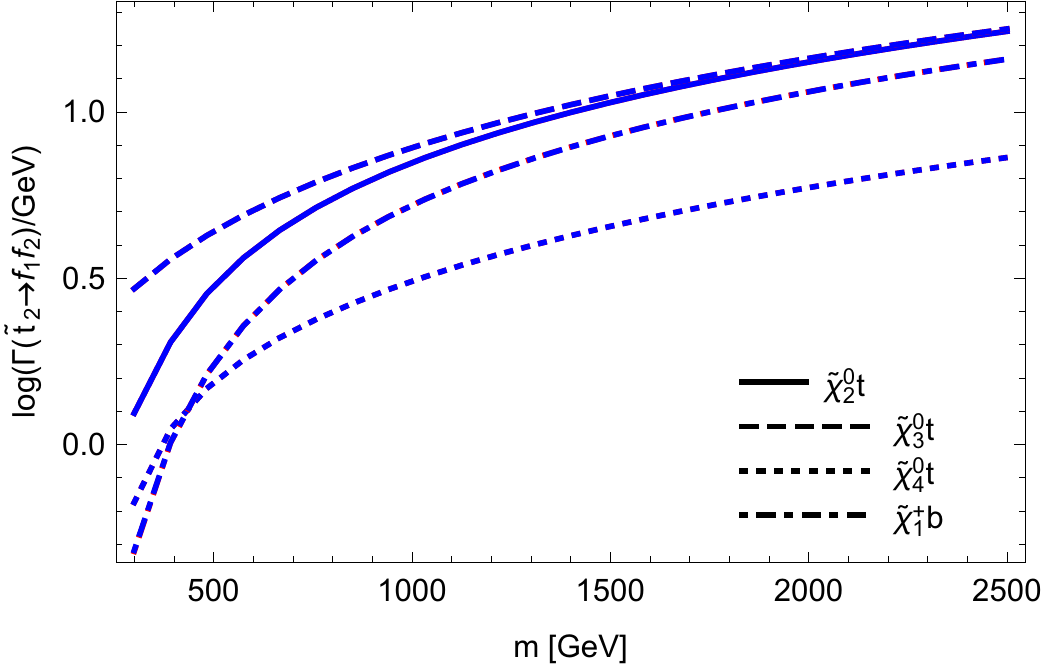} \hfill 
\includegraphics[width=0.45\linewidth]{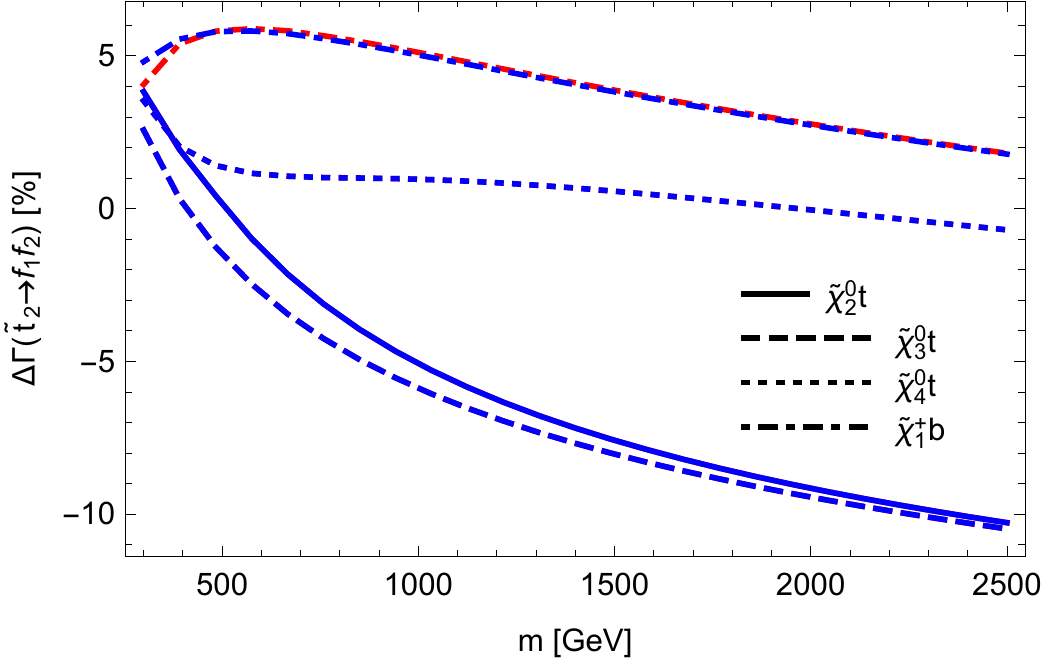}  \\
\includegraphics[width=0.45\linewidth]{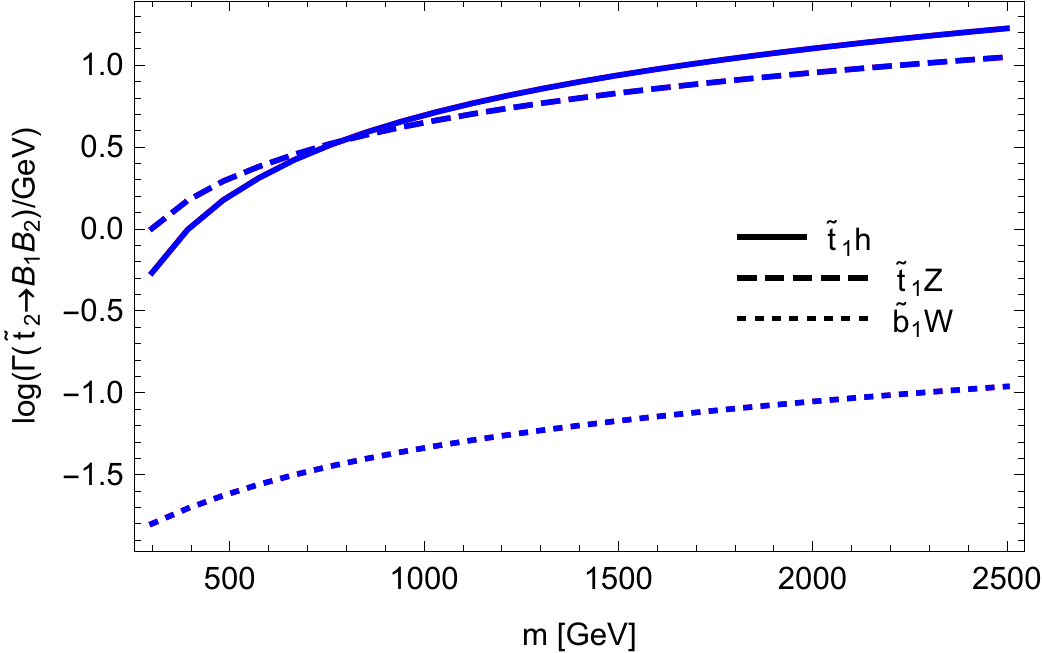} \hfill 
\includegraphics[width=0.45\linewidth]{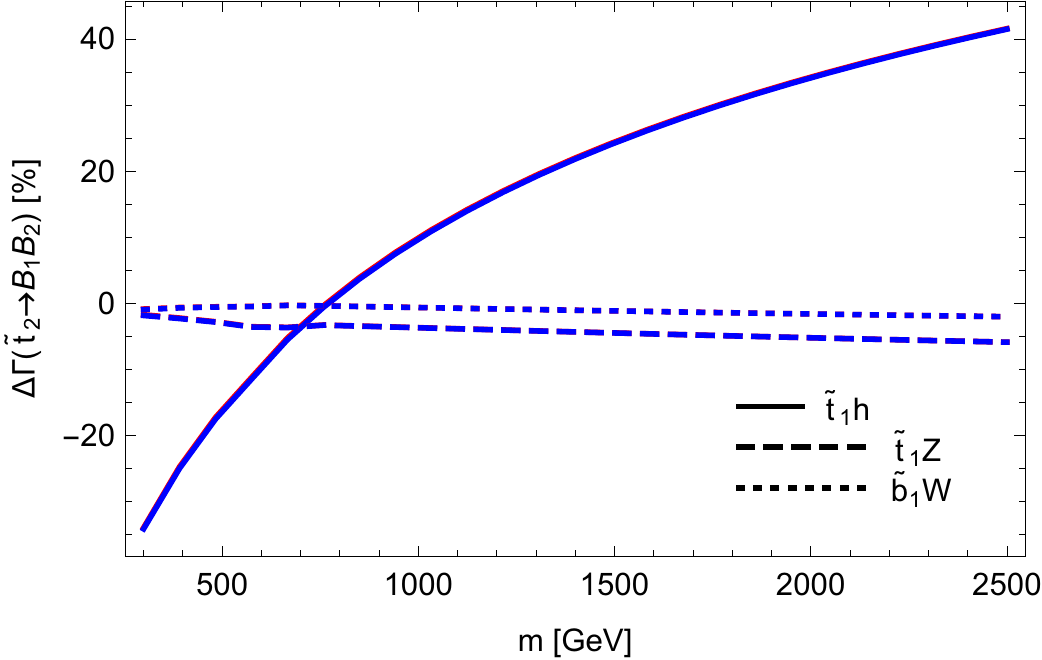} 
\caption{Comparison between \SARAH and \SFOLD for selected stop decays. On the left, the
loop-corrected partial widths are shown. On the right, the relative size of the loop 
correction is given. Blue lines are obtained with \SARAH, red lines with \SFOLD.}
\label{fig:SFOLD_stop}
\end{figure}
Next we turn to the comparison of the two-body decays of sfermions in the \mssm{}. For this purpose, we compared our results
against the public code \SFOLD 1.2.
We have applied several modifications to the code \SFOLD:
\begin{itemize}
 \item The variable to use loop-masses either in the loops or in the kinematics are set to $0$ in {\tt SFOLD.F}:
 \begin{lstlisting}[style=file]
  osextmassesOn = 0
  osloopmassesOn = 0
 \end{lstlisting}
 This is done to ensure that both codes use exactly the same masses everywhere. 
 \item We find a disagreement for the bremsstrahlung routines for $S\to S V$ decays. Therefore, we add to line 440 in {\tt Bremsstrahlung.F}
 of \SFOLD the terms:
\begin{lstlisting}[style=file]
  -2*g1**2*gt*gtC*Ii - 2*g1**2*gt*gtC*I0up1
\end{lstlisting}
 \item We find for $S\to SS$ and $S\to SV$ decays huge numerical loop-corrections that could even cause a negative width. 
 We could trace back the problem to diagrams with two massive vector bosons in the loop \footnote{In a private discussion with one author of \SFOLD
 the origin of the problem could not be identified. It might be 
 a numerical problem with the high-rank loop integrals which appear when performing the calculation in $R_\xi$ gauge.}.
 The problem is avoided by setting
 \begin{lstlisting}[style=file]
 Xipart1 = 0d0
 \end{lstlisting}
in {\tt Decay.F} of \SFOLD.
With this choice, it is no longer possible to change the gauge in \SFOLD, but the results are only valid in Feynman-'t Hooft gauge,
sufficient for our comparison.
\item We find that \SFOLD uses a different renormalisation prescription for the rotation matrices: it includes only the divergent parts 
for the counter-terms, while \SARAH calculates the counter-terms from the wavefunction renormalisation constants using \eqn{eq:CTrot}. 
In particular for $S\to SV$ decays this can induce large differences in the one-loop corrections. If the finite parts for the 
counter-terms of the rotation matrices are included, a cancellation between the wavefunction corrections and the counter-term correction appears 
which in sum gives much smaller one-loop corrections. Therefore, we added at the end of the file {\tt CalcRenConst99.F} of \SFOLD the following re-definitions 
of the counter-terms:
\begin{lstlisting}[style=file]
 Do i1=1,3
  Do i2=1,3
   dUSf1(:,:,i1,i2) = 0.25*MatMul(dZSf1(:,:,i1,i2) &
    & - Conjg(Transpose(dZSf1(:,:,i1,i2))),USf(:,:,i1,i2))
   dUSfC1(:,:,i1,i2) = Conjg(0.25*MatMul(dZSf1(:,:,i1,i2) &
    & - Conjg(Transpose(dZSf1(:,:,i1,i2))),USf(:,:,i1,i2)))
  End Do
 End Do
 dVCha1 = MatMul(dZChaL1 - Conjg(Transpose(dZChaL1)),VCha)/4
 dUCha1 = MatMul(dZChaR1 - Conjg(Transpose(dZChaR1)),UCha)/4     
 dZNeuRM1 = MatMul(dZNeuL1 - Conjg(Transpose(dZNeuL1)),ZNeu)/4
 dVChaC1 = Conjg(dVCha1)
 dUChaC1 = Conjg(dUCha1) 
 dZNeuRMC1 = Conjg(dZNeuRM1) 
\end{lstlisting}
\end{itemize}
\begin{figure}[t]
\includegraphics[width=0.45\linewidth]{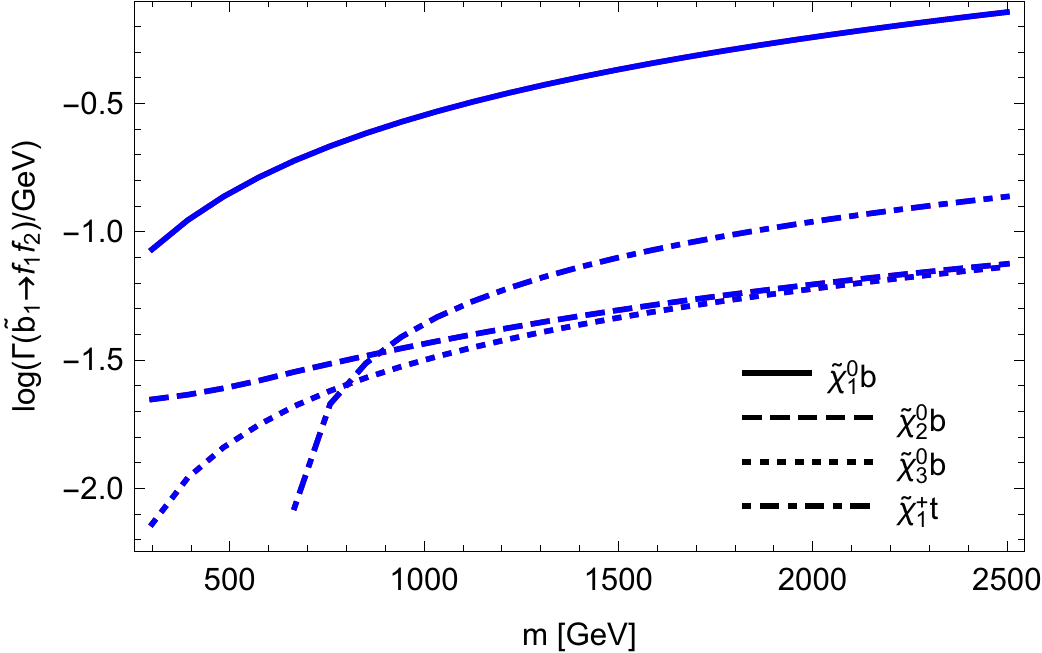} \hfill 
\includegraphics[width=0.45\linewidth]{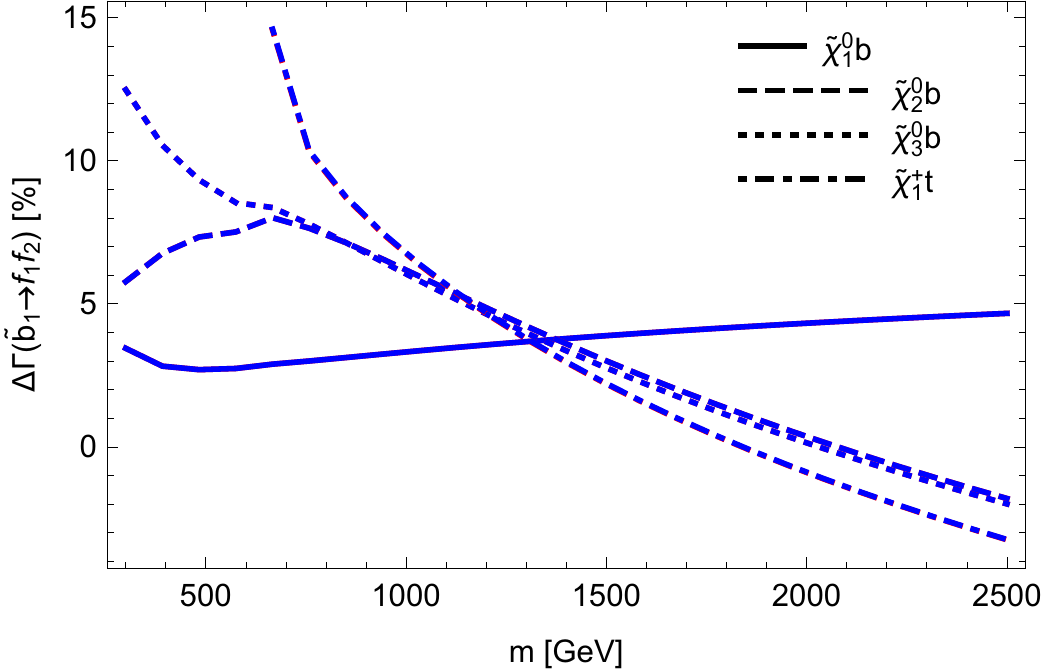}  \\
\includegraphics[width=0.45\linewidth]{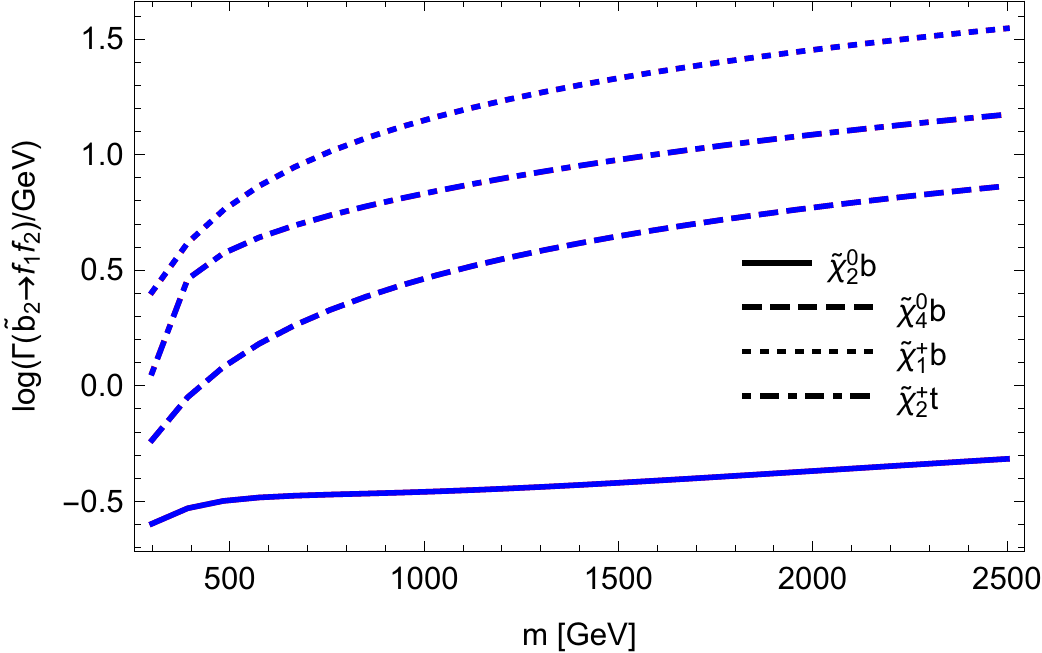} \hfill 
\includegraphics[width=0.45\linewidth]{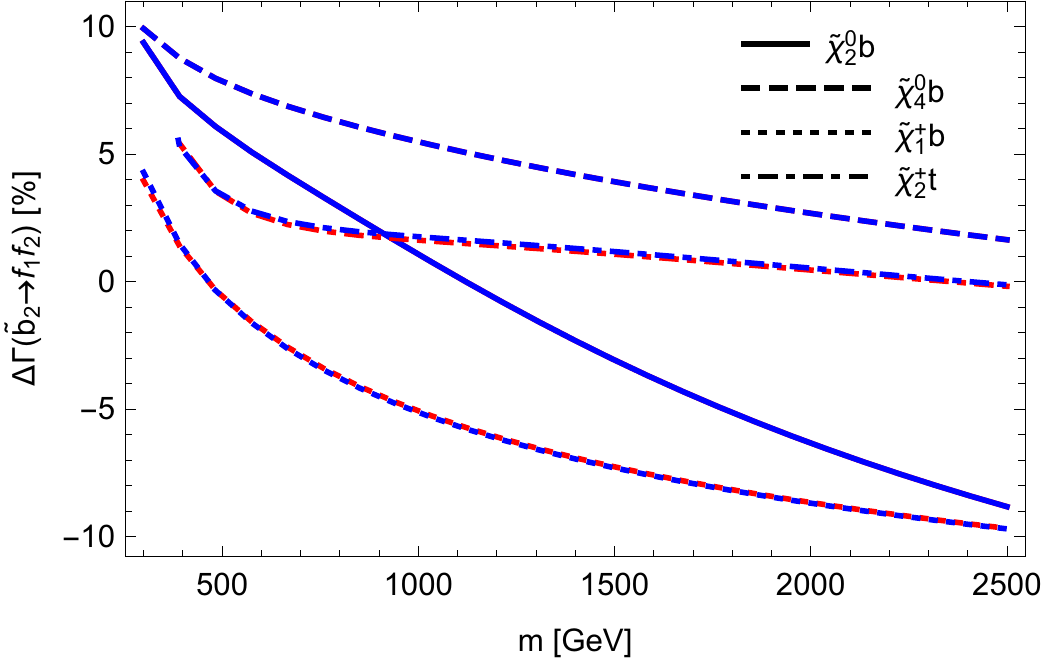}  \\
\includegraphics[width=0.45\linewidth]{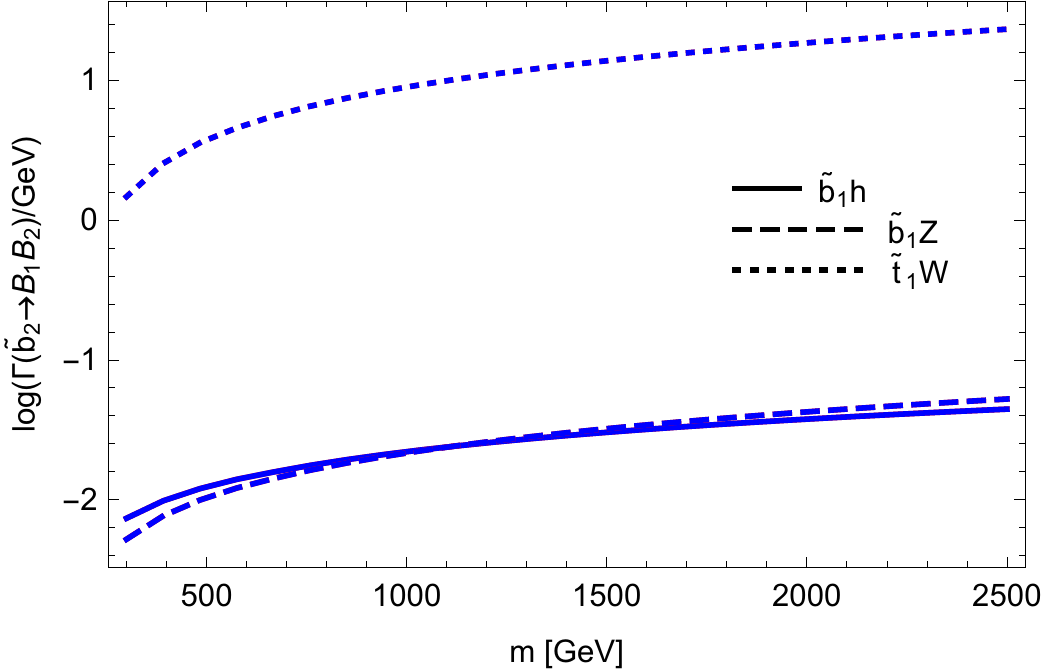} \hfill 
\includegraphics[width=0.45\linewidth]{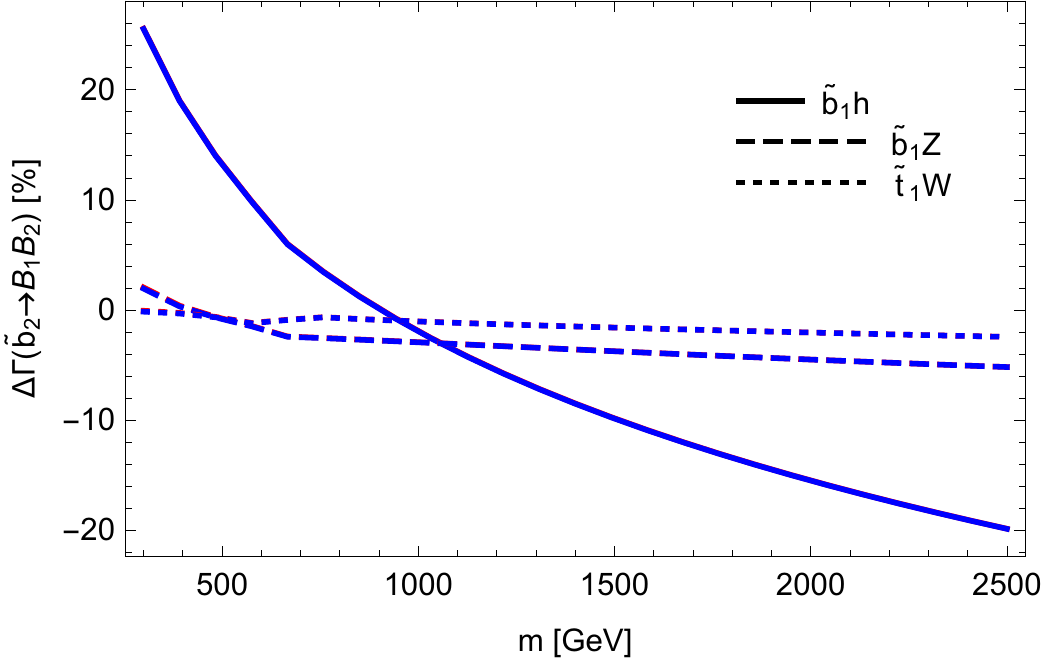} 
\caption{The same as \fig{fig:SFOLD_stop} for sbottom decays.}
\label{fig:SFOLD_sbottom1}
\end{figure}
The results for some representative decays for the light and heavy stop are shown in \fig{fig:SFOLD_stop}. Here and in the following 
we give the partial widths at \lo{} and \nlo{} as well as the relative size of the one-loop corrections defined as
\begin{equation}
\label{eq:deltagamma}
\Delta\Gamma = \frac{ \Gamma^{\text{\nlo{}}} -  \Gamma^{\text{\lo{}}} }{ \Gamma^{\text{\lo{}}}} \,.
\end{equation}

With our described adjustments we find an excellent agreement for the heavy stop decays into a Higgs or a gauge boson and a stop or sbottom. 
While the corrections for the decays into gauge bosons are comparably small and only of order of a few per-cent, the situation changes if the finite 
parts for the counter-terms described above are not included. In that case, i.e. when using \SFOLD out of the box,
the corrections for the decays with a $Z$ or $W$ boson in the final state can be a factor 
of $10$ larger. For the decays into a pair of fermions we also find very good agreement with only very small 
differences for small values of $m$. Similarly we show the results for the light and heavy sbottom decays in \fig{fig:SFOLD_sbottom1}. 
Here, the results are very similar to those of the stop decays. We do not add figures for stau or $\tau$-sneutrino decays, or the decays of first and second generation sfermions; they would look very similar to the ones for stop and sbottoms, only the overall size of the loop corrections being smaller. 
Thus, in total we found a very good agreement between \SARAH and \SFOLD for all kinds of two-body decays of sfermions. 

\subsubsection{Gluino decays in the \mssm{}: \SARAH vs. \FVSFOLD}
\begin{figure}[htb]
\includegraphics[width=0.49\linewidth]{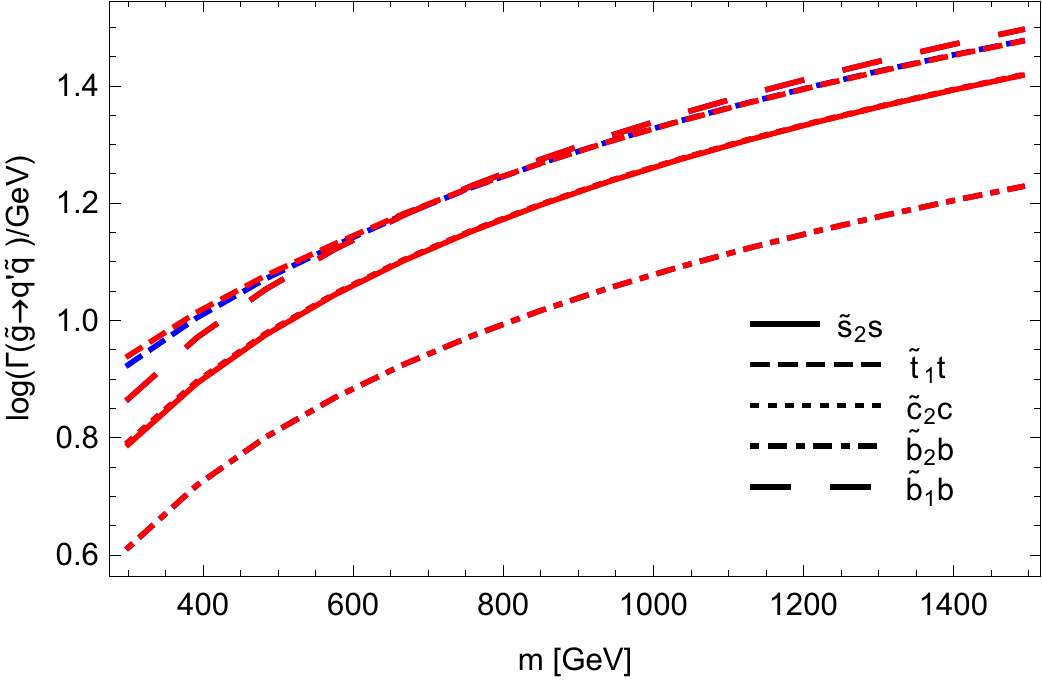} \hfill 
\includegraphics[width=0.49\linewidth]{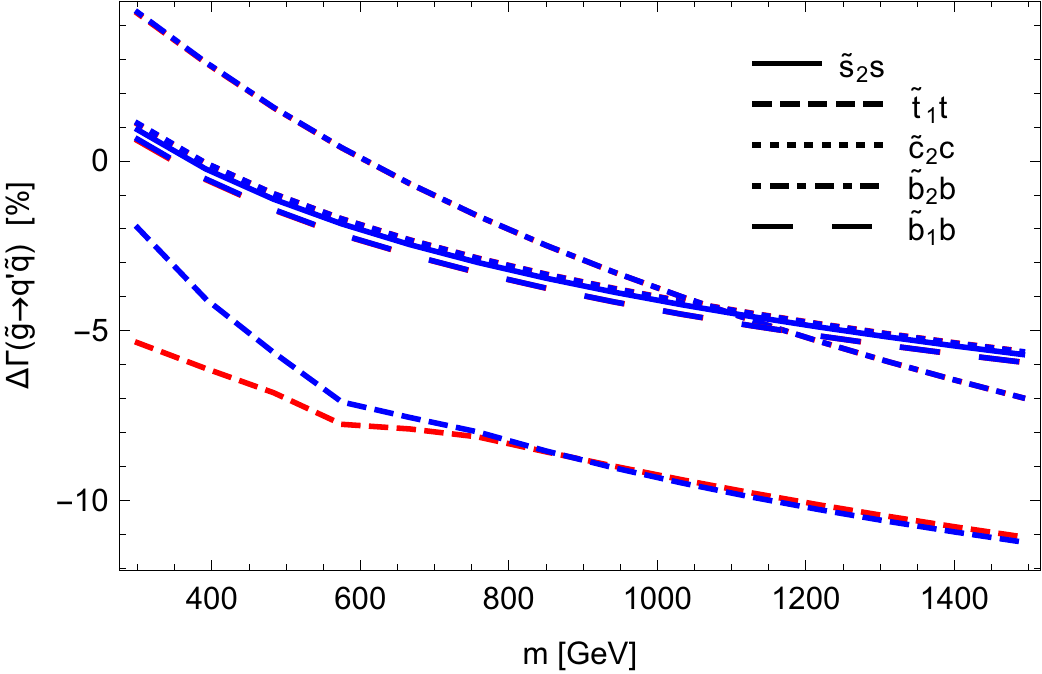} 
\caption{Comparison between \SARAH and \FVSFOLD for gluino decays. On the left, the
loop-corrected partial widths are shown. On the right, the relative size of the one-loop 
correction is given. Blue lines are obtained with \SARAH, red lines with \FVSFOLD.}
\label{fig:FVSfold}
\end{figure}
In this section we compare the decays of gluinos in the \mssm{} obtained with \SARAH and \SPheno against the results generated with the code \FVSFOLD.
We also performed similar adjustments in \FVSFOLD as done for \SFOLD for our comparison. However,
\FVSFOLD already includes the finite parts of the counter-terms of the squark rotation matrices, i.e. it was not necessary to add those.
Therefore without any larger adjustments, we find a very good agreement between \SARAH and \FVSFOLD as shown in \fig{fig:FVSfold}
Thus, \SARAH reproduces also the result of \citere{Eberl:2017pbu}, namely that the one-loop corrections to gluino decays reduce the
decay width by about $10$\%. 

\subsubsection{Heavy Higgs decays in the \mssm{}: \SARAH vs. \HFOLD}
\begin{figure}[hbt]
\includegraphics[width=0.44\linewidth]{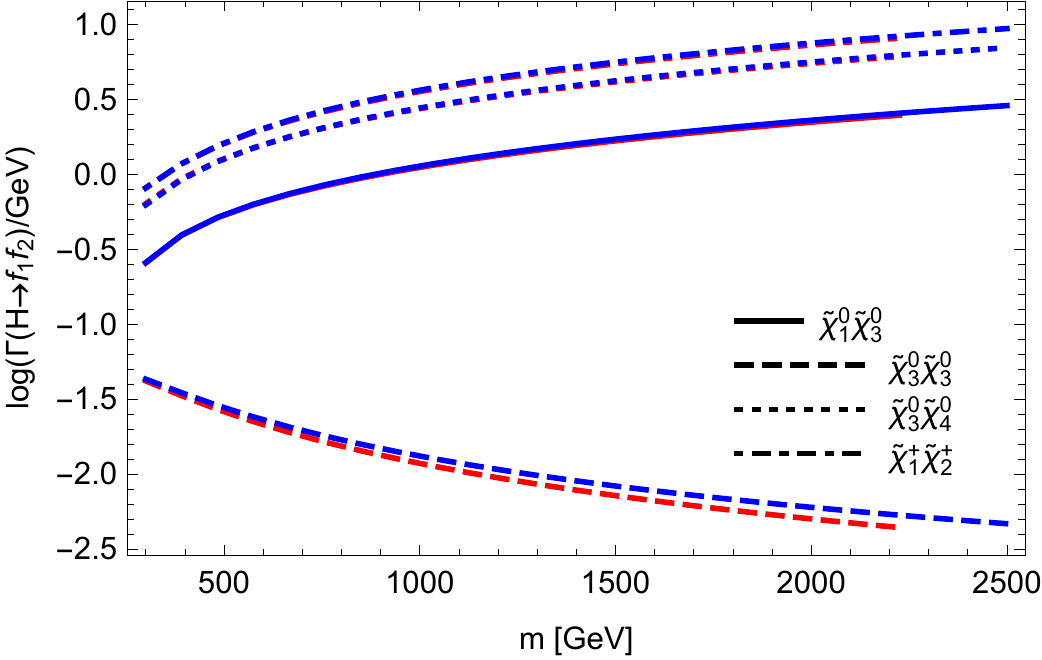} \hfill 
\includegraphics[width=0.44\linewidth]{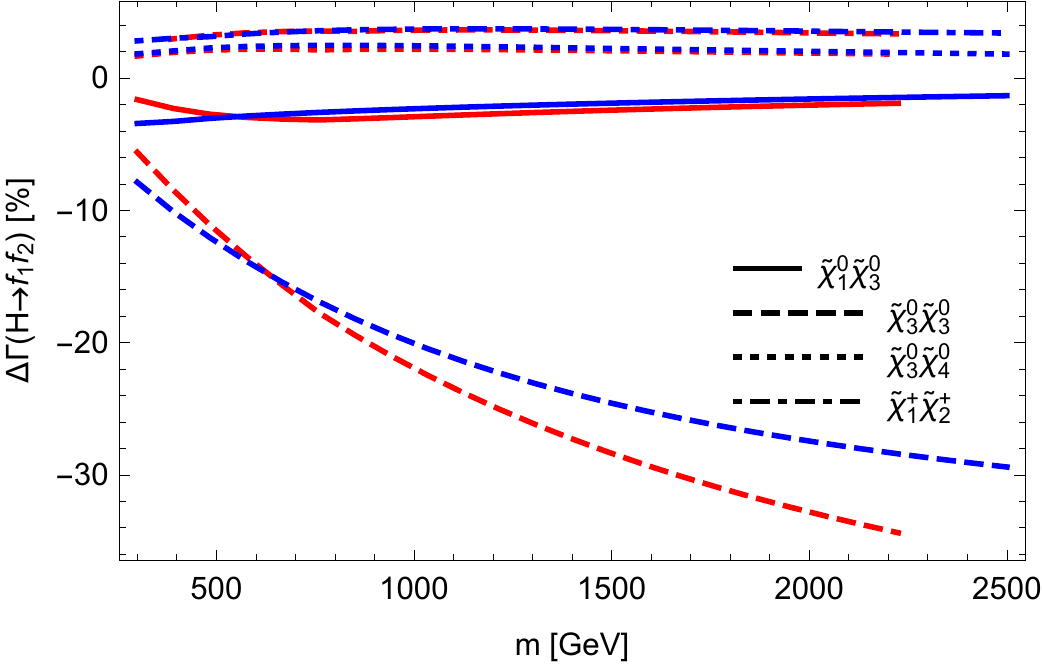}  \\
\includegraphics[width=0.44\linewidth]{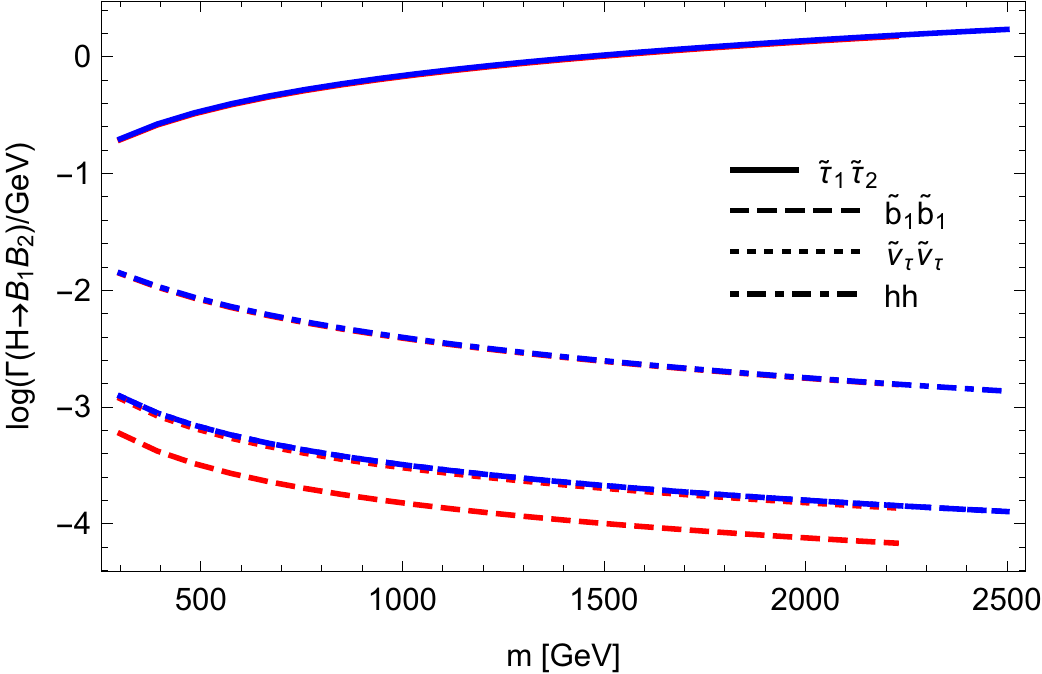} \hfill 
\includegraphics[width=0.44\linewidth]{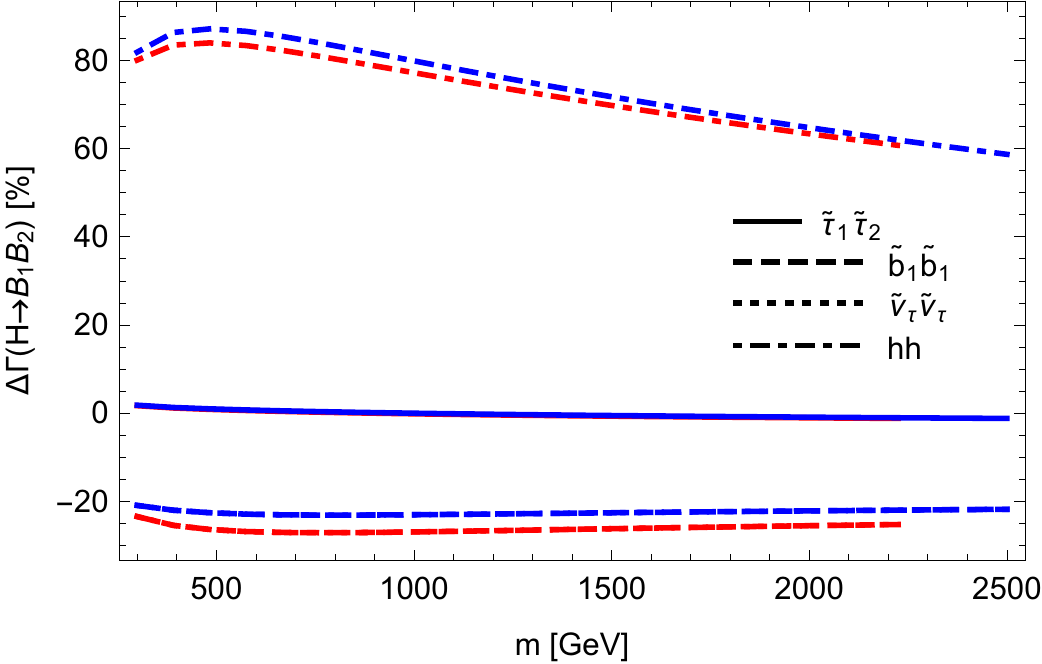}   \\
\includegraphics[width=0.44\linewidth]{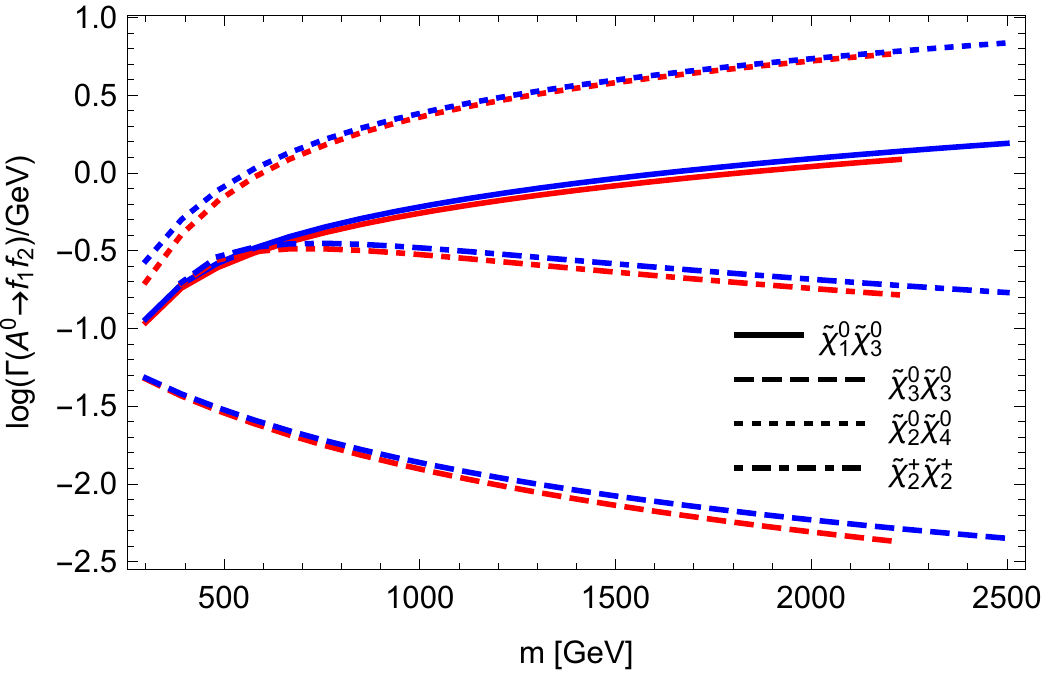} \hfill 
\includegraphics[width=0.44\linewidth]{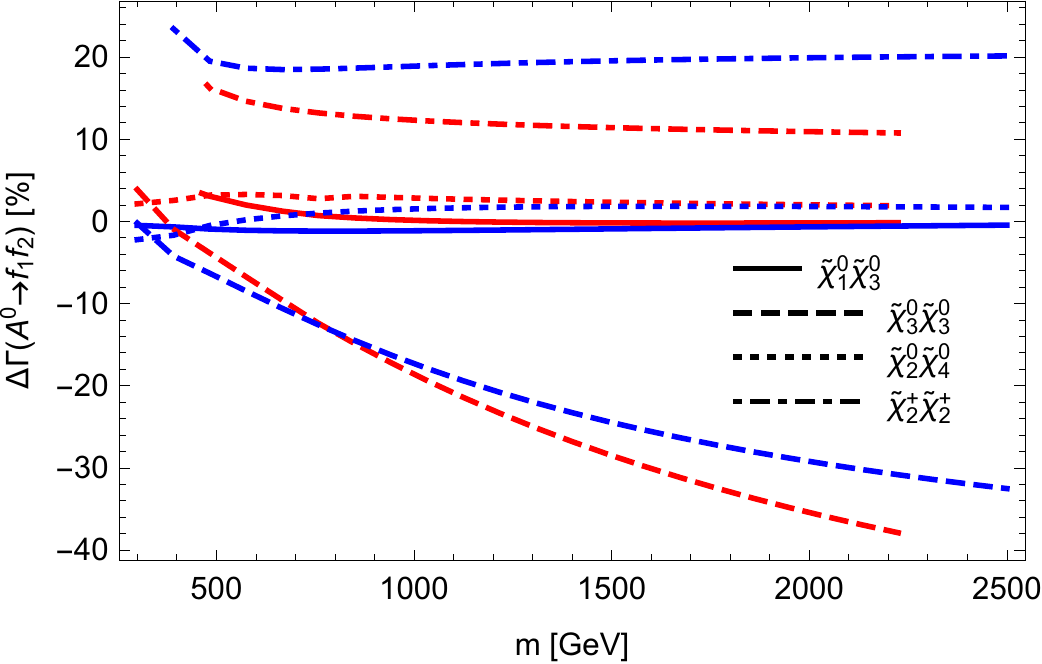}  
\caption{Comparison between \SARAH and \HFOLD for the heavy \cp{}-even and -odd Higgs. On the left, the
loop-corrected partial widths are shown. On the right, the relative size of the loop 
correction is given.  Blue lines are obtained with \SARAH, red lines with \HFOLD.}
\label{fig:HFOLD_H}
\end{figure}
\SARAH also makes predictions for the one-loop corrections of Higgs boson decays. However, it must be clearly stated that those predictions have to be interpreted with some care:
the automatised calculations are not yet optimised for the calculation of Higgs boson decays, in particular for the \sm{}-like Higgs boson. For such decays, 
we leave an appropriate definition of counter-terms, following our explanations in \sct{sec:setuploopdecays}, to future work. One reason is that for consistency
it will be necessary to use the counter-terms in the calculation of the mass spectrum as well. This is however not yet possible.
We want to stress that \SARAH already calculates the light Higgs into \sm{} particle decays by adapting higher-order corrections (even beyond \nlo{}) for the \sm{} and \mssm{} from literature.
Thus, the `old' results obtained with \SARAH are expected to be more accurate.

On the other hand, for the decays of heavy Higgs bosons, whose mass corrections are usually much smaller, and/or for decays into \bsm{} states the applied \nlo{} corrections are expected 
to work well, and the obtained results supersede the pure tree-level calculations often done for these decay modes. 
In order to validate these results, we compared them against the code \HFOLD which also makes predictions for the one-loop corrections of Higgs decays in the \mssm{}.
Here, we made the same adjustments as for \FVSFOLD: on-shell masses in loops and kinematics have been turned off.
In addition, we needed to turn off all improvements for the `old' calculation in \SPheno to obtain equivalent \lo{} results to \HFOLD.
The results are summarised in \fig{fig:HFOLD_H} where we compare our results for the decays of the neutral 
heavy Higgs states $H$ and $A^0$ into \susy{} particles and \sm{}-like Higgs bosons. Without further modifications we find that the predictions of the size of the one-loop 
corrections of both codes agree rather well in particular in the dominant decay modes. However, we find that for decay channels with small partial widths also
sizeable differences can be present. A detailed investigation of the remaining differences and also a comparison with other Higgs boson decay widths
calculations is left for a dedicated work. 
Such a future investigation should also focus on the detailed derivation and incorporation of the \Ufactors, that admix the Higgs bosons beyond tree-level.
This is particularly crucial when comparing to codes such as {\tt NMSSMCalc} or {\tt FeynHiggs}, where the definition of their $Z$-factors is different~\cite{Frank:2006yh,Baglio:2015noa}. We shortly discuss the relevance of \Ufactors{}
for Higgs boson decays in \sct{sec:externalufactors}.

\subsubsection{Radiative neutralino and gluino decays: \SARAH vs. \SPheno }
\label{sec:ComparisonSPheno}
\begin{figure}[hbt]
\includegraphics[width=0.5\linewidth]{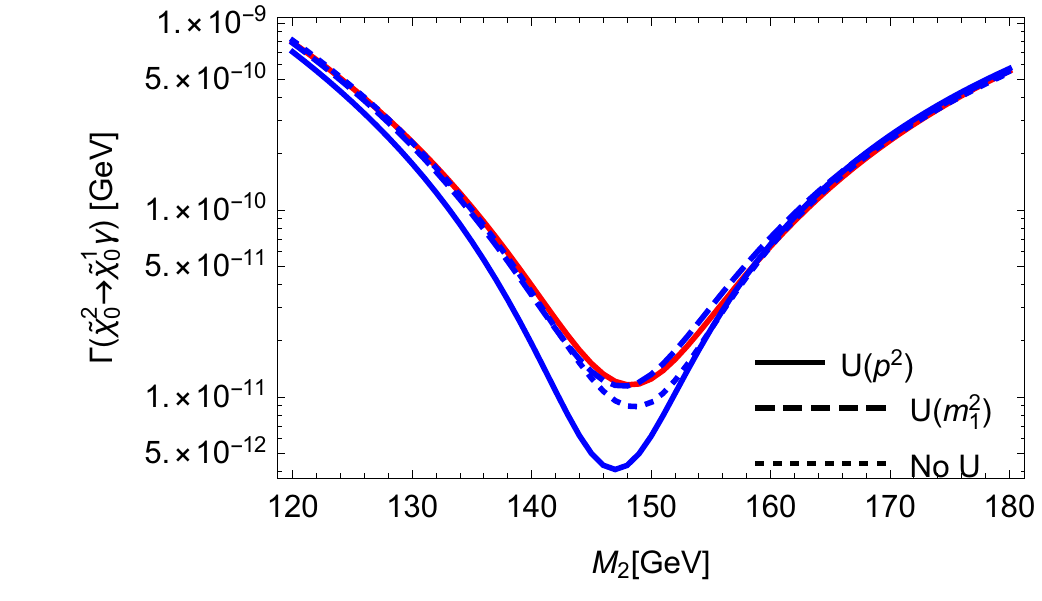} \hfill 
\includegraphics[width=0.5\linewidth]{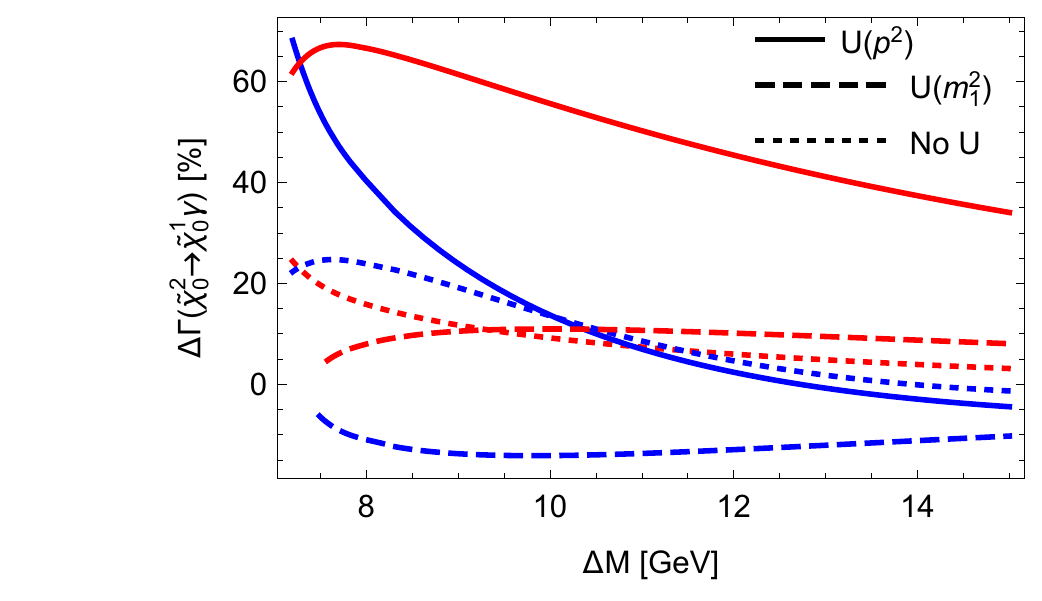}  \\
\includegraphics[width=0.5\linewidth]{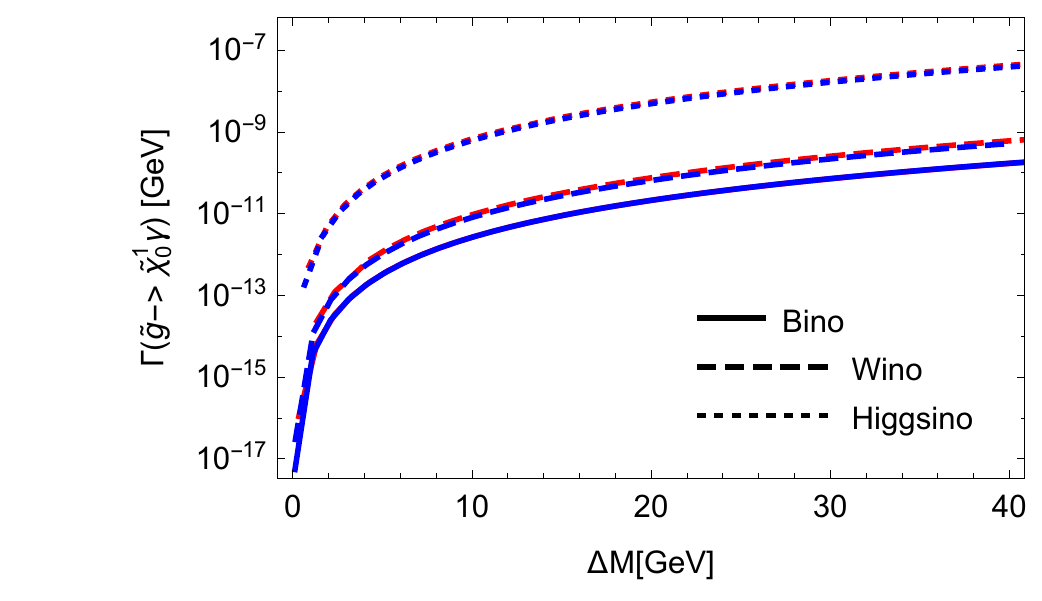} \hfill 
\includegraphics[width=0.5\linewidth]{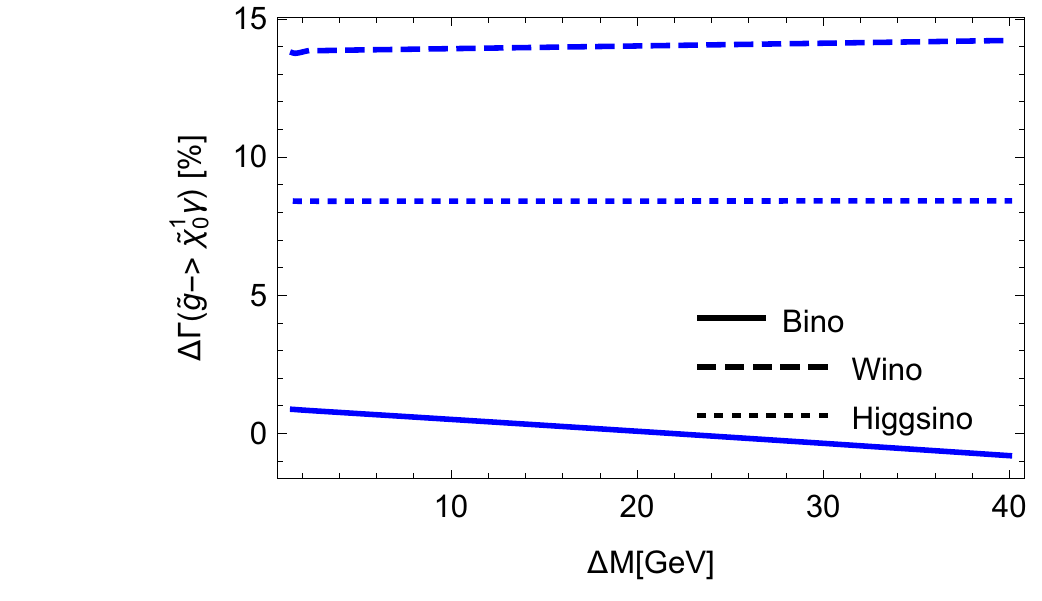} 
\caption{Comparison between \SARAH (blue) and \SPheno (red) for the 
loop-induced decays of the neutralino and gluino. 
On the left, the partial widths are shown.  Blue lines are obtained with \SARAH, red lines with \SPheno.
On the right, the relative difference between the codes as function of the mass 
splitting is shown (not $\Delta \Gamma$ from \eqn{eq:deltagamma}). In the case of the neutralino decay we show the impact of the 
\Ufactors, while for the gluino decays we compare the different kind of neutralinos.
The colour code for the upper right figure is:
blue for a bino \lsp{} and red for a wino \lsp{}.}
\label{fig:N2N1g}
\end{figure}
\SARAH does not only calculate the one-loop corrections to tree-level two-body decays, but also calculates the \lo{} result 
for loop-induced decay widths\footnote{Even if \SARAH will use the new routines to obtain also loop-induced decay width for Higgs states,
one should still use the old results for the diphoton and digluon rate. The latter also include the full model dependence at \lo{} but in
addition also higher-order \qcd{} corrections, see \citere{Staub:2016dxq}.} The most important applications for these routines are
radiatively induced decays of \bsm{} particles. The main candidates for such decays in the \mssm{} are $\tilde{\chi}^0_2 \to \tilde{\chi}^0_1 \gamma$
and $\tilde g \to \tilde{\chi}^0_1 g$. Those decays were already implemented in \SPheno based on the results of
\citere{Haber:1988px}. For our comparison, we choose parameter points with a light mass splitting between (i) a bino 
and wino \lsp{} and \nlsp{}, respectively; (ii) the gluino and all three kinds of neutralinos. For the case of the neutralino decay, the result for the
obtained width as a function of the wino mass parameter $M_2$ as well as the relative difference between \SPheno and \SARAH as
function of the mass splitting are shown in \fig{fig:N2N1g}. We show the \SARAH results for three different choices of the
\Ufactors:
(i) without \Ufactors{},
(ii) using the rotation matrices obtained with the momentum being the mass of the lightest neutralino in all vertices,
(iii) using $p^2$-dependent \Ufactors{}. The second
option corresponds to the procedure applied in \SPheno and thus we find a reasonable agreement within $10$\%. The results without \Ufactors{}
are very similar and only very close to the level crossing visible differences occur. However, when using the
$p^2$-dependent \Ufactors{}, the obtained width is significantly smaller. 
This is due to a cancellation between the vertex and wavefunction corrections, which is most efficient when including
the $p^2$ dependence in the \Ufactors{}.
For the decays of the gluino into a neutralino and gluon, we find very good agreement between \SPheno and \SARAH for all three kind of neutralinos,
see again \fig{fig:N2N1g}. Note that throughout the calculation of loop-induced decays loop-corrected masses are inserted.

\subsection{Impact of external \Ufactors}
\label{sec:externalufactors}
\begin{figure}[tb]
\includegraphics[height=5cm]{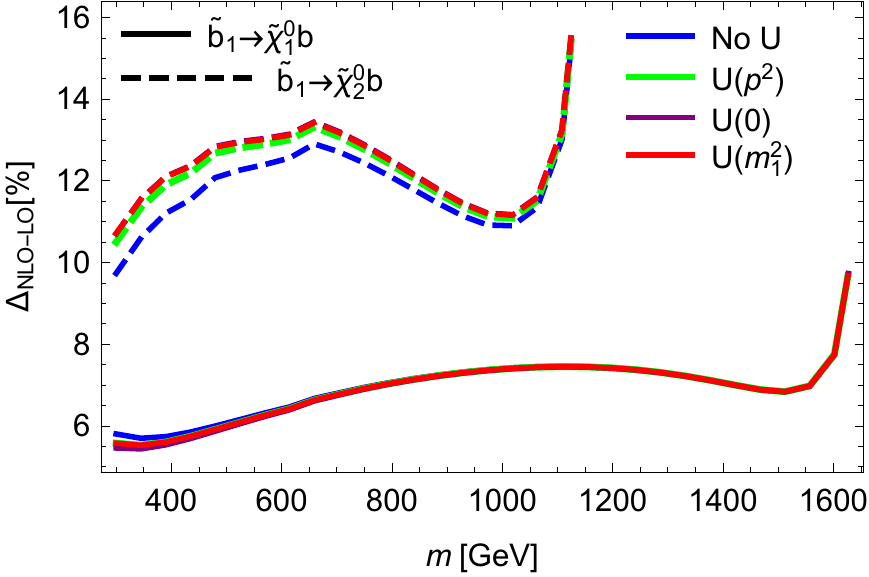} \hfill 
\includegraphics[height=5cm]{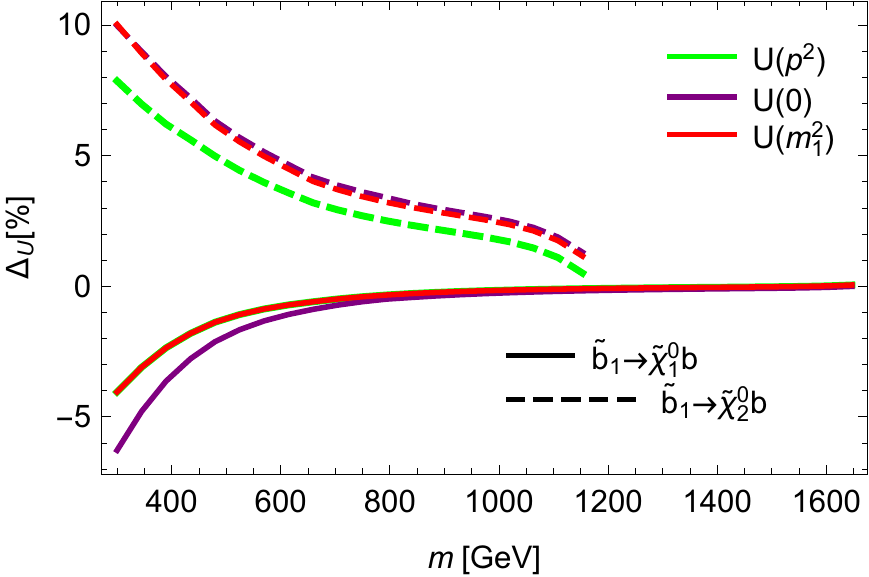} \\   
\includegraphics[height=5cm]{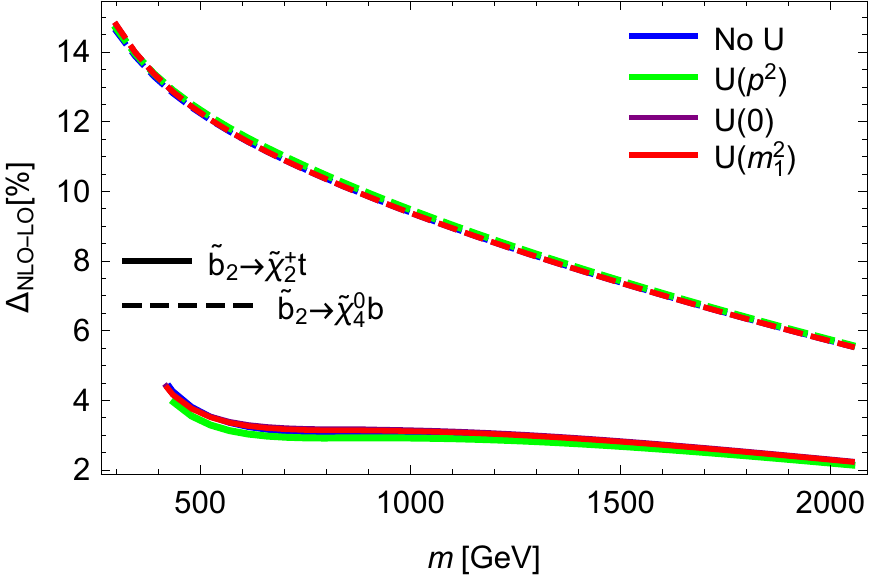} \hfill 
\includegraphics[height=5cm]{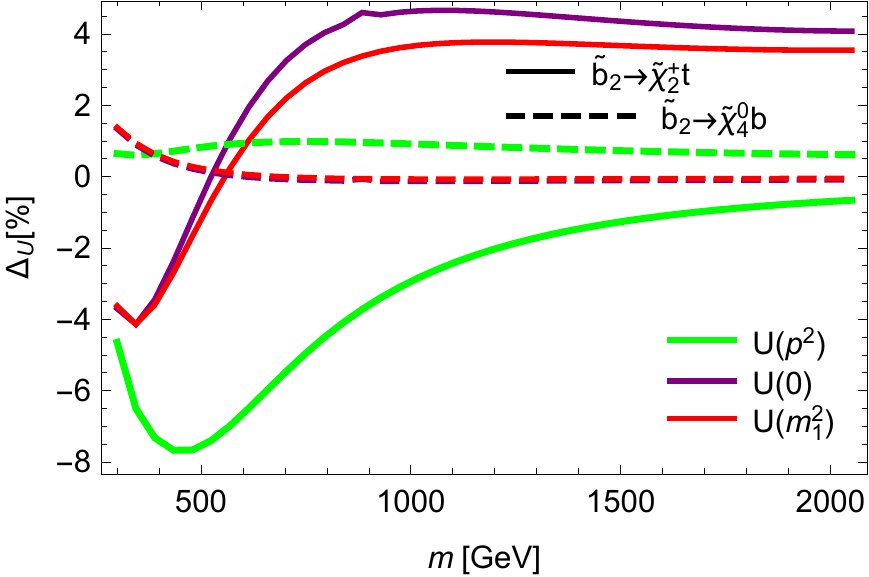} \\   
\caption{Impact of the external \Ufactors{} for sbottom decays. On the left side, we show the
relative \nlo{} correction when equal \Ufactors{} are applied at \lo{} and \nlo{}. On the right side, we show
the relative \nlo{} correction for different \Ufactors{} normalised to the relative \nlo{} correction
without \Ufactors{} defined in \eqn{eq:deltau}.}
\label{fig:Usbottom}
\end{figure}
\begin{figure}[tb]
\includegraphics[height=5cm]{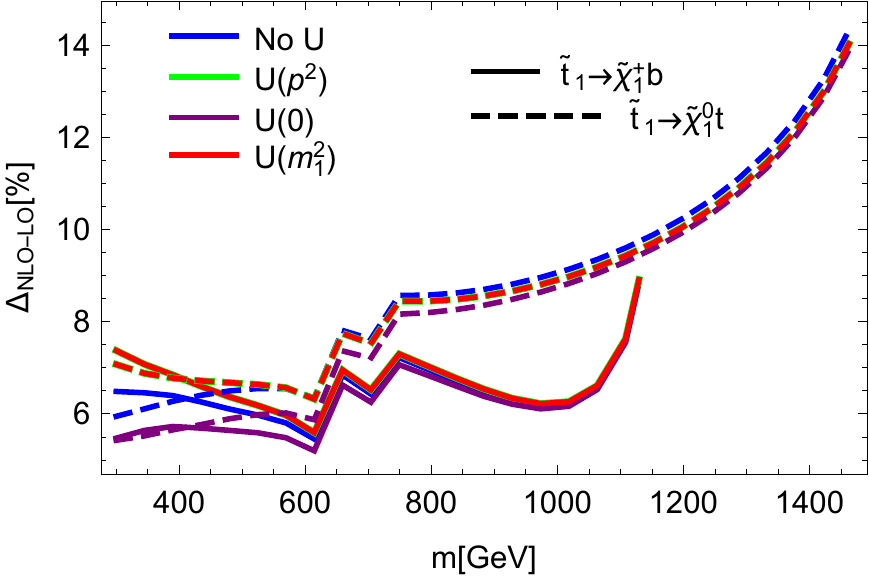} \hfill 
\includegraphics[height=5cm]{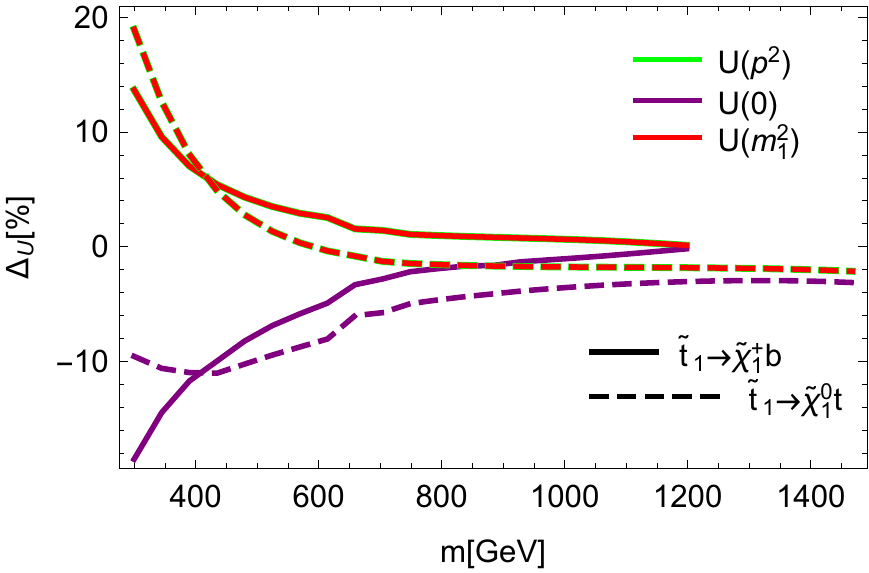} \\   
\includegraphics[height=5cm]{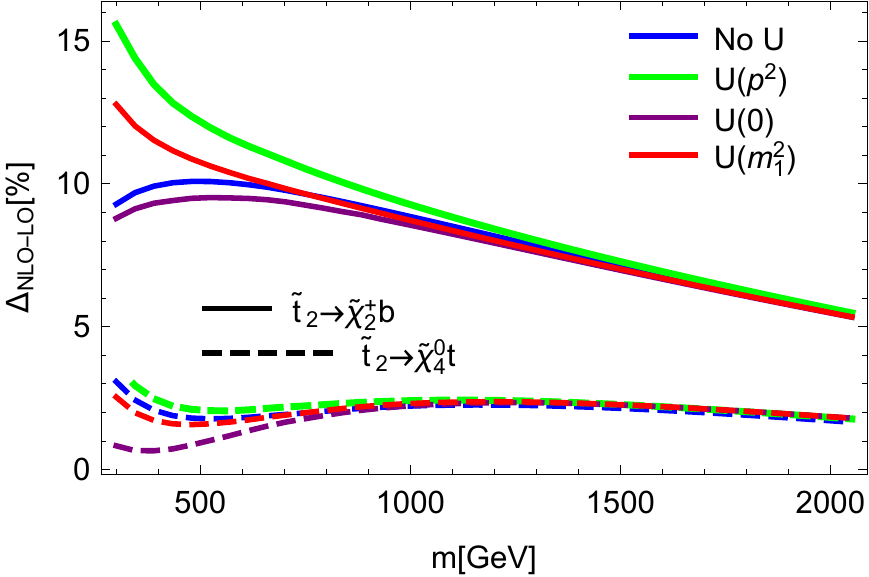} \hfill 
\includegraphics[height=5cm]{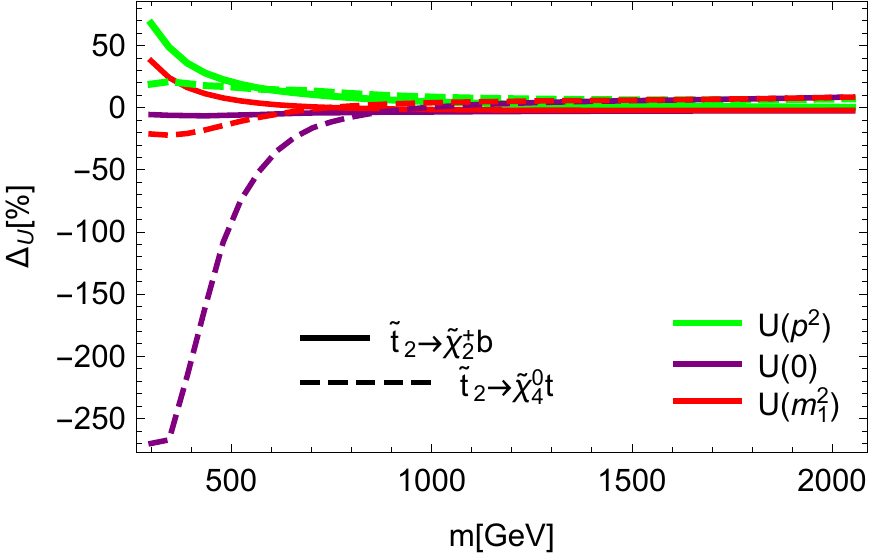} \\   
\caption{Same as \fig{fig:Usbottom} for decays of stops.}
 \label{fig:Ustop}
\end{figure}
\begin{figure}[tb]
\includegraphics[height=5cm]{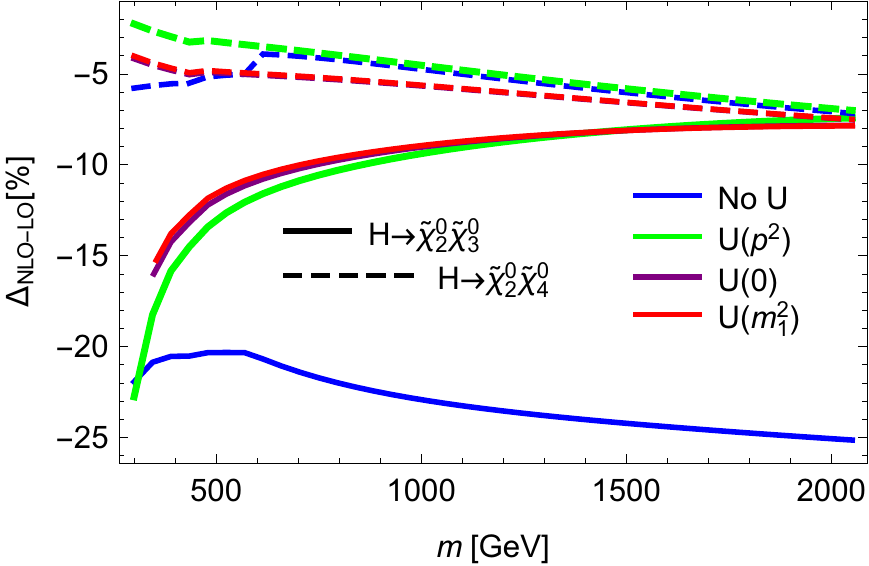} \hfill 
\includegraphics[height=5cm]{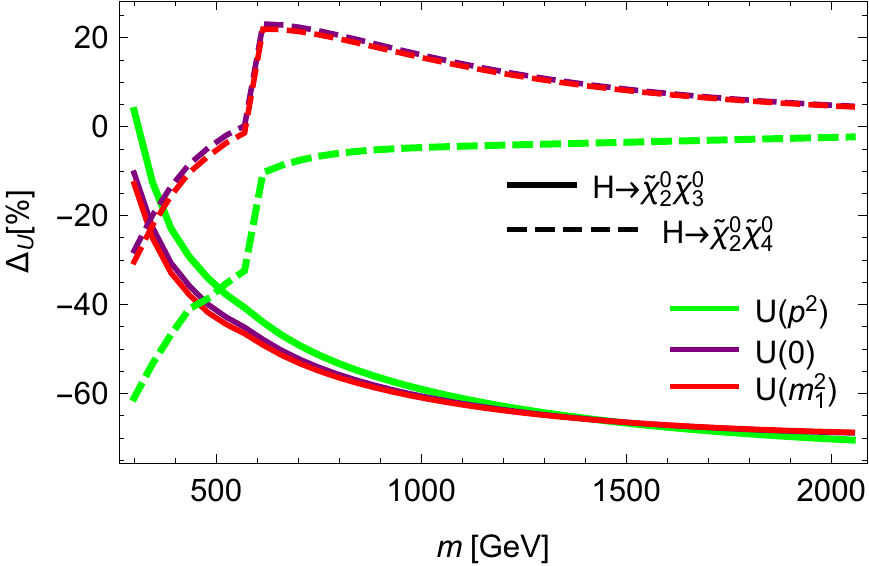} \\   
\caption{Same as \fig{fig:Usbottom} for decays of the heavy \cp{}-even Higgs boson.}
 \label{fig:UHiggs}
\end{figure}

Before we conclude, we want to give some impression of the numerical impact induced by the inclusion of \Ufactors.
For this purpose, we show in Figs.~\ref{fig:Usbottom}--\ref{fig:UHiggs} the size of the one-loop 
corrections for selected stop, sbottom and Higgs decays, respectively, using the three available options to calculate the \Ufactors.
We apply loop-corrected {\drbar{}} masses for all cases and particles, i.e. {\tt DECAYOPTIONS[1115]}$=$1. 
We focus on two effects: First we apply \Ufactors{} both at \lo{} and \nlo{} equally and, in the left figures, show the relative correction
induced by the \nlo{} corrections. Second, in the right figures, we show the effect of including the \Ufactors{}
in the \nlo{} calculation compared to the \nlo{} decay width calculation without external \Ufactors{}. More precisely, 
the shown value $\Delta_U$ is defined as
\begin{align}
\Delta_U = \left(\frac{ \Gamma_0^{\text{\nlo{}}} -  \Gamma_0^{\text{\lo{}}} }{ \Gamma_0^{\text{\lo{}}}}\right)^{-1} 
\left(\frac{ \Gamma_U^{\text{\nlo{}}} -  \Gamma_U^{\text{\lo{}}} }{ \Gamma_U^{\text{\lo{}}}} -  \frac{ \Gamma_0^{\text{\nlo{}}} -  \Gamma_0^{\text{\lo{}}} }{ \Gamma_0^{\text{\lo{}}}}\right)\,.
\label{eq:deltau}
\end{align}
Here, $\Gamma_0$ are the decay widths without applying \Ufactors{}. 
$\Delta_U$ encodes the difference in the relative correction factor from LO to NLO when applying
$U$-factors in contrast to not applying $U$-factors. It thus encodes the effect
of $U$-factors in the one-loop correction, factoring out their effect already
present at tree-level. Depending on the particle species the effect at tree-level
can already be pronounced, and thus was already included in previous \SARAH and \SPheno versions.
We therefore focus on the effect of the $U$-factors in the relative NLO correction.

For the sbottom decays into gauginos depicted in \fig{fig:Usbottom} the changes due to the inclusion of \Ufactors{} are moderate.
From the left figures it is apparent that the size of \nlo{} corrections is mostly independent of
the inclusion of \Ufactors{}. From the right figures we deduce that the effect of \Ufactors{}
on the relative \nlo{} correction remains below $10$\% for all choices. The reason is that the 
left-right mixing in the sbottom sector is in general small and nearly identical at tree- and loop-level.
Thus the $U$ matrices are almost diagonal.
This is different for the decays of stops shown in \fig{fig:Ustop} where the left-right mixing is more pronounced.
This mixing receives also a sizeable radiative correction which is encoded in the \Ufactors{}.
Consequently, there is also a larger sensitivity on how this matrix is calculated and incorporated
as shown in the right two figures.
We find that the results without momentum-dependence can differ from the
other two options by $30$\% for the considered decays. For the heavy stop, this effect is even more pronounced.
However for the decay width $\tilde{t}_2\to \tilde{\chi}_4^0t$, where the relative \nlo{} corrections encoded
in $\Delta U$ differ by more than $100$\%,
the absolute \nlo{} correction almost vanishes, as can be seen from the left figure.
Thus, in all examples for stop and sbottom decays,
the inclusion of \Ufactors{} gives only a moderate change in the relative \nlo{} corrections once (momentum-dependent)
\Ufactors{} are taken into account compared to the calculation without \Ufactors{}.

This is slightly different for the heavy Higgs decays shown in \fig{fig:UHiggs}.
As shown in the left figures all three options for the \Ufactors{} can alter the size of the
relative \nlo{} corrections significantly. In the right figures it is apparent that even for the relative \nlo{} 
correction differences of $50$\% and more
compared to the calculations without \Ufactors{} are
easily possible, a fact which is well known for Higgs bosons.
This shows the need to properly include these factors for Higgs boson decays even if the
radiative corrections to the masses are moderate and the particles are clearly separated in their masses.
Further studies for Higgs boson decays and a comparison of the \Ufactors{} to
$Z$-factors as discussed in \citeres{Frank:2006yh,Baglio:2015noa} are in order in future work.

\section{Conclusions}
\label{sec:conclusions}

In this paper we described a fully generic implementation of the calculation of two-body decay widths
at the full one-loop level in the \SARAH and \SPheno framework, which can be used
in a wide class of supported models. 
We presented the
necessary generic expressions for virtual and real corrections.
Wavefunction corrections are determined from on-shell conditions.
On the other hand, the parameters of the underlying model are by default
renormalised in a \drbar{} (or \msbar{}) scheme. 
We described how higher-order corrections for the external states can be taken 
into account. We also explained how we restore gauge invariance 
as well as ultraviolet and infra-red finiteness when setting the external 
masses to their loop-corrected values. We commented on the drawbacks compared to a
full on-shell approach which is model and process dependent. 

We have shown how the new features of \SARAH and \SPheno
can be used and how the user can implement own counter-terms
to be used for the calculation of two-body decay widths.
We studied the impact and relevance of such counter-terms for two examples in the \sm{},
namely the decay to the top-quark and the \sm{} Higgs boson decay into bottom quarks.
In addition, we compared our implementation for sfermion and gluino decays
within the \mssm{} against other
available codes, namely \SFOLD, \HFOLD and \FVSFOLD, which also
employ a \drbar{} renormalisation for the \mssm{} parameters.
After a few described adjustments in those codes we found an overall excellent agreement. 
For the \mssm{} and $R$-parity
violating models we also compared chargino and neutralino decays against \CNN,
which uses a full on-shell scheme for masses and couplings and
found numerically identical results.

The new extension is included in {\tt SARAH 4.11.0} and
makes it possible to study radiative corrections to
two-body decay modes in many different supersymmetric and non-supersymmetric models.
However, models with \cp{} violation and/or (additional) massive 
gauge bosons charged under U$(1)_{\rm em} \times \text{SU}(3)_c$ are not yet supported. This is left for 
future work. Other future extensions aim at necessary improvements to better handle Higgs boson decays,
in particular for the decays of the \sm{}-like Higgs boson to \sm{} particles and
the inclusion of external higher-order mass and mixing corrections.
Lastly the inclusion of decays of gauge bosons is in order, but left for future work.

\section*{Acknowledgements}
We are grateful to Dominik St\"ockinger for discussions at the early stage of this 
project. We thank Helmut Eberl for help when comparing to the codes \SFOLD and \HFOLD, 
and Thomas Hahn for his support concerning \FeynArts/\FormCalc.
SL acknowledges support by Deutsche Forschungsgemeinschaft
through the SFB~676 ``Particles, Strings and the Early Universe''. MDG acknowledges support from French state funds managed by the Agence Nationale de la Recherche (ANR), in the context of the LABEX
ILP (ANR-11-IDEX-0004-02, ANR-10-LABX-63), and the young researcher grant ``HiggsAutomator''
(ANR-15-CE31-0002). FS is supported by ERC Recognition Award ERC-RA-0008 of the Helmholtz Association.

\appendix
\section{Conventions and expressions for loop contributions}
\label{app:genericexp}

In this section we present our conventions for vertices and
the generic expressions for the loop contributions to the wavefunction corrections
and the vertex corrections. 
We factor out the Lorentz dependent part of the vertices and work with the following conventions
\begin{itemize}
 \item $FFS$ vertex ($\bar F_1 F_2 S$)
 \begin{align}
  c_L P_L + c_R P_R
 \end{align}
 \item $FFV$ vertex  ($\bar F_1 F_2 V^\mu$)
 \begin{align}
  \gamma_\mu (c_L P_L + c_R P_R)
 \end{align}
 \item $SSV$ vertex  ($S^*_1 S_2 V^\mu$)
 \begin{align}
  c (p^{S_1}_{\mu} - p^{S_2}_\mu) 
  \end{align}
 \item $SVV$ vertex  ($S_1 V_1^\nu V_2^\mu$)
 \begin{align}
  c g_{\mu\nu}
 \end{align}
 \item $VVV$ vertex ($V^1_\mu V^2_\nu V^3_\sigma$)
 \begin{align}
 c \left[ g_{\mu\nu} (p^{V_2}_\sigma - p^{V_1}_\sigma) + g_{\nu\sigma} (p^{V_3}_\mu - p^{V_2}_\mu) +g_{\mu\sigma}(p^{V_1}_\nu - p^{V_3}_\nu) \right]
 \end{align}
 \item $VVVV$ vertex ($V^1_\mu V^2_\nu V^3_\sigma V^4_\rho$)
 \begin{align}
 c_1 g_{\mu\nu} g_{\sigma\rho} + c_2 g_{\mu \sigma} g_{\nu \rho} + c_3 g_{\mu \rho} g_{\nu \sigma}\,.
 \end{align}
\end{itemize}
For the loop corrections
\SARAH inserts the various particle species of the model under consideration also
taking into account additional symmetry and colour factors, which are not depicted here. The various contributions
are then summed up using 
\begin{align}
\label{eq:vertexsum}
M^V_i=\sum_k \frac{1}{16\pi^2} M^{(k)}_i
\end{align} 
 and 
\begin{align}
\Pi=&\sum_k \frac{1}{16\pi^2}\Pi^{(k)} \,, \hspace{1cm} \dot\Pi=\sum_k \frac{1}{16\pi^2}\dot\Pi^{(k)}\\
\Sigma_X=&\sum_k \frac{1}{16\pi^2}\Sigma_X^{(k)} \,, \hspace{1cm} \dot\Sigma_X=\sum_k \frac{1}{16\pi^2}\dot\Sigma_X^{(k)}\,.
\end{align}
All results in the following are expressed in terms of Passarino-Veltman integrals. The scalar 
loop functions $A_0$, $B_0$ and $C_0$ are calculated numerically in \SPheno according 
to the standard recipe of \citere{tHooft:1978jhc}. Tensor integrals are related to the scalar 
functions according to the famous techniques developed in \citere{Passarino:1978jh}. Explicit expressions 
for  derivatives of two-point functions, $\dot B_0$, $\dot B_1$, which are used by \SPheno are given in the appendix 
of \citere{Liebler:2010bi}.
 
\subsection{Wavefunction corrections}

\subsubsection{Fermion}
\begin{figure}[h] 
\centering 
\includegraphics[width=0.5\linewidth]{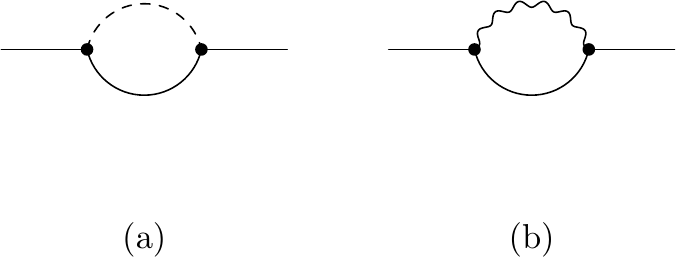} 
\caption{Generic one-loop diagrams for the fermion self-energy.}
\label{fig:WV_F}
\end{figure} 
\paragraph{(a) $FS$ diagram}

\begin{align}
\Sigma^{(a)}_L =& -2 c^1_L c^2_L B_1(p^2,m_1^2,m_2^2) \\
\Sigma^{(a)}_R =& -2 c^1_R c^2_R B_1(p^2,m_1^2,m_2^2) \\
\Sigma^{(a)}_{SL} =& 2 c^1_R c^2_L m_1 B_0(p^2,m_1^2,m_2^2) \\
\Sigma^{(a)}_{SR} =& 2 c^1_L c^2_R m_1 B_0(p^2,m_1^2,m_2^2) 
\end{align}

\begin{align}
\dot \Sigma^{(a)}_L =& -2 c^1_L c^2_L \dot B_1(p^2,m_1^2,m_2^2) \\
\dot \Sigma^{(a)}_R =& -2 c^1_R c^2_R \dot B_1(p^2,m_1^2,m_2^2) \\
\dot \Sigma^{(a)}_{SL} =& 2 c^1_R c^2_L m_1 \dot B_0(p^2,m_1^2,m_2^2) \\
\dot \Sigma^{(a)}_{SR} =& 2 c^1_L c^2_R m_1 \dot B_0(p^2,m_1^2,m_2^2) 
\end{align}

\paragraph{(b) $FV$ diagram}

\begin{align}
\Sigma^{(b)}_L =& -4 c^1_R c^2_R \left(B_1(p^2,m_1^2,m_2^2) + \frac12 r \right) \\
\Sigma^{(b)}_R =& -4 c^1_L c^2_L \left(B_1(p^2,m_1^2,m_2^2) + \frac12 r \right) \\
\Sigma^{(b)}_{SL} =& -8 c^1_L c^2_R m_1 \left(B_0(p^2,m_1^2,m_2^2) - \frac12 r \right) \\
\Sigma^{(b)}_{SR} =& -8 c^1_R c^2_L m_1 \left(B_0(p^2,m_1^2,m_2^2) - \frac12 r \right) 
\end{align}

\begin{align}
\dot \Sigma^{(b)}_L =& -4 c^1_R c^2_R \dot  B_1(p^2,m_1^2,m_2^2) \\
\dot \Sigma^{(b)}_R =& -4 c^1_L c^2_L \dot  B_1(p^2,m_1^2,m_2^2)  \\
\dot \Sigma^{(b)}_{SL} =& -8 c^1_L c^2_R m_1 \dot  B_0(p^2,m_1^2,m_2^2) \\
\dot \Sigma^{(b)}_{SR} =& -8 c^1_R c^2_L m_1 \dot  B_0(p^2,m_1^2,m_2^2)
\end{align}

For Majorana fermions, an additional overall factor $\frac12$ is present. 

\subsubsection{Scalar}
\begin{figure}[h] 
\centering 
\includegraphics[width=0.75\linewidth]{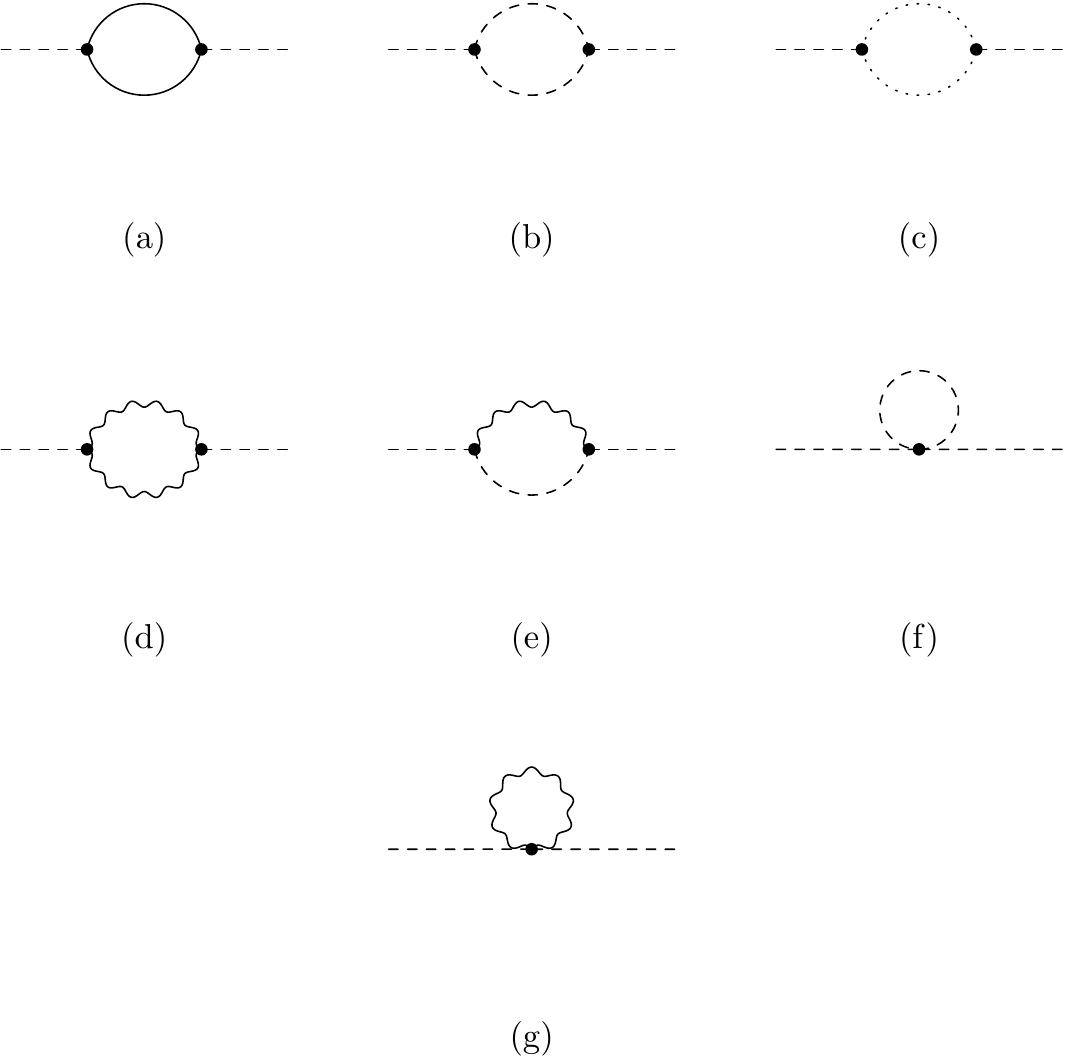}
\caption{Generic one-loop diagrams for the scalar self-energy.}
\label{fig:WV_S}
\end{figure} 

\paragraph{(a) $FF$ diagram}
\begin{align}
\Pi^{(a)}_{SS} = & (c^1_L c^2_L  + c^1_R c^2_R) G_0(p^2,m_1^2,m_2^2) -2 (c^1_L c^2_R + c^1_R c^2_L) B_0(p^2,m_1^2,m_2^2) \\
\dot \Pi^{(a)}_{SS} = & (c^1_L c^2_L  + c^1_R c^2_R) \dot G_0(p^2,m_1^2,m_2^2) -2 (c^1_L c^2_R + c^1_R c^2_L) \dot B_0(p^2,m_1^2,m_2^2)
\end{align}
with
\begin{align}
G_0(p^2,m_1^2,m_2^2) = & - A_0(m_1^2) - A_0(m_2^2) + (p^2 - m_1^2 - m_2^2) B_0(p^2,m_1^2,m_2^2) \\
\dot G_0(p^2,m_1^2,m_2^2) = & (p^2 - m_1^2 - m_2^2) \dot B_0(p^2,m_1^2,m_2^2) + B_0(p^2,m_1^2,m_2^2)
\end{align}
\paragraph{(b) $SS$ diagram}
\begin{align}
\Pi^{(b)}_{SS} = & c^1 c^2 B_0(p^2,m_1^2,m_2^2) \\
\dot \Pi^{(b)}_{SS} = & c^1 c^2 \dot B_0(p^2,m_1^2,m_2^2)
\end{align}

\paragraph{(c) $UU$ diagram}
\begin{align}
\Pi^{(c)}_{SS} = & -c^1 c^2 B_0(p^2,m_1^2,m_2^2) \\
\dot \Pi^{(c)}_{SS} = &  -c^1 c^2 \dot B_0(p^2,m_1^2,m_2^2)
\end{align}

\paragraph{(d) $VV$ diagram}
\begin{align}
\Pi^{(d)}_{SS} = & 4 c^1 c^2 \left(B_0(p^2,m_1^2,m_2^2) - \frac12 r \right) \\
\dot \Pi^{(d)}_{SS} = & 4 c^1 c^2 \dot B_0(p^2,m_1^2,m_2^2)
\end{align}

\paragraph{(e) $SV$ diagram}
\begin{align}
\Pi^{(b)}_{SS} = & c^1 c^2 F_0(p^2,m_1^2,m_2^2) \\
\dot \Pi^{(b)}_{SS} = & c^1 c^2 \dot F_0(p^2,m_1^2,m_2^2)
\end{align}
with 
\begin{align}
F_0(p^2,m_1^2,m_2^2) = & A_0(m_1^2)-2 A_0(m_2^2) - (2 p^2 + 2 m_1^2 - m_2^2) B_0(p^2,m_1^2,m_2^2) \\
\dot F_0(p^2,m_1^2,m_2^2) = & - 2 B_0(p^2,m_1^2,m_2^2) - (2 p^2 + 2 m_1^2 - m_2^2) \dot B_0(p^2,m_1^2,m_2^2)
\end{align}

\paragraph{(f) $S$ diagram}
\begin{align}
\Pi^{(f)}_{SS} = & - c^1 A_0(m_1^2) \\
\dot \Pi^{(f)}_{SS} = & 0
\end{align}

\paragraph{(g) $V$ diagram}
\begin{align}
\Pi^{(f)}_{SS} = & c^1 \left(A_0(m_1^2) - \frac{1}{2} r m_1^2\right) \\
\dot \Pi^{(f)}_{SS} = & 0
\end{align}

\subsubsection{Gauge boson}
\begin{figure}[h] 
\centering 
\includegraphics[width=0.75\linewidth]{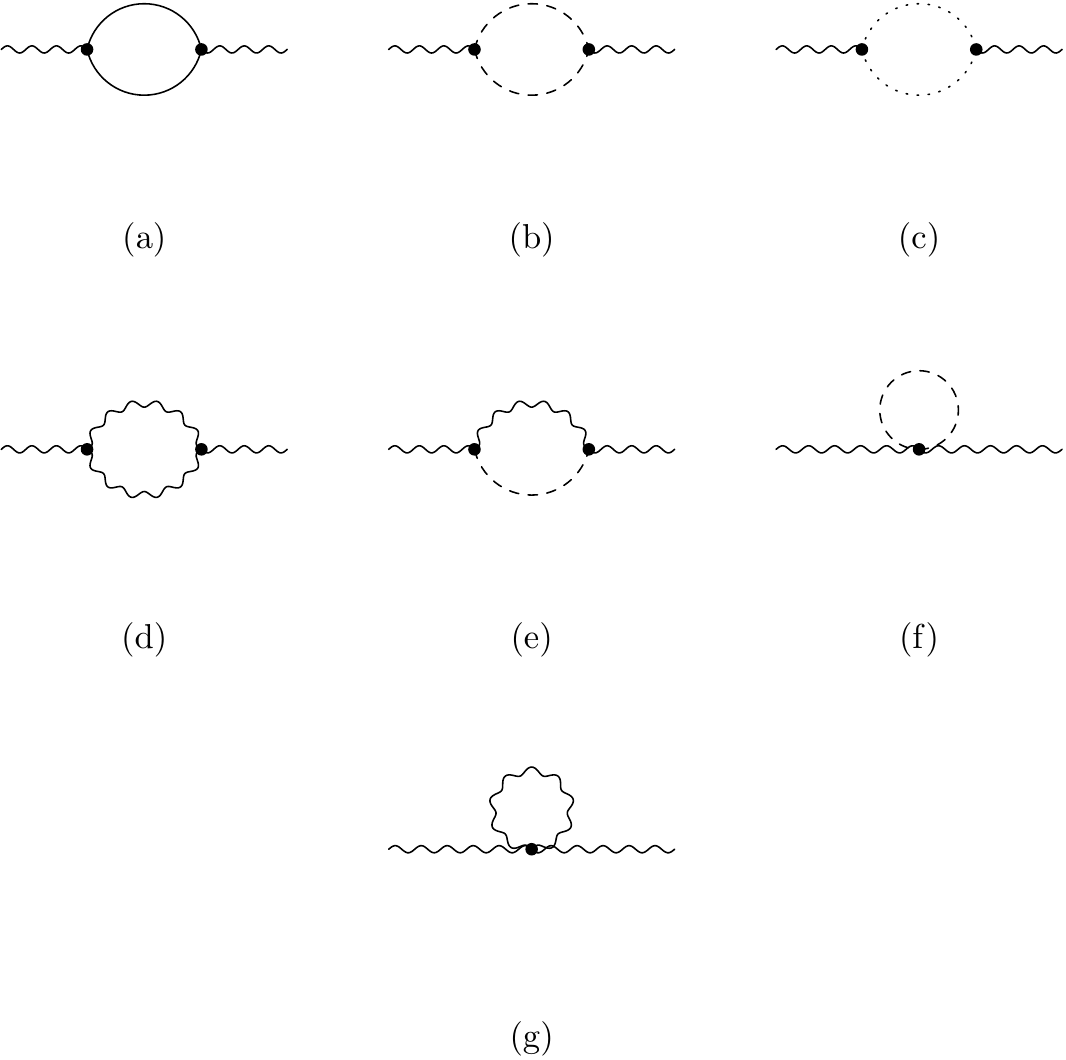} 
\caption{Generic one-loop diagrams for the gauge boson self-energy.}
\label{fig:WV_V}
\end{figure} 

\paragraph{(a) $FF$ diagram}
\begin{align}
\Pi^{(a)}_{VV} = & (c^1_L c^2_L + c^1_R c^2_R) H_0(p^2,m_1^2,m_2^2) + 4(c^1_L c^2_R + c^1_R c^2_L) B_0(p^2,m_1^2,m_2^2) \\
\dot \Pi^{(a)}_{VV} = &  (c^1_L c^2_L + c^1_R c^2_R) \dot H_0(p^2,m_1^2,m_2^2) + 4(c^1_L c^2_R + c^1_R c^2_L)\dot B_0(p^2,m_1^2,m_2^2) \\
\end{align}
with
\begin{align}
H_0(p^2,m_1^2,m_2^2) = &  4 B_{00}(p^2,m_1^2,m_2^2) - A_0(m_1^2) - A_0(m_2^2)\nn\\& + (p^2 - m_1^2 - m_2^2) B_0(p^2,m_1^2,m_2^2)   \\
\dot H_0(p^2,m_1^2,m_2^2) = & 4 \dot B_{00}(p^2,m_1^2,m_2^2) + (p^2 - m_1^2 - m_2^2) \dot B_0(p^2,m_1^2,m_2^2) + B_0(p^2,m_1^2,m_2^2)
\end{align}

\paragraph{(b) $SS$ diagram}
\begin{align}
\Pi^{(b)}_{VV} = & -4 c^1 c^2 B_{00}(p^2,m_1^2,m_2^2)   \\
\dot \Pi^{(b)}_{VV} = & -4 c^1 c^2 \dot B_{00}(p^2,m_1^2,m_2^2) 
\end{align}

\paragraph{(c) $UU$ diagram}
\begin{align}
\Pi^{(c)}_{VV} = & c^1 c^2 B_{00}(p^2,m_1^2,m_2^2)   \\
\dot \Pi^{(c)}_{VV} = &  c^1 c^2 \dot B_{00}(p^2,m_1^2,m_2^2) 
\end{align}

\paragraph{(d) $VV$ diagram}
\begin{align}
\Pi^{(d)}_{VV} = & -c^1 c^2 V_0(p^2,m_1^2,m_2^2)   \\
\dot \Pi^{(d)}_{VV} = & - c^1 c^2 \dot V_0(p^2,m_1^2,m_2^2) 
\end{align}
with
\begin{align}
V_0(p^2,m_1^2,m_2^2) = & 10 B_{00}(p^2,m_1^2,m_2^2) + (m_1^2+m_2^2 + 4p^2) B_0(p^2,m_1^2,m_2^2) \nn \\ 
&  + A_0(m_1^2) + A_0(m_2^2) - 2r \left(m_1^2-m_2^2 - \frac{1}{3} p^2 \right) \\
\dot V_0(p^2,m_1^2,m_2^2) = & 10 \dot B_{00}(p^2,m_1^2,m_2^2) \nn\\& + (m_1^2+m_2^2 + 4p^2) \dot B_0(p^2,m_1^2,m_2^2) + 4 B_0(p^2,m_1^2,m_2^2)
\end{align}

\paragraph{(e) $SV$ diagram}
\begin{align}
\Pi^{(e)}_{VV} = &  c^1 c^2 B_0(p^2,m_1^2,m_2^2)    \\
\dot \Pi^{(e)}_{VV} = &  c^1 c^2 \dot B_0(p^2,m_1^2,m_2^2) 
\end{align}

\paragraph{(f) $S$ diagram}
\begin{align}
\Pi^{(f)}_{VV} = &  c^1 A_0(m_1^2)   \\
\dot \Pi^{(f)}_{VV} = & 0
\end{align}

\paragraph{(g) $V$ diagram}
\begin{align}
\Pi^{(f)}_{VV} = &  - (4 c^1_1 + c^1_2 + c^1_3) A_0(m_1^2) + 2 m_1^2 c^1_1 r  \\
\dot \Pi^{(f)}_{VV} = & 0
\end{align}

\subsection{Vertex corrections}

\subsubsection{Fermion to fermion and scalar decays} 
\begin{figure}[h] 
\centering 
\includegraphics[width=0.75\linewidth]{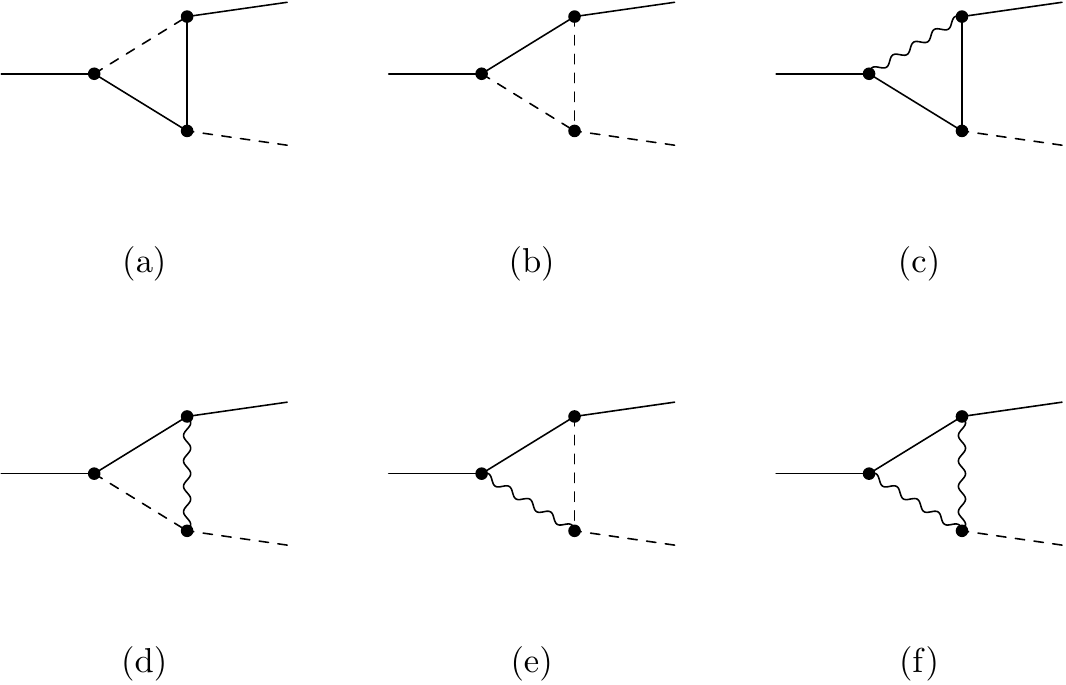} 
\caption{Generic diagrams contributing to $F\to FS$ decays.}
\label{fig:FFS}
\end{figure} 
\paragraph{(a) $SFF$ diagram  } 
\begin{align} 
M^{(a)}_1 = &i \Big[B_0 c^1_R c^2_R c^3_L + C_1 c^1_R c^2_R c^3_L p_0^2 - C_2 c^1_L c^2_L c^3_R p_0 p_1 - C_0 c^1_R c^2_R c^3_L p_1^2 - C_1 c^1_R c^2_R c^3_L p_1^2 \nn \\ & 
- C_2 c^1_R c^2_R c^3_L p_1^2 + C_0 c^1_R c^2_R c^3_L m_1^2 + C_1 c^1_L c^2_R c^3_L p_0 m_2 - C_1 c^1_R c^2_L c^3_R p_1 m_2   \nn \\ & 
- C_2 c^1_R c^2_L c^3_R p_1 m_2 +(C_1 c^1_L c^2_R c^3_R p_0 - (C_1 + C_2) c^1_R c^2_L c^3_L p_1 \nn\\&
+ C_0 (c^1_L c^2_R c^3_R p_0 - c^1_R c^2_L c^3_L p_1 + c^1_R c^2_R c^3_R m_2)) m_3\Big]\\ 
M^{(a)}_2 = & M^{(a)}_1 | L \leftrightarrow R  
\end{align} 
with $B_i = B_i(p_2^2, m_2^2, m_3^2)$ and $C_i = C_i(p_2^2, p_0^2, p_1^2, m_3^2, m_2^2, m_1^2)$. 

\paragraph{(b) $FSS$ diagram  } 
\begin{align} 
M^{(b)}_1 = &-i c^3 (C_2 c^1_L c^2_R p_0 + C_1 c^1_R c^2_L p_1 - C_0 c^1_R c^2_R m_1)\\ 
M^{(b)}_2 = & M^{(b)}_1 | L \leftrightarrow R  
\end{align} 
with $C_i = C_i(p_1^2, p_2^2, p_0^2, m_1^2, m_3^2, m_2^2)$.

\paragraph{(c) $VFF$ diagram  } 
\begin{align} 
M^{(c)}_1 = &-2 i \Big[2 B_0 c^1_R c^2_L c^3_R - c^1_L c^2_L p_0 (C_1 c^3_R m_2 + (C_0 + C_1) c^3_L m_3)   \nn \\ & 
+ c^1_R (- c^2_L c^3_R (r + 2 C_1 (- p_0^2 + p_1^2) + C_2 (p_0^2 + 3 p_1^2 - p_2^2) + 2 C_0 (p_1 - m_1) (p_1 + m_1))  \nn \\ &
+ 2 C_0 c^2_L c^3_L m_2 m_3 + c^2_R p_1 ((C_1 + C_2) c^3_L m_2 + (C_0 + C_1 + C_2) c^3_R m_3))\Big]\\ 
M^{(c)}_2 = & M^{(c)}_1 | L \leftrightarrow R  
\end{align} 
with $B_i = B_i(p_2^2, m_2^2, m_3^2)$ and $C_i = C_i(p_2^2, p_0^2, p_1^2, m_3^2, m_2^2, m_1^2)$.

\paragraph{(d) $FSV$ diagram  } 
\begin{align} 
M^{(d)}_1 = &-i c^3 \Big[B_0 c^1_R c^2_L + 2 C_1 c^1_R c^2_L p_0^2 + 2 C_2 c^1_R c^2_L p_0^2 -2 C_1 c^1_L c^2_R p_0 p_1 - C_2 c^1_L c^2_R p_0 p_1 + C_1 c^1_R c^2_L p_1^2\nn \\ & 
 -2 C_1 c^1_R c^2_L p_2^2 - ((2 C_0 + C_2) c^1_L c^2_L p_0 + (- C_0 + C_1) c^1_R c^2_R p_1) m_1 + C_0 c^1_R c^2_L m_1^2\Big]\\ 
M^{(d)}_2 = & M^{(d)}_1 | L \leftrightarrow R  
\end{align} 
with $B_i = B_i(p_2^2, m_2^2, m_3^2)$ and $C_i = C_i(p_1^2, p_2^2, p_0^2, m_1^2, m_3^2, m_2^2)$.

\paragraph{(e) $FVS$ diagram  } 
\begin{align} 
M^{(e)}_1 = &i c^3 \Big[B_0 c^1_R c^2_R + C_2 c^1_R c^2_R p_0^2 - C_1 c^1_L c^2_L p_0 p_1 -2 C_2 c^1_L c^2_L p_0 p_1 + 2 C_1 c^1_R c^2_R p_1^2 + 2 C_2 c^1_R c^2_R p_1^2 \nn \\ &
-2 C_2 c^1_R c^2_R p_2^2 - ((- C_0 + C_2) c^1_L c^2_R p_0 + (2 C_0 + C_1) c^1_R c^2_L p_1) m_1 + C_0 c^1_R c^2_R m_1^2\Big]\\ 
M^{(e)}_2 = & M^{(e)}_1 | L \leftrightarrow R  
\end{align} 
with $B_i = B_i(p_2^2, m_3^2, m_2^2)$ and $C_i = C_i(p_1^2, p_2^2, p_0^2, m_1^2, m_3^2, m_2^2)$. 

\paragraph{(f) $FVV$ diagram  } 
\begin{align} 
M^{(f)}_1 = &2 i c^3 (C_2 c^1_L c^2_L p_0 + C_1 c^1_R c^2_R p_1 + 2 C_0 c^1_R c^2_L m_1)\\ 
M^{(f)}_2 = & M^{(f)}_1 | L \leftrightarrow R  
\end{align} 
with $C_i = C_i(p_1^2, p_2^2, p_0^2, m_1^2, m_3^2, m_2^2)$.

\subsubsection{Fermion to fermion and gauge boson decays} 
\begin{figure}[h] 
\centering 
\includegraphics[width=0.75\linewidth]{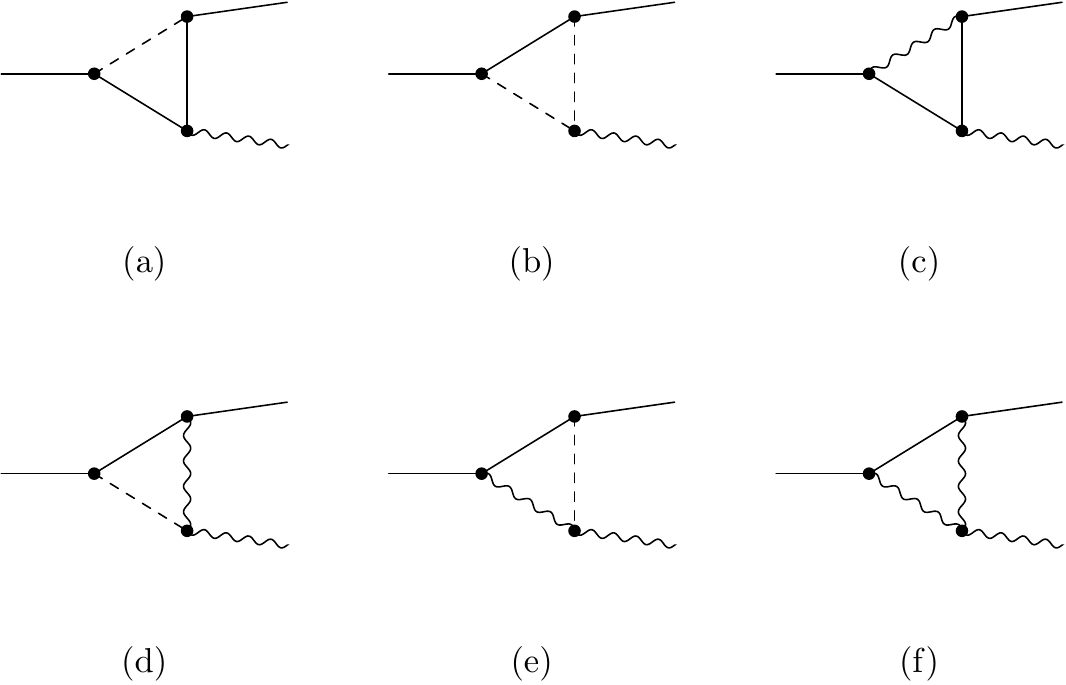} 
\caption{Generic diagrams contributing to $F\to FV$ decays.}
\label{fig:FFV}
\end{figure} 
\paragraph{(a) $SFF$ diagram  } 
\begin{align} 
M^{(a)}_1 = &-i \Big[B_0 c^1_R c^2_L c^3_R -2 C_{00} c^1_R c^2_L c^3_R - C_0 c^1_R c^2_L c^3_R p_0^2 - C_1 c^1_R c^2_L c^3_R p_0^2 - C_2 c^1_R c^2_L c^3_R p_0^2 \nn \\ & 
+ C_2 c^1_L c^2_R c^3_L p_0 p_1 + C_1 c^1_R c^2_L c^3_R p_1^2 + C_0 c^1_R c^2_L c^3_R m_1^2 - C_0 c^1_L c^2_L c^3_R p_0 m_2  \nn \\ &
- C_1 c^1_L c^2_L c^3_R p_0 m_2 - C_2 c^1_L c^2_L c^3_R p_0 m_2 - C_0 c^1_R c^2_R c^3_L p_1 m_2 - C_1 c^1_R c^2_R c^3_L p_1 m_2  \nn \\ &
+ (C_1 c^1_L c^2_L c^3_L p_0 + C_2 c^1_L c^2_L c^3_L p_0 + C_1 c^1_R c^2_R c^3_R p_1 - C_0 c^1_R c^2_L c^3_L m_2) m_3\Big]\\ 
M^{(a)}_2 = & M^{(a)}_1 | L \leftrightarrow R \\ 
M^{(a)}_3 = &2 i \Big[C_{12} (- c^1_L c^2_R c^3_L p_0 + c^1_R c^2_L c^3_R p_1) + c^2_R (- C_{22} c^1_L c^3_L p_0 + (C_0 + C_1) c^1_R c^3_L m_2 \nn \\ & 
+  C_2 c^3_L (- c^1_L p_0 + c^1_R m_2) - C_1 c^1_R c^3_R m_3)\Big]\\ 
M^{(a)}_4 = & M^{(a)}_3 | L \leftrightarrow R  
\end{align} 
with $B_i = B_i(p_2^2, m_2^2, m_3^2)$ and $C_i = C_i(p_2^2, p_1^2, p_0^2, m_2^2, m_3^2, m_1^2)$. 

\paragraph{(b) $FSS$ diagram  } 
\begin{align} 
M^{(b)}_1 = &2 i C_{00} c^1_R c^2_L c^3\\ 
M^{(b)}_2 = & M^{(b)}_1 | L \leftrightarrow R \\ 
M^{(b)}_3 = &-2 i c^3 \Big[(C_2 + C_{22}) c^1_L c^2_R p_0 + (C_1 + C_{11}) c^1_R c^2_L p_1  \nn \\ &
+ C_{12} (c^1_L c^2_R p_0 + c^1_R c^2_L p_1) - (C_0 + C_1 + C_2) c^1_R c^2_R m_1\Big]\\ 
M^{(b)}_4 = & M^{(b)}_3 | L \leftrightarrow R  
\end{align} 
with $C_i = C_i(p_1^2, p_2^2, p_0^2, m_1^2, m_3^2, m_2^2)$. 

\paragraph{(c) $VFF$ diagram  } 
\begin{align} 
M^{(c)}_1 = &-i \Big[2 B_0 c^1_L c^2_L c^3_L -4 C_{00} c^1_L c^2_L c^3_L - c^1_L c^2_L c^3_L r -2 C_0 c^1_L c^2_L c^3_L p_0^2 -2 C_1 c^1_L c^2_L c^3_L p_0^2 \nn \\ & 
-4 C_2 c^1_L c^2_L c^3_L p_0^2 -2 C_2 c^1_R c^2_R c^3_R p_0 p_1 + 2 C_1 c^1_L c^2_L c^3_L p_1^2 -2 C_2 c^1_L c^2_L c^3_L p_1^2  \nn \\ &
+ 2 C_2 c^1_L c^2_L c^3_L p_2^2 + 2 C_0 c^1_L c^2_L c^3_L m_1^2 -2 C_0 c^1_L c^2_L c^3_R m_2 m_3\Big]\\ 
M^{(c)}_2 = & M^{(c)}_1 | L \leftrightarrow R \\ 
M^{(c)}_3 = &-4 i \Big[C_{22} c^1_R c^2_R c^3_R p_0 + C_{12} (c^1_R c^2_R c^3_R p_0 - c^1_L c^2_L c^3_L p_1)\nn\\& + C_2 c^1_L (- c^2_L c^3_L p_1 + c^2_R c^3_R m_2 + c^2_R c^3_L m_3)\Big]\\ 
M^{(c)}_4 = & M^{(c)}_3 | L \leftrightarrow R  
\end{align} 
with $B_i = B_i(p_2^2, m_2^2, m_3^2)$ and $C_i = C_i(p_2^2, p_1^2, p_0^2, m_2^2, m_3^2, m_1^2)$.

\paragraph{(d) $FSV$ diagram  } 
\begin{align} 
M^{(d)}_1 = &-i c^3 (- C_2 c^1_L c^2_L p_0 + C_1 c^1_R c^2_R p_1 + C_0 c^1_R c^2_L m_1)\\ 
M^{(d)}_2 = & M^{(d)}_1 | L \leftrightarrow R \\ 
M^{(d)}_3 = &-2 i C_1 c^1_R c^2_R c^3\\ 
M^{(d)}_4 = & M^{(d)}_3 | L \leftrightarrow R  
\end{align} 
with $C_i = C_i(p_1^2, p_2^2, p_0^2, m_1^2, m_3^2, m_2^2)$.

\paragraph{(e) $FVS$ diagram  } 
\begin{align} 
M^{(e)}_1 = &-i c^3 (C_2 c^1_R c^2_L p_0 - C_1 c^1_L c^2_R p_1 + C_0 c^1_L c^2_L m_1)\\ 
M^{(e)}_2 = & M^{(e)}_1 | L \leftrightarrow R \\ 
M^{(e)}_3 = &-2 i C_2 c^1_L c^2_R c^3\\ 
M^{(e)}_4 = & M^{(e)}_3 | L \leftrightarrow R  
\end{align} 
with $C_i = C_i(p_1^2, p_2^2, p_0^2, m_1^2, m_3^2, m_2^2)$. 

\paragraph{(f) $FVV$ diagram  } 
\begin{align} 
M^{(f)}_1 = &i c^3 \Big[2 B_0 c^1_L c^2_L + 4 C_{00} c^1_L c^2_L - c^1_L c^2_L r + 2 C_1 c^1_L c^2_L p_0^2 + 3 C_2 c^1_L c^2_L p_0^2 + 3 C_1 c^1_R c^2_R p_0 p_1 \nn \\ &
+ 3 C_2 c^1_R c^2_R p_0 p_1 + 3 C_1 c^1_L c^2_L p_1^2 + 2 C_2 c^1_L c^2_L p_1^2 -2 C_1 c^1_L c^2_L p_2^2 -2 C_2 c^1_L c^2_L p_2^2 \nn \\ &
+ 3 C_0 (c^1_R c^2_L p_0 + c^1_L c^2_R p_1) m_1 
+ 2 C_0 c^1_L c^2_L m_1^2\Big]\\ 
M^{(f)}_2 = & M^{(f)}_1 | L \leftrightarrow R \\ 
M^{(f)}_3 = &-2 i c^3 \Big[- C_1 c^1_R c^2_R p_0 + 2 C_{12} c^1_R c^2_R p_0 + 2 C_{22} c^1_R c^2_R p_0 + 2 C_{11} c^1_L c^2_L p_1 \nn \\ &
 + 2 C_{12} c^1_L c^2_L p_1 - C_2 c^1_L c^2_L p_1 + 3 (C_1 + C_2) c^1_L c^2_R m_1\Big]\\ 
M^{(f)}_4 = & M^{(f)}_3 | L \leftrightarrow R  
\end{align} 
with $B_i = B_i(p_2^2, m_2^2, m_3^2)$ and $C_i = C_i(p_1^2, p_2^2, p_0^2, m_1^2, m_3^2, m_2^2)$.

\subsubsection{Scalar to two fermion decays} 
\begin{figure}[h] 
\centering 
\includegraphics[width=0.75\linewidth]{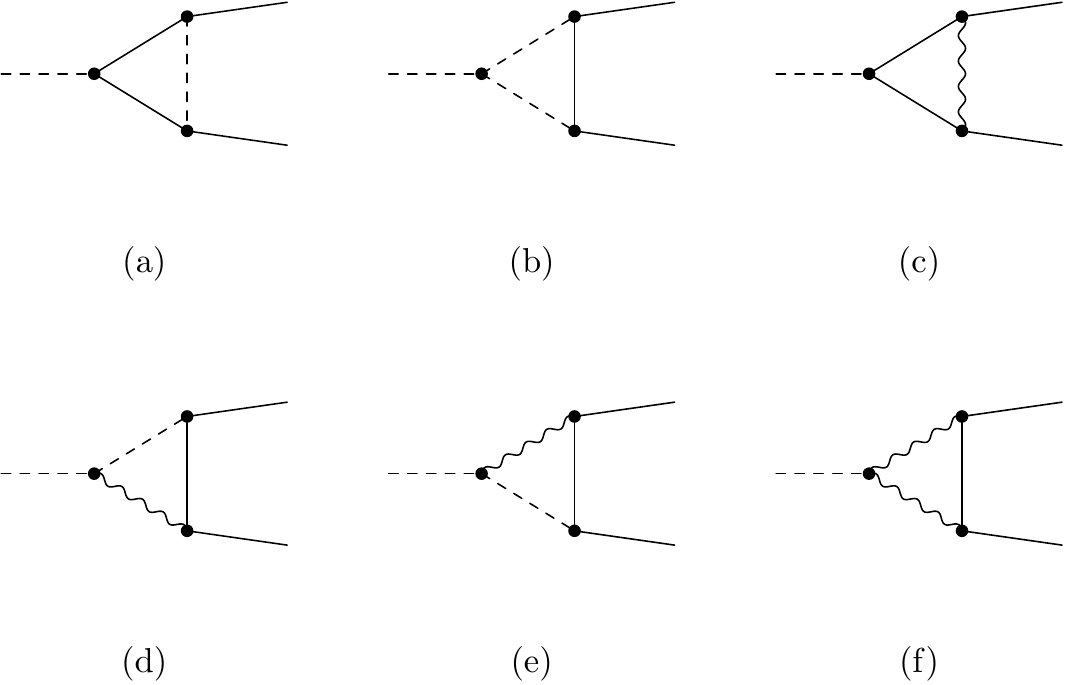} 
\caption{Generic diagrams contributing to $S\to FF$ decays.}
\label{fig:SFF}
\end{figure} 
\paragraph{(a) $FFS$ diagram  } 
\begin{align} 
M^{(a)}_1 = &i \Big[B_0 c^1_L c^2_R c^3_R + C_2 c^1_L c^2_R c^3_R p_1^2 - C_2 c^1_R c^2_L c^3_L p_1 p_2 - C_0 c^1_L c^2_L c^3_R p_1 m_1 - C_2 c^1_L c^2_L c^3_R p_1 m_1 \nn \\ &
+ C_0 c^1_R c^2_R c^3_L p_2 m_1 + C_0 c^1_L c^2_R c^3_R m_1^2 - C_2 c^1_R c^2_L c^3_R p_1 m_2 + C_0 c^1_R c^2_R c^3_R m_1 m_2  \nn \\ &
+ C_1 (c^1_L c^2_R c^3_R p_0^2 - c^1_L c^2_L c^3_R p_1 m_1 + c^1_R c^2_R c^3_L p_2 m_1 - c^1_R c^2_L c^3_R p_1 m_2 + c^1_L c^2_R c^3_L p_2 m_2)\Big]\\ 
M^{(a)}_2 = & M^{(a)}_1 | L \leftrightarrow R  
\end{align} 
with $B_i = B_i(p_2^2, m_2^2, m_3^2)$ and $C_i = C_i(p_0^2, p_2^2, p_1^2, m_1^2, m_2^2, m_3^2)$.

\paragraph{(b) $SSF$ diagram  } 
\begin{align} 
M^{(b)}_1 = &-i c^1 (C_1 c^2_L c^3_R p_1 + C_2 c^2_R c^3_L p_2 - C_0 c^2_R c^3_R m_3)\\ 
M^{(b)}_2 = & M^{(b)}_1 | L \leftrightarrow R  
\end{align}
with $C_i = C_i(p_1^2, p_0^2, p_2^2, m_3^2, m_1^2, m_2^2)$.

\paragraph{(c) $FFV$ diagram  } 
\begin{align} 
M^{(c)}_1 = &-2 i \Big[2 B_0 c^1_R c^2_L c^3_R - (C_0 + C_1) c^1_L c^2_L c^3_L p_2 m_1 + c^1_L c^3_R ((C_1 + C_2) c^2_R p_1 + 2 C_0 c^2_L m_1) m_2 \nn \\ &
+ c^1_R ((C_0 + C_1 + C_2) c^2_R c^3_R p_1 m_1 + c^2_L (c^3_R (- r + 2 C_1 p_0^2 + C_2 (p_0^2 + p_1^2 - p_2^2) \nn \\ & + 2 C_0 m_1^2) - C_1 c^3_L p_2 m_2))\Big]\\ 
M^{(c)}_2 = & M^{(c)}_1 | L \leftrightarrow R  
\end{align} 
with $B_i = B_i(p_2^2, m_2^2, m_3^2)$ and $C_i = C_i(p_0^2, p_2^2, p_1^2, m_1^2, m_2^2, m_3^2)$.

\paragraph{(d) $SVF$ diagram  } 
\begin{align} 
M^{(d)}_1 = &i c^1 \Big[B_0 c^2_R c^3_R - C_2 c^2_R c^3_R p_0^2 - C_0 c^2_R c^3_R p_1^2 + C_2 c^2_R c^3_R p_1^2 - C_1 c^2_L c^3_L p_1 p_2 -2 C_2 c^2_L c^3_L p_1 p_2\nn \\ &
+ C_0 c^2_R c^3_R m_1^2 - ((2 C_0 + C_1) c^2_L c^3_R p_1 + (- C_0 + C_2) c^2_R c^3_L p_2) m_3\Big]\\ 
M^{(d)}_2 = & M^{(d)}_1 | L \leftrightarrow R  
\end{align} 
with $B_i = B_i(p_2^2, m_3^2, m_2^2)$ and $C_i = C_i(p_1^2, p_0^2, p_2^2, m_3^2, m_1^2, m_2^2)$.

\paragraph{(e) $VSF$ diagram  } 
\begin{align} 
M^{(e)}_1 = &-i c^1 \Big[B_0 c^2_L c^3_R + C_2 c^2_L c^3_R p_0^2 - C_0 c^2_L c^3_R p_1^2 - C_2 c^2_L c^3_R p_1^2 - C_2 c^2_R c^3_L p_1 p_2 + C_2 c^2_L c^3_R p_2^2 \nn \\ &
+ C_0 c^2_L c^3_R m_1^2 + C_0 c^2_R c^3_R p_1 m_3 -2 C_0 c^2_L c^3_L p_2 m_3 - C_2 c^2_L c^3_L p_2 m_3 \nn \\ &
- C_1 (c^2_L c^3_R (2 p_0^2 + p_1^2 -2 p_2^2) + c^2_R p_1 (2 c^3_L p_2 + c^3_R m_3))\Big]\\ 
M^{(e)}_2 = & M^{(e)}_1 | L \leftrightarrow R  
\end{align} 
with $B_i = B_i(p_2^2, m_3^2, m_2^2)$ and $C_i = C_i(p_1^2, p_0^2, p_2^2, m_3^2, m_1^2, m_2^2)$.

\paragraph{(f) $VVF$ diagram  } 
\begin{align} 
M^{(f)}_1 = &2 i c^1 (C_1 c^2_R c^3_R p_1 + C_2 c^2_L c^3_L p_2 + 2 C_0 c^2_L c^3_R m_3)\\ 
M^{(f)}_2 = & M^{(f)}_1 | L \leftrightarrow R  
\end{align} 
with $C_i = C_i(p_1^2, p_0^2, p_2^2, m_3^2, m_1^2, m_2^2)$.

\subsubsection{Scalar to two scalar decays} 
\begin{figure}[h] 
\centering 
\includegraphics[width=0.90\linewidth]{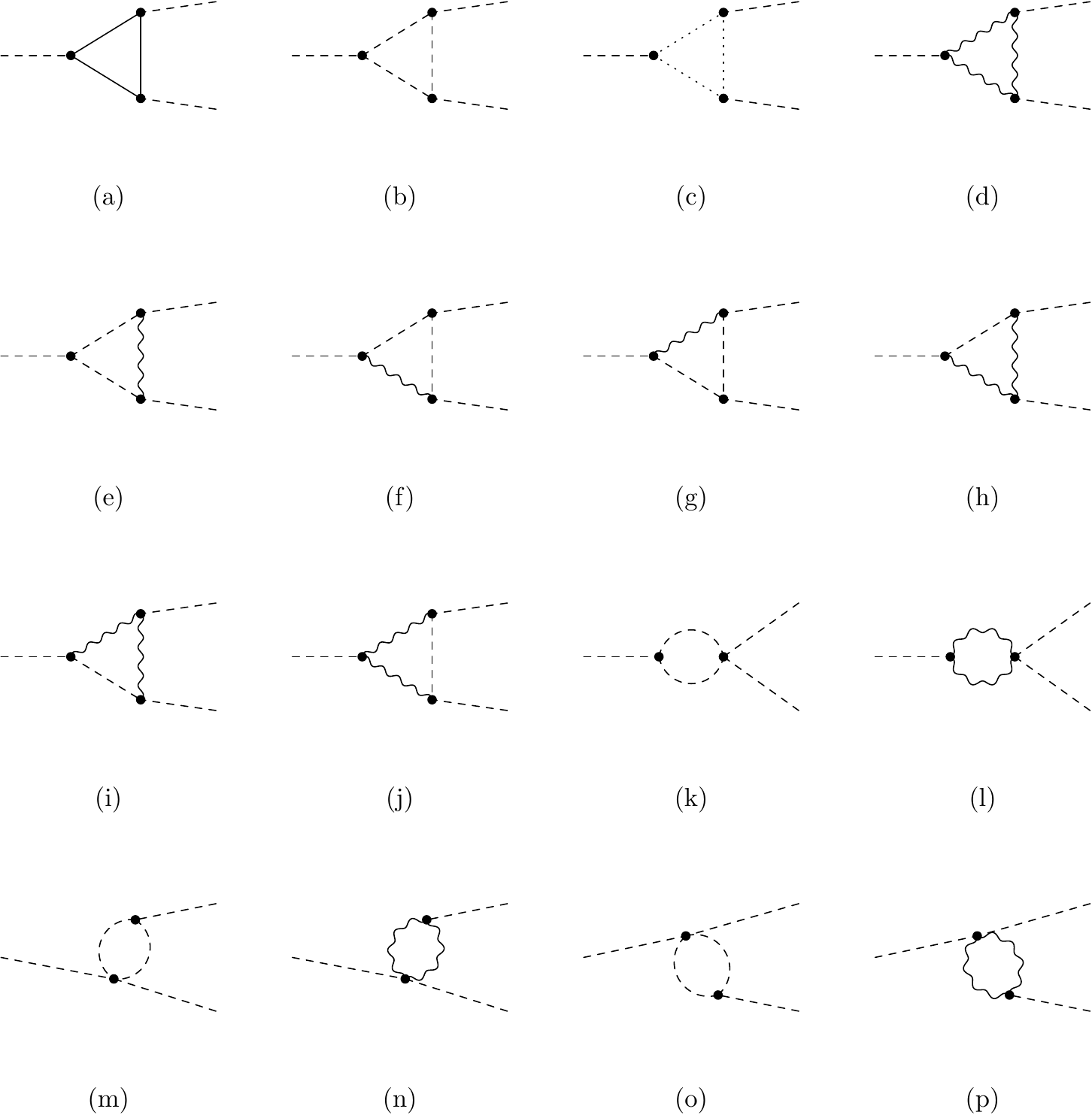} 
\caption{Generic diagrams contributing to $S\to SS$ decays.}
\label{fig:SSS}
\end{figure} 
\paragraph{(a) $FFF$ diagram  } 
\begin{align} 
M^{(a)} = &i \Big[C_1 c^1_R c^2_R c^3_L p_0^2 m_1 + 3 C_2 c^1_R c^2_R c^3_L p_0^2 m_1 + C_1 c^1_L c^2_L c^3_R p_0^2 m_1  \nn \\ &
+ 3 C_2 c^1_L c^2_L c^3_R p_0^2 m_1 + 3 C_1 c^1_R c^2_R c^3_L p_1^2 m_1 + C_2 c^1_R c^2_R c^3_L p_1^2 m_1 + 3 C_1 c^1_L c^2_L c^3_R p_1^2 m_1 \nn \\ &
+ C_2 c^1_L c^2_L c^3_R p_1^2 m_1 - C_1 c^1_R c^2_R c^3_L p_2^2 m_1 - C_2 c^1_R c^2_R c^3_L p_2^2 m_1 - C_1 c^1_L c^2_L c^3_R p_2^2 m_1\nn \\ &
- C_2 c^1_L c^2_L c^3_R p_2^2 m_1 + C_2 c^1_L c^2_R c^3_L p_0^2 m_2 + C_2 c^1_R c^2_L c^3_R p_0^2 m_2 + 2 C_1 c^1_L c^2_R c^3_L p_1^2 m_2 \nn \\ &
+ C_2 c^1_L c^2_R c^3_L p_1^2 m_2 + 2 C_1 c^1_R c^2_L c^3_R p_1^2 m_2 + C_2 c^1_R c^2_L c^3_R p_1^2 m_2 - C_2 c^1_L c^2_R c^3_L p_2^2 m_2\nn \\&
- C_2 c^1_R c^2_L c^3_R p_2^2 m_2 + (c^1_R c^2_L c^3_L + c^1_L c^2_R c^3_R) (2 C_2 p_0^2 + C_1 (p_0^2 + p_1^2 - p_2^2)) m_3 \nn \\ &
+ 2 B_0 (c^1_R (c^2_R c^3_L m_1 + c^2_L c^3_R m_2 + c^2_L c^3_L m_3) + c^1_L (c^2_L c^3_R m_1 + c^2_R c^3_L m_2 + c^2_R c^3_R m_3))\nn \\ &
+ C_0 m_1 (c^1_R (c^2_R c^3_L (p_0^2 + p_1^2 - p_2^2 + 2 m_1^2) + 2 c^2_R c^3_R m_2 m_3 + 2 c^2_L m_1 (c^3_R m_2 + c^3_L m_3)) \nn \\ &
+ c^1_L (c^2_L c^3_R (p_0^2 + p_1^2 - p_2^2 + 2 m_1^2) + 2 c^2_L c^3_L m_2 m_3+ 2 c^2_R m_1 (c^3_L m_2 + c^3_R m_3)))\Big]
\end{align} 
with $B_i = B_i(p_2^2, m_2^2, m_3^2)$ and $C_i = C_i(p_1^2, p_2^2, p_0^2, m_1^2, m_3^2, m_2^2)$.

\paragraph{(b) $SSS$ diagram  } 
\begin{align} 
M^{(b)} = &-i C_0(p_1^2, p_2^2, p_0^2, m_1^2, m_3^2, m_2^2)  c^1 c^2 c^3 
\end{align} 

\paragraph{(c) $UUU$ diagram  } 
\begin{align} 
M^{(c)} = &i C_0(p_1^2, p_2^2, p_0^2, m_1^2, m_3^2, m_2^2)  c^1 c^2 c^3 
\end{align} 

\paragraph{(d) $VVV$ diagram  } 
\begin{align} 
M^{(d)} = &4 i C_0(p_1^2, p_2^2, p_0^2, m_1^2, m_3^2, m_2^2)  c^1 c^2 c^3 
\end{align} 

\paragraph{(e) $SSV$ diagram  } 
\begin{align} 
M^{(e)} = &-i c^1 c^2 c^3 \Big[B_0 + (C_1 + C_2) (p^2_0 - p^2_1)\nn\\& + ( C_2 - C_1) p_2^2 + C_0 (p_2^2 - p_0^2  + m_1^2)\Big]
\end{align} 
with $B_i = B_i(p_2^2, m_2^2, m_3^2)$ and $C_i = C_i(p_1^2, p_2^2, p_0^2, m_1^2, m_3^2, m_2^2)$.

\paragraph{(f) $SVS$ diagram  } 
\begin{align} 
M^{(f)} = &-i c^1 c^2 c^3 \Big[B_0 - (C_1 + C_2) (p^2_0 - p^2_1)\nn\\& + (C_1 - C_2) p_2^2 + C_0 (- p_1^2 + p_2^2 + m_1^2)\Big]
\end{align} 
with $B_i = B_i(p_2^2, m_3^2, m_2^2)$ and $C_i = C_i(p_1^2, p_2^2, p_0^2, m_1^2, m_3^2, m_2^2)$.

\paragraph{(g) $VSS$ diagram  } 
\begin{align} 
M^{(g)} = &-i c^1 c^2 c^3 \Big[B_0 + C_1 p_0^2 + 3 C_2 p_0^2 + 3 C_1 p_1^2\nn\\& + C_2 p_1^2 - (C_1 + C_2) p_2^2 + C_0 (2 (p_0^2 + p_1^2 - p_2^2) + m_1^2)\Big]
\end{align} 
with $B_i = B_i(p_2^2, m_2^2, m_3^2)$ and $C_i = C_i(p_1^2, p_2^2, p_0^2, m_1^2, m_3^2, m_2^2)$.

\paragraph{(h) $SVV$ diagram  } 
\begin{align} 
M^{(h)} = &\frac{i}{2} c^1 c^2 c^3 \Big[2 B_0 - C_1 p_0^2 -3 C_2 p_0^2 -3 C_1 p_1^2\nn\\& - C_2 p_1^2 + (C_1 + C_2) p_2^2 + C_0 (p_0^2 + p_1^2 - p_2^2 + 2 m_1^2)\Big]
\end{align}
with $B_i = B_i(p_2^2, m_2^2, m_3^2)$ and $C_i = C_i(p_1^2, p_2^2, p_0^2, m_1^2, m_3^2, m_2^2)$.

\paragraph{(i) $VSV$ diagram  } 
\begin{align} 
M^{(i)} = &\frac{i}{2} c^1 c^2 c^3 \Big[2 B_0 + 2 C_1 (2 p_0^2 + p_1^2 -2 p_2^2)\nn\\& + C_2 (7 p_0^2 - p_1^2 + p_2^2) + 2 C_0 (3 p_0^2 - p_1^2 + p_2^2 + m_1^2)\Big]
\end{align} 
with $B_i = B_i(p_2^2, m_2^2, m_3^2)$ and $C_i = C_i(p_1^2, p_2^2, p_0^2, m_1^2, m_3^2, m_2^2)$. 

\paragraph{(j) $VVS$ diagram  } 
\begin{align} 
M^{(j)} = &\frac{i}{2} c^1 c^2 c^3 \Big[2 B_0 - C_1 p_0^2 + 2 C_2 p_0^2 + 7 C_1 p_1^2 + 4 C_2 p_1^2  \nn \\ & + (C_1 -4 C_2) p_2^2 + 2 C_0 (- p_0^2 + 3 p_1^2 + p_2^2 + m_1^2)\Big]
\end{align} 
with $B_i = B_i(p_2^2, m_3^2, m_2^2)$ and $C_i = C_i(p_1^2, p_2^2, p_0^2, m_1^2, m_3^2, m_2^2)$.

\paragraph{(k) $SS$ diagram  } 
\begin{align} 
M^{(k)} = &\frac{i}{2} B_0(p_0^2, m_1^2, m_2^2)  c^1 c^2 
\end{align} 

\paragraph{(l) $VV$ diagram  } 
\begin{align} 
M^{(l)} = &i c^1 c^2 (2 B_0(p_0^2, m_1^2, m_2^2)  - r) 
\end{align} 

\paragraph{(m) $SS$ diagram  } 
\begin{align} 
M^{(m)} = &\frac{i}{2} B_0(p_1^2, m_1^2, m_2^2)  c^1 c^2 
\end{align} 

\paragraph{(n) $VV$ diagram  } 
\begin{align} 
M^{(n)} = &i c^1 c^2 (2 B_0(p_1^2, m_1^2, m_2^2) - r) 
\end{align} 

\paragraph{(o) $SS$ diagram  } 
\begin{align} 
M^{(o)} = &\frac{i}{2} B_0(p_2^2, m_1^2, m_2^2)  c^1 c^3 
\end{align} 

\paragraph{(p) $VV$ diagram  } 
\begin{align} 
M^{(p)} = &i c^1 c^3 (2 B_0(p_2^2, m_1^2, m_2^2)  - r) 
\end{align} 

\subsubsection{Scalar to scalar and gauge boson decays} 
\begin{figure}[h] 
\centering 
\includegraphics[width=0.90\linewidth]{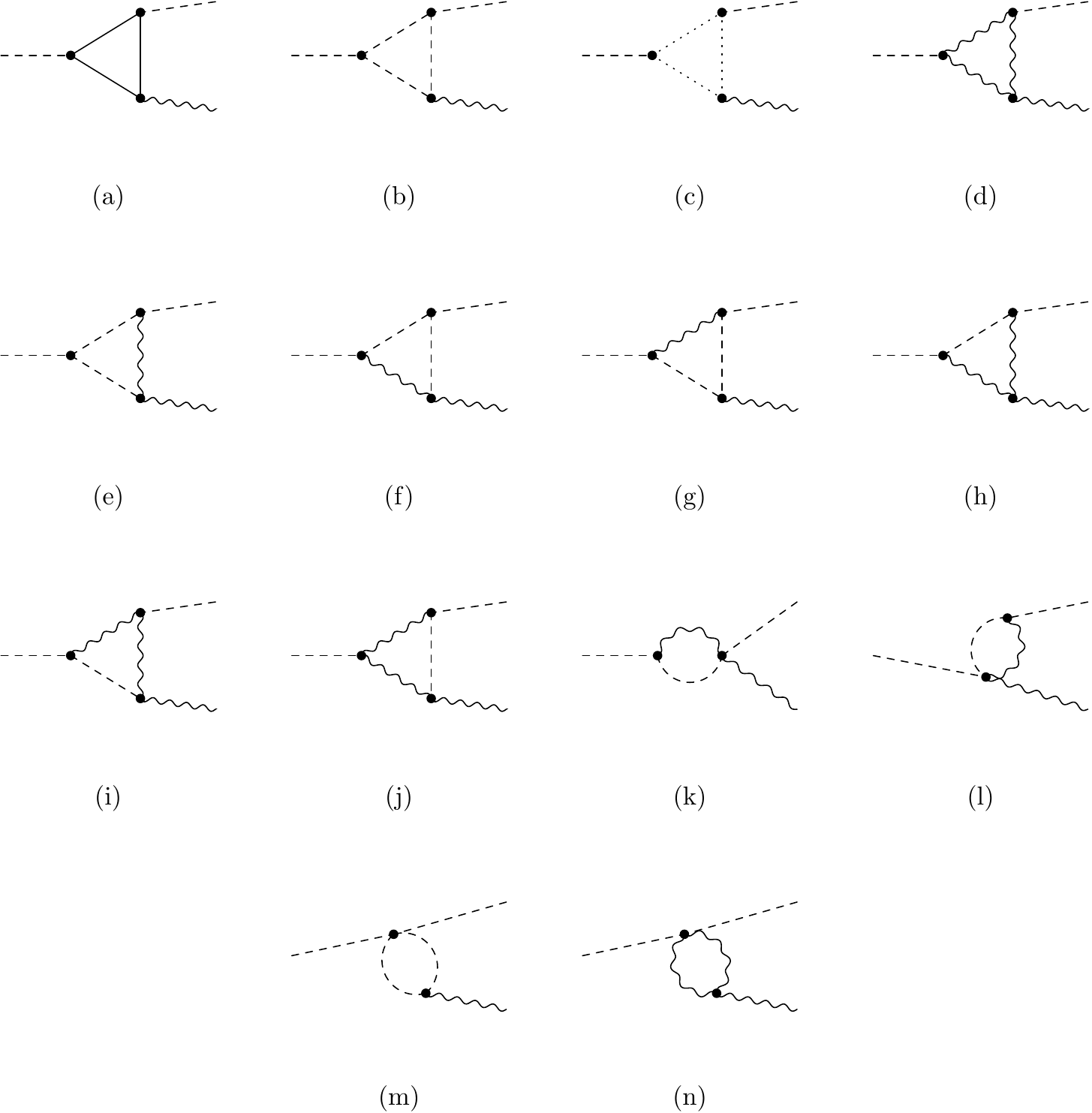} 
\caption{Generic diagrams contributing to $S\to SV$ decays.}
\label{fig:SSV}
\end{figure} 
\paragraph{(a) $FFF$ diagram  } 
\begin{align} 
M^{(a)} = &-2 i \Big[B_0 c^1_L c^2_R c^3_L + B_0 c^1_R c^2_L c^3_R + C_2 c^1_L c^2_R c^3_L p_0^2 + C_2 c^1_R c^2_L c^3_R p_0^2  \nn \\ &
+ C_1 c^1_L c^2_R c^3_L p_1^2 + C_1 c^1_R c^2_L c^3_R p_1^2 + 2 C_0 c^1_L c^2_R c^3_L m_1^2 + C_1 c^1_L c^2_R c^3_L m_1^2  \nn \\ &
+ C_2 c^1_L c^2_R c^3_L m_1^2 + 2 C_0 c^1_R c^2_L c^3_R m_1^2 + C_1 c^1_R c^2_L c^3_R m_1^2 + C_2 c^1_R c^2_L c^3_R m_1^2  \nn \\ &
+ C_0 c^1_R c^2_R c^3_L m_1 m_2 + C_1 c^1_R c^2_R c^3_L m_1 m_2 + C_2 c^1_R c^2_R c^3_L m_1 m_2 + C_0 c^1_L c^2_L c^3_R m_1 m_2\nn\\&
+ C_1 c^1_L c^2_L c^3_R m_1 m_2 + C_2 c^1_L c^2_L c^3_R m_1 m_2 + ((C_0 + C_1 + C_2) (c^1_L c^2_L c^3_L + c^1_R c^2_R c^3_R) m_1 \nn \\ &
+ (C_1 + C_2) (c^1_R c^2_L c^3_L + c^1_L c^2_R c^3_R) m_2) m_3\Big]
\end{align} 
with $B_i = B_i(p_2^2, m_2^2, m_3^2)$ and $C_i = C_i(p_1^2, p_2^2, p_0^2, m_1^2, m_3^2, m_2^2)$.

\paragraph{(b) $SSS$ diagram  } 
\begin{align} 
M^{(b)} = &-2 i (C_0 + C_1 + C_2) c^1 c^2 c^3 
\end{align} 
with $C_i = C_i(p_1^2, p_2^2, p_0^2, m_1^2, m_3^2, m_2^2)$.

\paragraph{(c) $UUU$ diagram  } 
\begin{align} 
M^{(c)} = &i (C_0 + C_1 + C_2) c^1 c^2 c^3
\end{align} 
with $C_i = C_i(p_1^2, p_2^2, p_0^2, m_1^2, m_3^2, m_2^2)$.

\paragraph{(d) $VVV$ diagram  } 
\begin{align} 
M^{(d)} = &-6 i (C_0 + C_1 + C_2) c^1 c^2 c^3
\end{align} 
with $C_i = C_i(p_1^2, p_2^2, p_0^2, m_1^2, m_3^2, m_2^2)$.

\paragraph{(e) $SSV$ diagram  } 
\begin{align} 
M^{(e)} = &-i (C_0 - C_1 - C_2) c^1 c^2 c^3
\end{align} 
with $C_i = C_i(p_1^2, p_2^2, p_0^2, m_1^2, m_3^2, m_2^2)$.

\paragraph{(f) $SVS$ diagram  } 
\begin{align} 
M^{(f)} = &i (C_0 - C_1 - C_2) c^1 c^2 c^3
\end{align} 
with $C_i = C_i(p_1^2, p_2^2, p_0^2, m_1^2, m_3^2, m_2^2)$.

\paragraph{(g) $VSS$ diagram  } 
\begin{align} 
M^{(g)} = &-2 i c^1 c^2 c^3 \Big[4 C_{00} + 5 C_2 p_0^2 + 3 C_{22} p_0^2 + 3 C_2 p_1^2 + C_{22} p_1^2 + 4 C_{12} (p_0^2 + p_1^2)  \nn \\ &
-2 C_{12} p_2^2 -3 C_2 p_2^2 - C_{22} p_2^2 + (C_0 + C_1 + C_2) m_1^2 + 2 C_0 p_0^2 + 3 C_1 p_0^2 + C_{11} p_0^2 \nn \\ &
+ 2 C_0 p_1^2 + 5 C_1 p_1^2 + 3 C_{11} p_1^2 - (2 C_0 + 3 C_1 + C_{11}) p_2^2\Big]
\end{align} 
with $C_i = C_i(p_1^2, p_2^2, p_0^2, m_1^2, m_3^2, m_2^2)$.

\paragraph{(h) $SVV$ diagram  } 
\begin{align} 
M^{(h)} = &-i c^1 c^2 c^3 \Big[ 4 B_0 -4 C_{00} + C_1 p_0^2 -4 C_{12} p_0^2 - C_2 p_0^2 -3 C_{22} p_0^2 - C_1 p_1^2  \nn \\ &
-4 C_{12} p_1^2 + C_2 p_1^2 - C_{22} p_1^2 + (-2 (C_1 - C_{12} + C_2) + C_{22}) p_2^2 \nn\\& + C_0 (p_2^2 + 4 m_1^2) - C_{11} (p_0^2 + 3 p_1^2 - p_2^2)\Big]
\end{align} 
with $B_i = B_i(p_2^2, m_2^2, m_3^2)$ and $C_i = C_i(p_1^2, p_2^2, p_0^2, m_1^2, m_3^2, m_2^2)$.

\paragraph{(i) $VSV$ diagram  } 
\begin{align} 
M^{(i)} = &-i (2 C_0 + C_1 + C_2) c^1 c^2 c^3
\end{align} 
with $C_i = C_i(p_1^2, p_2^2, p_0^2, m_1^2, m_3^2, m_2^2)$.

\paragraph{(j) $VVS$ diagram  } 
\begin{align} 
M^{(j)} = &i (2 C_0 + C_1 + C_2) c^1 c^2 c^3
\end{align} 
with $C_i = C_i(p_1^2, p_2^2, p_0^2, m_1^2, m_3^2, m_2^2)$.

\paragraph{(k) $SV$ diagram  } 
\begin{align} 
M^{(k)} = &-i c^1 c^2 (B_0 - B_1)
\end{align} 
with $B_i = B_i(p_0^2, m_1^2, m_2^2)$.

\paragraph{(l) $SV$ diagram  } 
\begin{align} 
M^{(l)} = &i c^1 c^2 (B_0 - B_1)
\end{align} 
with $B_i = B_i(p_1^2, m_1^2, m_2^2)$.

\paragraph{(m) $SS$ diagram  } 
\begin{align} 
M^{(m)} = 0
\end{align}

\paragraph{(n) $VV$ diagram  } 
\begin{align} 
M^{(n)} = 0
\end{align}

\subsubsection{Scalar to two gauge boson decays} 
\begin{figure}[h] 
\centering 
\includegraphics[width=0.90\linewidth]{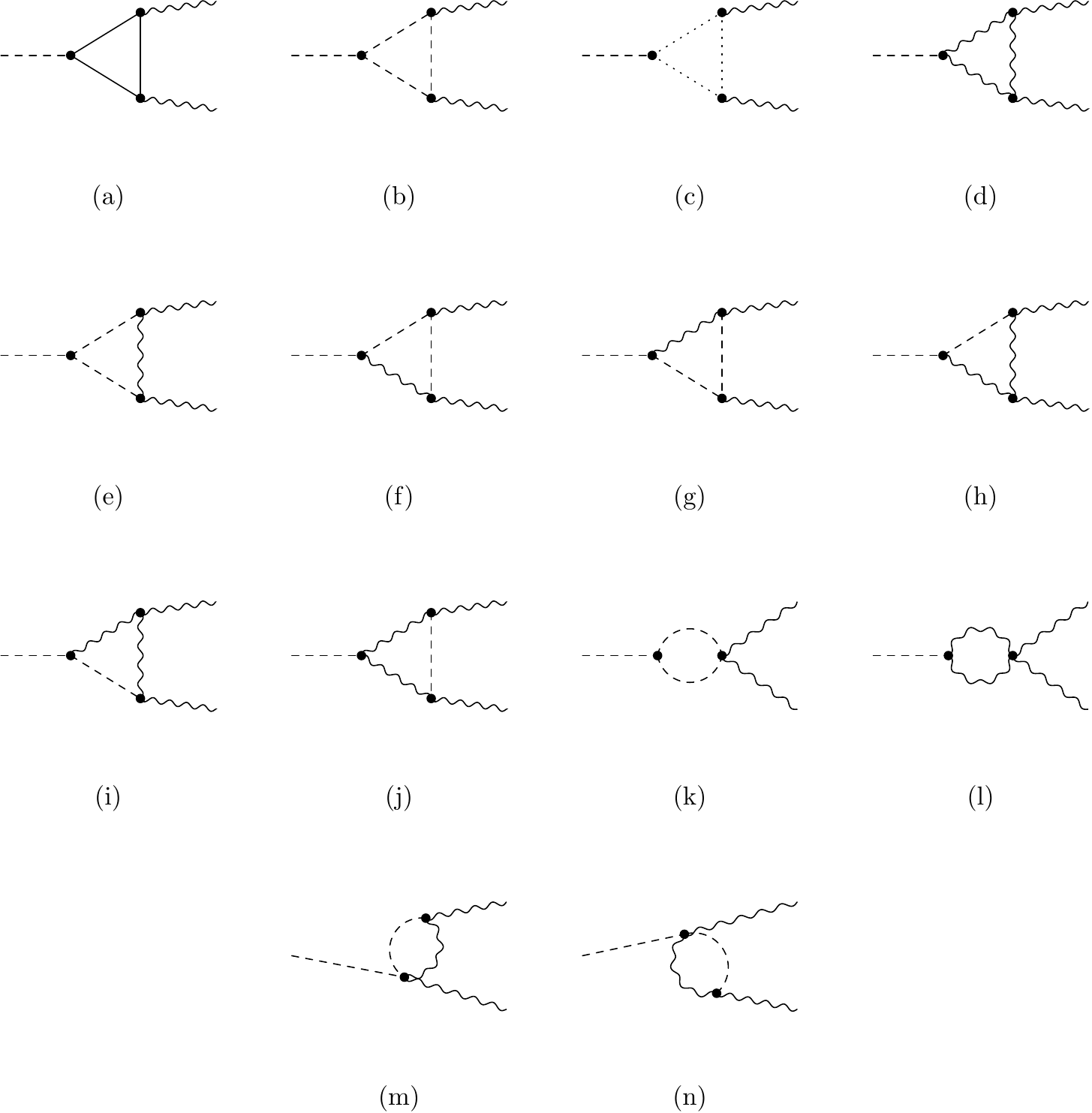} 
\caption{Generic diagrams contributing to $S\to VV$ decays.}
\label{fig:SVV}
\end{figure} 
\paragraph{(a) $FFF$ diagram  } 
\begin{align} 
M^{(a)}_1 = &-i \Big[2 B_0 c^1_L c^2_L c^3_L m_1 + 2 B_0 c^1_R c^2_R c^3_R m_1 + C_0 c^1_L c^2_L c^3_L p_0^2 m_1 + C_1 c^1_L c^2_L c^3_L p_0^2 m_1  \nn \\ &
+ 3 C_2 c^1_L c^2_L c^3_L p_0^2 m_1 + C_0 c^1_R c^2_R c^3_R p_0^2 m_1 + C_1 c^1_R c^2_R c^3_R p_0^2 m_1 + 3 C_2 c^1_R c^2_R c^3_R p_0^2 m_1 \nn \\ &
+ C_0 c^1_L c^2_L c^3_L p_1^2 m_1 + 3 C_1 c^1_L c^2_L c^3_L p_1^2 m_1 + C_2 c^1_L c^2_L c^3_L p_1^2 m_1 + C_0 c^1_R c^2_R c^3_R p_1^2 m_1 \nn \\ &
 + 3 C_1 c^1_R c^2_R c^3_R p_1^2 m_1 + C_2 c^1_R c^2_R c^3_R p_1^2 m_1 - C_0 c^1_L c^2_L c^3_L p_2^2 m_1 - C_1 c^1_L c^2_L c^3_L p_2^2 m_1\nn \\ &
  - C_2 c^1_L c^2_L c^3_L p_2^2 m_1 - C_0 c^1_R c^2_R c^3_R p_2^2 m_1 - C_1 c^1_R c^2_R c^3_R p_2^2 m_1 - C_2 c^1_R c^2_R c^3_R p_2^2 m_1 \nn\\&
+ 2 C_0 c^1_L c^2_L c^3_L m_1^3 + 2 C_0 c^1_R c^2_R c^3_R m_1^3 + 2 B_0 c^1_R c^2_L c^3_L m_2 + 2 B_0 c^1_L c^2_R c^3_R m_2  \nn \\ &
+ C_2 c^1_R c^2_L c^3_L p_0^2 m_2 + C_2 c^1_L c^2_R c^3_R p_0^2 m_2 + 2 C_1 c^1_R c^2_L c^3_L p_1^2 m_2 + C_2 c^1_R c^2_L c^3_L p_1^2 m_2 \nn \\ &
 + 2 C_1 c^1_L c^2_R c^3_R p_1^2 m_2 + C_2 c^1_L c^2_R c^3_R p_1^2 m_2 - C_2 c^1_R c^2_L c^3_L p_2^2 m_2 - C_2 c^1_L c^2_R c^3_R p_2^2 m_2  \nn \\ &
+ 2 C_0 c^1_R c^2_L c^3_L m_1^2 m_2 + 2 C_0 c^1_L c^2_R c^3_R m_1^2 m_2\nn\\&
-4 C_{00} (c^1_L c^2_L c^3_L m_1 + c^1_R c^2_R c^3_R m_1 + c^1_R c^2_L c^3_L m_2 + c^1_L c^2_R c^3_R m_2) \nn \\ &
- ((c^1_L c^2_R c^3_L + c^1_R c^2_L c^3_R) (2 B_0 + 2 C_2 p_0^2 + C_1 (p_0^2 + p_1^2 - p_2^2) + 2 C_0 m_1^2)\nn \\ &
+ 2 C_0 (c^1_R c^2_R c^3_L + c^1_L c^2_L c^3_R) m_1 m_2) m_3\Big]\\ 
M^{(a)}_2 = &2 i \Big[C_0 (c^1_L c^2_L c^3_L + c^1_R c^2_R c^3_R) m_1 + (2 C_{12} + 3 C_2 + 2 C_{22}) (c^1_L c^2_L c^3_L + c^1_R c^2_R c^3_R) m_1 \nn \\ &
+ (2 C_{12} + C_2 + 2 C_{22}) (c^1_R c^2_L c^3_L + c^1_L c^2_R c^3_R) m_2 + C_1 (c^1_L c^2_L c^3_L m_1 + c^1_R c^2_R c^3_R m_1 \nn \\ & - c^1_L c^2_R c^3_L m_3 - c^1_R c^2_L c^3_R m_3)\Big]
\end{align} 
with $B_i = B_i(p_2^2, m_2^2, m_3^2)$ and $C_i = C_i(p_1^2, p_2^2, p_0^2, m_1^2, m_3^2, m_2^2)$.

\paragraph{(b) $SSS$ diagram  } 
\begin{align} 
M^{(b)}_1 = &4 i C_{00} c^1 c^2 c^3\\ 
M^{(b)}_2 = &4 i (C_{12} + C_2 + C_{22}) c^1 c^2 c^3 
\end{align} 
with $C_i = C_i(p_1^2, p_2^2, p_0^2, m_1^2, m_3^2, m_2^2)$.

\paragraph{(c) $UUU$ diagram  } 
\begin{align} 
M^{(c)}_1 = &i C_{00} c^1 c^2 c^3\\ 
M^{(c)}_2 = &i (C_{12} + C_2 + C_{22}) c^1 c^2 c^3 
\end{align} 
with $C_i = C_i(p_1^2, p_2^2, p_0^2, m_1^2, m_3^2, m_2^2)$.

\paragraph{(d) $VVV$ diagram  } 
\begin{align} 
M^{(d)}_1 = &-\frac{i}{2} c^1 c^2 c^3 \Big[4 B_0 + 20 C_{00} -4 r + C_1 p_0^2 + 4 C_2 p_0^2 + 5 C_1 p_1^2 + 2 C_2 p_1^2 \nn \\ &
- (C_1 + 2 C_2) p_2^2 + C_0 (-4 p_0^2 + 6 p_1^2 + 4 (p_2^2 + m_1^2))\Big]\\ 
M^{(d)}_2 = &-i (5 C_0 + C_1 + 10 (C_{12} + C_2 + C_{22})) c^1 c^2 c^3 
\end{align} 
with $B_i = B_i(p_2^2, m_2^2, m_3^2)$ and $C_i = C_i(p_1^2, p_2^2, p_0^2, m_1^2, m_3^2, m_2^2)$.

\paragraph{(e) $SSV$ diagram  } 
\begin{align} 
M^{(e)}_1 = &i C_0 c^1 c^2 c^3\\ 
M^{(e)}_2 = & 0 \\ 
\end{align} 
with $C_i=C_i(p_1^2, p_2^2, p_0^2, m_1^2, m_3^2, m_2^2)$.

\paragraph{(f) $SVS$ diagram  } 
\begin{align} 
M^{(f)}_1 = &2 i C_{00} c^1 c^2 c^3\\ 
M^{(f)}_2 = &2 i (C_{12} - C_2 + C_{22}) c^1 c^2 c^3 
\end{align} 
with $C_i=C_i(p_1^2, p_2^2, p_0^2, m_1^2, m_3^2, m_2^2)$.

\paragraph{(g) $VSS$ diagram  } 
\begin{align} 
M^{(g)}_1 = &2 i C_{00} c^1 c^2 c^3\\ 
M^{(g)}_2 = &2 i (2 C_0 + 2 C_1 + C_{12} + 3 C_2 + C_{22}) c^1 c^2 c^3 
\end{align} 
with $C_i=C_i(p_1^2, p_2^2, p_0^2, m_1^2, m_3^2, m_2^2)$.

\paragraph{(h) $SVV$ diagram  } 
\begin{align} 
M^{(h)}_1 = &i c^1 c^2 c^3 \Big[B_0 - C_{00} - (C_1 + C_2) (p^2_0 - p^2_1) \nn \\ & + (C_1 - C_2) p_2^2 + C_0 (- p_1^2 + p_2^2 + m_1^2)\Big]\\ 
M^{(h)}_2 = &i (4 C_1 - C_{12} + C_2 - C_{22}) c^1 c^2 c^3 
\end{align} 
with $B_i = B_i(p_2^2, m_2^2, m_3^2)$ and $C_i = C_i(p_1^2, p_2^2, p_0^2, m_1^2, m_3^2, m_2^2)$.

\paragraph{(i) $VSV$ diagram  } 
\begin{align} 
M^{(i)}_1 = &i c^1 c^2 c^3 \Big[B_0 - C_{00} + C_1 p_0^2 + 3 C_2 p_0^2 + 3 C_1 p_1^2 + C_2 p_1^2 \nn \\ & - (C_1 + C_2) p_2^2 + C_0 (2 (p_0^2 + p_1^2 - p_2^2) + m_1^2)\Big]\\ 
M^{(i)}_2 = &-i (2 C_0 -2 C_1 + C_{12} + 3 C_2 + C_{22}) c^1 c^2 c^3 
\end{align} 
with $B_i = B_i(p_2^2, m_2^2, m_3^2)$ and $C_i = C_i(p_1^2, p_2^2, p_0^2, m_1^2, m_3^2, m_2^2)$.

\paragraph{(j) $VVS$ diagram  } 
\begin{align} 
M^{(j)}_1 = &-i C_0(p_1^2, p_2^2, p_0^2, m_1^2, m_3^2, m_2^2)  c^1 c^2 c^3\\ 
M^{(j)}_2 = & 0  
\end{align} 

\paragraph{(k) $SS$ diagram  } 
\begin{align} 
M^{(k)}_1 = &\frac{i}{2} B_0(p_0^2, m_1^2, m_2^2)  c^1 c^2\\ 
M^{(k)}_2 = & 0  
\end{align} 

\paragraph{(l) $VV$ diagram  } 
\begin{align} 
M^{(l)}_1 = &i c^1 c^2 (3 B_0(p_0^2, m_1^2, m_2^2)  - r)\\ 
M^{(l)}_2 = & 0  
\end{align} 

\paragraph{(m) $SV$ diagram  } 
\begin{align} 
M^{(m)}_1 = &-i B_0(p_1^2, m_1^2, m_2^2)  c^1 c^2\\ 
M^{(m)}_2 = & 0  
\end{align} 

\paragraph{(n) $SV$ diagram  } 
\begin{align} 
M^{(n)}_1 = &-i B_0(p_2^2, m_1^2, m_2^2)  c^1 c^3\\ 
M^{(n)}_2 = & 0 
\end{align} 

\section{Expressions for real corrections}
\label{app:realcorr}

Following the notation introduced in \sct{sec:realcorrections}, we present here the formulas
employed for the calculation of real corrections (massless gauge boson emission) to the six generic decays under consideration. Common to all processes is the calculation of the group theory factors, which we describe in the first subsection.

\subsection{Group theory factors}

For real three-body decays, we start with a two-body matrix element $\mathcal{M}_0$ where a particle having
four-momentum $p_0$ decays to two others with four-momenta $p_1,p_2$, and we can attach a photon
in \emph{four} ways: either to the incoming or outgoing legs, or to the vertex itself.
This latter case is only possible when the tree-level vertex involves two scalars and a massive gauge boson;
in \SARAH such vertices are stored in the form
\begin{align}
\mathcal{L} \supset -ic_{ija} (S_i \partial_\mu \ov{S}_j - \ov{S}_j \partial_\mu S_i ) V^{a\,\mu} \rightarrow c_{ija} S_i \ov{S}_j  V^{a\,\mu}  (p_\mu(S_i) - p_\mu (\ov{S}_j))\,.
\label{EQ:SSVcoupling}\end{align}
If we attach a massless gauge boson $A_\mu^b$ to the above term then we can have the coupling
\begin{align}
\lagr \supset c_{ijab} S_i \ov{S}_j  V^{a\,\mu} A_\mu^b\,.
\label{EQ:SSVVcoupling}\end{align}
We should now understand that the indices for $V,A$ correspond to the \emph{unbroken} gauge symmetry under which $A_\mu^b$ transforms. 
By insisting on gauge invariance up to linear order we find three sets of conditions:
\begin{align}
0=& T_{ii'}^c (S_i)c_{i'ja} + T_{jj'}^c (\ov{S}_j)c_{ij'a} + T_{a a'}^c (V) c_{ija'} \\
=& T_{ii'}^c (S_i)c_{i'jab} + T_{jj'}^c (\ov{S}_j)c_{ij'ab} + T_{a a'}^c (V) c_{ija'b} + i f^{cbb'} c_{ijab'} \\
=& c_{ijab} + c_{ij'a} T^c_{j'j} (\ov{S}_j) - c_{i'ja} T^c_{i'i} (S_i)\,. \label{EQ:GaugeQuartic}
\end{align}
Summation is only implied over the primed indices. The first two equations arise from requiring simple gauge
invariance of the individual vertices, while the third involves the derivatives of the gauge transformation
cancelling against the $A_\mu^a \rightarrow A_\mu^a + \partial_\mu \Lambda^a + \ldots$ part.
Importantly, \eqn{EQ:GaugeQuartic} completely determines $c_{ijab}$.

If the gauge group is that of the \sm{}, then the quartic coupling can only be relevant for a $W$ boson
and a photon, for which case the first two equations simply become charge conservation and \eqn{EQ:GaugeQuartic} implies
\begin{align}
c_{ijW^-A} =& (Q_{S_i} - Q_{\bar{S}_j}) c_{ijW^-} = (2 Q_{S_i} - 1)c_{ijW^-}\,.
\end{align}

The above logic can also be used to compute the group theory factors for the Bremsstrahlung decay cross-sections,
where a particle with momentum $p_0$ decays to final states with momenta $p_1, p_2$ and a photon/gluon with momentum $k$. 

We first compute all of the relevant processes using only the primitive vertices stripped of group theory factors,
and split up the squared amplitude depending on which leg the photon or gluon propagator has been attached.
We first make a general definition of the coupling $c_{ijk} $ in the same was as we did in \eqn{EQ:SSVcoupling}
and \eqn{EQ:SSVVcoupling} and strip off any Lorentz indices, so for fields (scalars, fermions or vectors)
$\Phi_i$ in representations of the gauge group $R_0 \rightarrow R_1, R_2$ we have the coupling
\begin{align}
\lagr \supset c_{ijk} \Phi_i \ov{\Phi}_j \ov{\Phi}_k \times \mathrm{Lorentz\ part}\,.
\end{align}
Then the matrix element is 
\begin{align}
(\mathcal{M}_{1})_{j_0 j_1 j_2 a} =& g  \sum_{j^\prime_1} T^a(R_0)_{j_0^\prime j_0} c_{j^\prime_0 j_1 j_2} 
\frac{f_0(p_i)}{-2p_0 \cdot k} + g  \sum_{j^\prime_1} T^a(R_1)_{j_1 j^\prime_1 } c_{j_0 j^\prime_1 j_2}  \frac{f_1(p_i)}{2p_1 \cdot k} \nn\\
&+  g  \sum_{j^\prime_2} T^a(R_2)_{j_2 j^\prime_2 } c_{j_0 j_1 j_2^\prime}  \frac{f_2(p_i)}{2p_2 \cdot k} + c_{j_0 j_1 j_2 a} f_3(p_i) \,,
\end{align}
where the functions $f_i^\mu (p_i)$ may have spinor or other Lorentz indices, and include the wavefunction factors,
as appropriate to the final states; and the index $i$ labels the leg to which the photon/gluon is attached:
$0$ for incoming, $1,2$ for the outgoing legs and $3$ to denote the vertex in the case of scalar to scalar plus gauge boson decays.
We have explicitly included the intermediate propagators to show the potential infra-red divergences -- and also
because they allow easy identification of the appropriate diagram. 

Then, defining $x_0 \equiv - 2p_0 \cdot k, x_1 \equiv 2 p_1 \cdot k, x_2 \equiv 2p_2 \cdot k, x_3 \equiv 1$, a generic squared matrix element can be expressed as
\begin{align}
|\mathcal{M}_1|^2 \equiv & \sum_{i,j=0}^3 \tilde{C}_{ij} \sum_{\rm spins} \frac{f_i f_j^*}{x_i x_j}\,.
\end{align}
Now all of the group-theory factors are encoded in $\tilde{C}_{ij}$. However, while $\tilde{C}_{ji} = \tilde{C}_{ij}^*$, it is also
clear from the hermeticity of the generators that the $\tilde{C}_{ij}$ are real -- for example using $d(R_0)$ as the dimension of representation $R_0$
\begin{align}
(\tC_{12})^* =& \frac{1}{d(R_0)} \bigg( \sum_{i,j,k,i',j'} c_{i'jk} T^a (R_0)_{i'i} T^a(R_1)^*_{jj'} c_{ij'k}^* \bigg)^* \nn\\
=& \frac{1}{d(R_0)}  \sum_{i,j,k,i',j'} c_{i'jk}^* T^a (R_0)_{ii'} T^a(R_1)_{j'j} c_{ij'k}  \nn\\ 
=&  \frac{1}{d(R_0)}  \sum_{i,j,k,i',j'}  c_{i'j k}  T^a (R_0)_{i'i} T^a(R_1)_{jj'} c_{ij'k}^* = \tC_{12}
\end{align}
and therefore we can write 
\begin{equation}
|\mathcal{M}_1|^2 = \sum_{i \le j}^3 \tC_{ij} \omega_{ij}, \qquad \omega_{ij} \equiv \left\{ \begin{array}{cl} \sum_{\rm spins} \frac{f_i f_j^* + f_j f_i^*}{x_i x_j}, & i < j \\
\sum_{\rm spins} \frac{f_i f_i^*}{x_i x_j}, & i = j \end{array}\right.\,. 
\end{equation}
Using the gauge invariance of the Lagrangian terms, we obtain for $i<3$:
\begin{align}
0=&\tC_{0j} - \tC_{1j} - \tC_{2j} = \tC_{i0} - \tC_{i1} - \tC_{i2} \nn\\
=& \tC_{i3} - \tC_{i0} - \tC_{i1} = \tC_{3i} - \tC_{0i} - \tC_{1i}\,,
\end{align}
where for $S\rightarrow SV$ decays we define the heavy gauge boson as particle $2$. These are \emph{almost}
enough to completely determine the colour factor of the amplitude in terms of the two-body decay factor $C$
\begin{align}
c_{ijk} c^*_{i'jk} \equiv C \delta_{ii'}\,,
\end{align}
because we can write
\begin{align}
\tC_{ii} =& C_2 (R_i) C\,.
\end{align}
These conditions are sufficient to determine all of the group factors in terms of $C$, the quadratic Casimir $C_2(R_i)$,
and one remaining colour factor. However, we can also use the gauge invariance of the Lagrangian term to \emph{second order} to obtain
\begin{align}
0=& \tC_{00} + \tC_{11} + \tC_{22} - 2\tC_{01} -2\tC_{02} + 2\tC_{12} = -\tC_{00} + \tC_{11} + \tC_{22} + 2\tC_{12}\,.
\end{align}
Hence all of the group theory factors are proportional to $C$, and we can therefore define 
\begin{align}
C_{ij} \equiv \frac{\tC_{ij}}{C} 
\end{align}
giving
\begin{align}
C_{ii} =& C_{2}(R_i)\,, \qquad i<3  \nn\\
C_{12} =& \frac{1}{2} ( C_2(R_0) - C_2(R_1) - C_2(R_2)) \nn\\
C_{02} =& \frac{1}{2} ( C_2(R_0) - C_2(R_1) + C_2(R_2)) \nn\\
C_{01} =& \frac{1}{2} ( C_2(R_0) + C_2(R_1) - C_2(R_2)) \nn\\
C_{i3} = C_{3i} =& C_{i1} + C_{i2}\,, \qquad i<3 \nn\\
C_{33} =& (C_2 (R_0) + C_2(R_1))C  + 2 C_{01} = 2C_2(R_0) + 2C_2(R_1) - C_2(R_2)
\label{EQ:GeneralCFactors}\end{align}
and we can write for a generic process
\begin{align}
\Gamma \propto \Omega \equiv \sum_{i \le j}^3 C_{ij} \Omega_{ij}\,.
\end{align}

In the case of a U$(1)$ gauge boson (the photon), we have $C_{ij} = Q_i Q_j$ and we define $Q_3 \equiv Q_0 + Q_1.$ This also follows from the above expressions, for example
\begin{align}
C_{12} =& \frac{1}{2} ( Q_0^2 - Q_1^2 - Q_2^2) = \frac{1}{2} ( (Q_1 + Q_2)^2 - Q_1^2 - Q_2^2) = Q_1 Q_2\,.
\end{align}

\subsection{$F\to FS$}

The real corrections for the fermionic decays $F\to FS$ are given by
\begin{align}
\Gamma_{F\to FS+\gamma/g}=\frac{c_g^2}{(4\pi)^3m_X}C'_S\left[(c_Lc_L^*+c_Rc_R^*)\Omega^{LL}+(c_Lc_R^*+c_Rc_L^*)\Omega^{LR}\right]\,.
\end{align}
The coupling $c_g$ is the electromagnetic coupling $e$ or the strong coupling $g_s$ for photon and gluon emission, respectively.
$c_L$ and $c_R$ are the left- and right-handed coupling of the tree-level vertex of $F\to FS$.
We identify $F_{\text{in}}=X$, $F_{\text{out}}=Y_1$ and $S=Y_2$ with the indices $0,1$ and $2$, respectively.
The individual contributions $\Omega_{ij}^{AB}$ are given by:
{\allowdisplaybreaks\begin{align}
\Omega_{00}^{LR} =& -8 I_{00} m_X^3 m_{Y_1}\\
\Omega_{01}^{LR} =& -8 I_0 m_X m_{Y_1} - 8 I_1 m_X m_{Y_1} + I_{01} (-8 m_X^3 m_{Y_1} - 8 m_X m_{Y_1}^3 + 8 m_X m_{Y_1} m_{Y_2}^2)\\
\Omega_{02}^{LR} =& 8 I_0 m_X m_{Y_1} + I_{02} (8 m_X^3 m_{Y_1} - 8 m_X m_{Y_1}^3 + 8 m_X m_{Y_1} m_{Y_2}^2)\\
\Omega_{11}^{LR} =& -8 I_{11} m_X m_{Y_1}^3\\
\Omega_{12}^{LR} =& -8 I_1 m_X m_{Y_1} + I_{12} (8 m_X^3 m_{Y_1} - 8 m_X m_{Y_1}^3 - 8 m_X m_{Y_1} m_{Y_2}^2)\\
\Omega_{22}^{LR} =& -8 I_{2} m_X m_{Y_1} - 8 I_{22} m_X m_{Y_1} m_{Y_2}^2\\
\Omega_{00}^{LL} =& 2 I + I_0 (-2 m_X^2 + 2 m_{Y_1}^2 - 2 m_{Y_2}^2) + I_{00} (-4 m_X^4 - 4 m_X^2 m_{Y_1}^2 + 4 m_X^2 m_{Y_2}^2)\\
\Omega_{01}^{LL} =& -8 I - 2 I_0^1 - 2 I_1^0 + I_0 (-2 m_X^2 - 6 m_{Y_1}^2 + 6 m_{Y_2}^2) + I_1 (-6 m_X^2 - 2 m_{Y_1}^2 + 6 m_{Y_2}^2)\\\nn
&+ I_{01} (-4 m_X^4 - 8 m_X^2 m_{Y_1}^2 - 4 m_{Y_1}^4 + 8 m_X^2 m_{Y_2}^2 + 8 m_{Y_1}^2 m_{Y_2}^2 - 4 m_{Y_2}^4)\\
\Omega_{02}^{LL} =& 4 I - 2 I_0^2 + I_0 (2 m_X^2 + 6 m_{Y_1}^2 - 6 m_{Y_2}^2) + 4 I_{2} m_{Y_2}^2\\\nn
& + I_{02} (4 m_X^4 - 4 m_{Y_1}^4 + 8 m_{Y_1}^2 m_{Y_2}^2 - 4 m_{Y_2}^4)\\
\Omega_{11}^{LL} =& 2 I + I_1 (2 m_X^2 - 2 m_{Y_1}^2 - 2 m_{Y_2}^2) + I_{11} (-4 m_X^2 m_{Y_1}^2 - 4 m_{Y_1}^4 + 4 m_{Y_1}^2 m_{Y_2}^2)\\
\Omega_{12}^{LL} =& -4 I + 2 I_1^2 - 4 I_{2} m_{Y_2}^2 + I_1 (-6 m_X^2 - 2 m_{Y_1}^2 + 6 m_{Y_2}^2) \\\nn
&+ I_{12} (4 m_X^4 - 4 m_{Y_1}^4 - 8 m_X^2 m_{Y_2}^2 + 4 m_{Y_2}^4)\\
\Omega_{22}^{LL} =& 4 I + I_{2} (-4 m_X^2 - 4 m_{Y_1}^2 + 8 m_{Y_2}^2) + I_{22} (-4 m_X^2 m_{Y_2}^2 - 4 m_{Y_1}^2 m_{Y_2}^2 + 4 m_{Y_2}^4)
\end{align}}
We also implemented real corrections for which $Y_1$ is taken to be massless, if $Y_1$ is charge and colour neutral.

\subsection{$S\to FF$}

The real corrections for the scalar decays $S\to FF$ are given by
\begin{align}
\Gamma_{S\to FF+\gamma/g}=\frac{c_g^2}{(4\pi)^3m_X}C'_S\left[(c_Lc_L^*+c_Rc_R^*)\Omega^{LL}+(c_Lc_R^*+c_Rc_L^*)\Omega^{LR}\right]\,.
\end{align}
The notation follows the notation introduced in the previous subsection.
However, we now identify $S=X$, $F=Y_1$ and $F=Y_2$ with the indices $0,1$ and $2$, respectively.
The individual contributions $\Omega_{ij}^{AB}$ are given by:
{\allowdisplaybreaks\begin{align}
\Omega_{00}^{LR} =& 16 I_{00} m_X^2 m_{Y_1} m_{Y_2}\\
\Omega_{01}^{LR} =& -16 I_0 m_{Y_1} m_{Y_2} - 16 I_1 m_{Y_1} m_{Y_2} + I_{01} (-16 m_X^2 m_{Y_1} m_{Y_2} - 16 m_{Y_1}^3 m_{Y_2} + 16 m_{Y_1} m_{Y_2}^3)\\
\Omega_{02}^{LR} =& 16 I_0 m_{Y_1} m_{Y_2} + I_{02} (16 m_X^2 m_{Y_1} m_{Y_2} - 16 m_{Y_1}^3 m_{Y_2} + 16 m_{Y_1} m_{Y_2}^3)\\
\Omega_{11}^{LR} =& 16 I_{11} m_{Y_1}^3 m_{Y_2}\\
\Omega_{12}^{LR} =& 16 I_1 m_{Y_1} m_{Y_2} + I_{12} (-16 m_X^2 m_{Y_1} m_{Y_2} + 16 m_{Y_1}^3 m_{Y_2} + 16 m_{Y_1} m_{Y_2}^3)\\
\Omega_{22}^{LR} =& 16 I_{2} m_{Y_1} m_{Y_2} + 16 I_{22} m_{Y_1} m_{Y_2}^3\\
\Omega_{00}^{LL} =& -8 I_0 m_X^2 + I_{00} (-8 m_X^4 + 8 m_X^2 m_{Y_1}^2 + 8 m_X^2 m_{Y_2}^2)\\
\Omega_{01}^{LL} =& 4 I + 4 I_1^0 + I_1 (12 m_X^2 - 4 m_{Y_1}^2 - 12 m_{Y_2}^2) + I_0 (-8 m_{Y_1}^2 - 8 m_{Y_2}^2)\\\nn
&  + I_{01} (8 m_X^4 - 8 m_{Y_1}^4 - 16 m_X^2 m_{Y_2}^2 + 8 m_{Y_2}^4)\\
\Omega_{02}^{LL} =& -4 I - 4 I_2^0 + I_{2} (-12 m_X^2 + 12 m_{Y_1}^2 + 4 m_{Y_2}^2) + I_0 (8 m_{Y_1}^2 + 8 m_{Y_2}^2)\\\nn
&  + I_{02} (-8 m_X^4 + 16 m_X^2 m_{Y_1}^2 - 8 m_{Y_1}^4 + 8 m_{Y_2}^4)\\
\Omega_{11}^{LL} =& I_1 (4 m_X^2 + 4 m_{Y_1}^2 - 4 m_{Y_2}^2) + I_{11} (-8 m_X^2 m_{Y_1}^2 + 8 m_{Y_1}^4 + 8 m_{Y_1}^2 m_{Y_2}^2)\\
\Omega_{12}^{LL} =& 8 I + 4 I_1^2 + 4 I_2^1 + I_{2} (-12 m_X^2 + 12 m_{Y_1}^2 + 4 m_{Y_2}^2) + I_1 (-12 m_X^2 + 4 m_{Y_1}^2\\\nn
&  + 12 m_{Y_2}^2) + I_{12} (8 m_X^4 - 16 m_X^2 m_{Y_1}^2 + 8 m_{Y_1}^4 - 16 m_X^2 m_{Y_2}^2 + 16 m_{Y_1}^2 m_{Y_2}^2 + 8 m_{Y_2}^4)\\
\Omega_{22}^{LL} =& I_{2} (4 m_X^2 - 4 m_{Y_1}^2 + 4 m_{Y_2}^2) + I_{22} (-8 m_X^2 m_{Y_2}^2 + 8 m_{Y_1}^2 m_{Y_2}^2 + 8 m_{Y_2}^4)
\end{align}}
We also implemented real corrections where one of the final state fermions can be massless.
Again this fermion has to be charge and colour neutral.

\subsection{$F\to FV$}

The real corrections for the fermionic decays $F\to FV$ are given by
\begin{align}
\Gamma_{F\to FV+\gamma/g}=\frac{c_g^2}{(4\pi)^3m_{Y_2}^2m_X}C'_S\left[(c_Lc_L^*+c_Rc_R^*)\Omega^{LL}+(c_Lc_R^*+c_Rc_L^*)\Omega^{LR}\right]\,.
\end{align}
The coupling $c_g$ is the electromagnetic coupling $e$ or the strong coupling $g_s$ for photon and gluon emission, respectively.
$c_L$ and $c_R$ are the left- and right-handed coupling of the tree-level vertex of $F\to FV$.
The final state gauge boson $Y_2$ is fixed to one of the two heavy gauge bosons, $Z$ or $W$.
Clearly only the photon couples to the $W$ boson, such that we subsequently present results
independently for $F\to FW+\gamma$ on the one side and $F\to FZ+\gamma/g$ and $F\to FW+g$
on the other side.
For gluon emission $C_{00} = C_{11} = C_{01} = C_2(R_F), C_{i2} =0$; for photon emission $C_{ij}=Q_iQ_j$, where
we again identify $F_{\text{in}}=X$ and $F_{\text{out}}=Y_1$ with indices $0$ and $1$, respectively.

For $F\to FW+\gamma$ the individual contributions are given by:
{\allowdisplaybreaks\begin{align}\nn
\Omega_{00}^{LR} =& -4 I m_X m_{Y_1} - 4 I_0^2 m_X m_{Y_1} + 24 I_0 m_X m_{Y_1} m_{Y_2}^2\\\nn
&  + 24 I_{2} m_X m_{Y_1} m_{Y_2}^2 + 24 I_{00} m_X^3 m_{Y_1} m_{Y_2}^2 + 24 I_{02} m_X^3 m_{Y_1} m_{Y_2}^2\\
&  - 24 I_{02} m_X m_{Y_1}^3 m_{Y_2}^2 + 24 I_{02} m_X m_{Y_1} m_{Y_2}^4 + 24 I_{22} m_X m_{Y_1} m_{Y_2}^4\\\nn
\Omega_{01}^{LR} =& 12 I m_X m_{Y_1} + 4 I_0^1 m_X m_{Y_1} + 4 I_0^2 m_X m_{Y_1} - 48 I_{2} m_X m_{Y_1} m_{Y_2}^2\\\nn
&  - 48 I_{02} m_X^3 m_{Y_1} m_{Y_2}^2 + 48 I_{01} m_X m_{Y_1}^3 m_{Y_2}^2 + 48 I_{02} m_X m_{Y_1}^3 m_{Y_2}^2\\
&  - 48 I_{22} m_X m_{Y_1} m_{Y_2}^4\\\nn
\Omega_{11}^{LR} =& 4 I_1^0 m_X m_{Y_1} + 24 I_1 m_X m_{Y_1} m_{Y_2}^2 + 24 I_{2} m_X m_{Y_1} m_{Y_2}^2\\\nn
&  + 24 I_{01} m_X^3 m_{Y_1} m_{Y_2}^2 + 24 I_{02} m_X^3 m_{Y_1} m_{Y_2}^2 - 24 I_{01} m_X m_{Y_1}^3 m_{Y_2}^2\\\nn
&  - 24 I_{02} m_X m_{Y_1}^3 m_{Y_2}^2 + 24 I_{11} m_X m_{Y_1}^3 m_{Y_2}^2 - 24 I_{01} m_X m_{Y_1} m_{Y_2}^4\\
&  - 24 I_{02} m_X m_{Y_1} m_{Y_2}^4 + 24 I_{22} m_X m_{Y_1} m_{Y_2}^4\\\nn
\Omega_{00}^{LL} =& 2 I m_X^2 + 2 I_0^2 m_X^2 - 4 I_0 m_X^4 - 4 I_{2} m_X^4 - 4 I_{00} m_X^6 - 4 I_{02} m_X^6 + 2 I m_{Y_1}^2 + 2 I_0^2 m_{Y_1}^2\\\nn
& + 8 I_0 m_X^2 m_{Y_1}^2 + 8 I_{2} m_X^2 m_{Y_1}^2 + 8 I_{00} m_X^4 m_{Y_1}^2 + 12 I_{02} m_X^4 m_{Y_1}^2 - 4 I_0 m_{Y_1}^4 - 4 I_{2} m_{Y_1}^4\\\nn
& - 4 I_{00} m_X^2 m_{Y_1}^4 - 12 I_{02} m_X^2 m_{Y_1}^4 + 4 I_{02} m_{Y_1}^6 + 4 I m_{Y_2}^2 + 4 I_0^2 m_{Y_2}^2 - 8 I_2^1 m_{Y_2}^2\\\nn
&  - 8 I_{22}^{01} m_{Y_2}^2 - 4 I_0 m_X^2 m_{Y_2}^2 - 4 I_{2} m_X^2 m_{Y_2}^2 - 4 I_{00} m_X^4 m_{Y_2}^2 - 8 I_{02} m_X^4 m_{Y_2}^2\\\nn
&  - 4 I_{22} m_X^4 m_{Y_2}^2 - 4 I_0 m_{Y_1}^2 m_{Y_2}^2 - 4 I_{2} m_{Y_1}^2 m_{Y_2}^2 - 4 I_{00} m_X^2 m_{Y_1}^2 m_{Y_2}^2\\\nn
&  + 8 I_{02} m_X^2 m_{Y_1}^2 m_{Y_2}^2 + 8 I_{22} m_X^2 m_{Y_1}^2 m_{Y_2}^2 - 4 I_{22} m_{Y_1}^4 m_{Y_2}^2 + 8 I_0 m_{Y_2}^4\\\nn
&  + 8 I_{2} m_{Y_2}^4 + 8 I_{00} m_X^2 m_{Y_2}^4 + 4 I_{02} m_X^2 m_{Y_2}^4 - 4 I_{22} m_X^2 m_{Y_2}^4 - 12 I_{02} m_{Y_1}^2 m_{Y_2}^4\\
&  - 4 I_{22} m_{Y_1}^2 m_{Y_2}^4 + 8 I_{02} m_{Y_2}^6 + 8 I_{22} m_{Y_2}^6\\\nn
\Omega_{01}^{LL} =&-6 I m_X^2 - 2 I_0^1 m_X^2 - 2 I_0^2 m_X^2 + 8 I_{2} m_X^4 + 8 I_{02} m_X^6 - 6 I m_{Y_1}^2 - 2 I_0^1 m_{Y_1}^2\\\nn
&  - 2 I_0^2 m_{Y_1}^2 - 16 I_{2} m_X^2 m_{Y_1}^2 - 8 I_{01} m_X^4 m_{Y_1}^2 - 24 I_{02} m_X^4 m_{Y_1}^2 + 8 I_{2} m_{Y_1}^4\\\nn
&  + 16 I_{01} m_X^2 m_{Y_1}^4 + 24 I_{02} m_X^2 m_{Y_1}^4 - 8 I_{01} m_{Y_1}^6 - 8 I_{02} m_{Y_1}^6 + 4 I m_{Y_2}^2 - 4 I_0^1 m_{Y_2}^2\\\nn
&  - 4 I_0^2 m_{Y_2}^2 + 16 I_2^0 m_{Y_2}^2 + 16 I_2^1 m_{Y_2}^2 + 16 I_{22}^{01} m_{Y_2}^2 + 8 I_{2} m_X^2 m_{Y_2}^2\\\nn
&  + 8 I_{02} m_X^4 m_{Y_2}^2 + 8 I_{22} m_X^4 m_{Y_2}^2 + 8 I_{2} m_{Y_1}^2 m_{Y_2}^2 - 8 I_{01} m_X^2 m_{Y_1}^2 m_{Y_2}^2\\\nn
&  - 16 I_{22} m_X^2 m_{Y_1}^2 m_{Y_2}^2 - 8 I_{01} m_{Y_1}^4 m_{Y_2}^2 - 8 I_{02} m_{Y_1}^4 m_{Y_2}^2 + 8 I_{22} m_{Y_1}^4 m_{Y_2}^2\\\nn
&  - 16 I_{2} m_{Y_2}^4 - 16 I_{02} m_X^2 m_{Y_2}^4 + 8 I_{22} m_X^2 m_{Y_2}^4 + 16 I_{01} m_{Y_1}^2 m_{Y_2}^4\\
&  + 16 I_{02} m_{Y_1}^2 m_{Y_2}^4 + 8 I_{22} m_{Y_1}^2 m_{Y_2}^4 - 16 I_{22} m_{Y_2}^6\\\nn
\Omega_{11}^{LL} =& -2 I_1^0 m_X^2 - 4 I_1 m_X^4 - 4 I_{2} m_X^4 - 4 I_{01} m_X^6 - 4 I_{02} m_X^6 - 2 I_1^0 m_{Y_1}^2\\\nn
&  + 8 I_1 m_X^2 m_{Y_1}^2 + 8 I_{2} m_X^2 m_{Y_1}^2 + 12 I_{01} m_X^4 m_{Y_1}^2 + 12 I_{02} m_X^4 m_{Y_1}^2 - 4 I_{11} m_X^4 m_{Y_1}^2\\\nn
&  - 4 I_1 m_{Y_1}^4 - 4 I_{2} m_{Y_1}^4 - 12 I_{01} m_X^2 m_{Y_1}^4 - 12 I_{02} m_X^2 m_{Y_1}^4 + 8 I_{11} m_X^2 m_{Y_1}^4\\\nn
&  + 4 I_{01} m_{Y_1}^6 + 4 I_{02} m_{Y_1}^6 - 4 I_{11} m_{Y_1}^6 - 8 I m_{Y_2}^2 - 4 I_1^0 m_{Y_2}^2 - 16 I_2^0 m_{Y_2}^2\\\nn
&  - 8 I_2^1 m_{Y_2}^2 - 8 I_{22}^{01} m_{Y_2}^2 - 4 I_1 m_X^2 m_{Y_2}^2 - 4 I_{2} m_X^2 m_{Y_2}^2 - 4 I_{22} m_X^4 m_{Y_2}^2\\\nn
&  - 4 I_1 m_{Y_1}^2 m_{Y_2}^2 - 4 I_{2} m_{Y_1}^2 m_{Y_2}^2 - 8 I_{01} m_X^2 m_{Y_1}^2 m_{Y_2}^2 - 8 I_{02} m_X^2 m_{Y_1}^2 m_{Y_2}^2\\\nn
&  - 4 I_{11} m_X^2 m_{Y_1}^2 m_{Y_2}^2 + 8 I_{22} m_X^2 m_{Y_1}^2 m_{Y_2}^2 + 8 I_{01} m_{Y_1}^4 m_{Y_2}^2 + 8 I_{02} m_{Y_1}^4 m_{Y_2}^2\\\nn
&  - 4 I_{11} m_{Y_1}^4 m_{Y_2}^2 - 4 I_{22} m_{Y_1}^4 m_{Y_2}^2 + 8 I_1 m_{Y_2}^4 + 8 I_{2} m_{Y_2}^4 + 12 I_{01} m_X^2 m_{Y_2}^4\\\nn
&  + 12 I_{02} m_X^2 m_{Y_2}^4 - 4 I_{22} m_X^2 m_{Y_2}^4 - 4 I_{01} m_{Y_1}^2 m_{Y_2}^4 - 4 I_{02} m_{Y_1}^2 m_{Y_2}^4\\
&  + 8 I_{11} m_{Y_1}^2 m_{Y_2}^4 - 4 I_{22} m_{Y_1}^2 m_{Y_2}^4 - 8 I_{01} m_{Y_2}^6 - 8 I_{02} m_{Y_2}^6 + 8 I_{22} m_{Y_2}^6
\end{align}}

For $F\to FZ+\gamma/g$ and $F\to FW+g$ the individual contributions are given by:
{\allowdisplaybreaks\begin{align}
\Omega_{00}^{LR} =&  24 I_{00} m_X^3 m_{Y_1} m_{Y_2}^2\\\nn
\Omega_{01}^{LR} =& 8 I m_X m_{Y_1} + 4 I_0^1 m_X m_{Y_1} + 4 I_1^0 m_X m_{Y_1} + 24 I_0 m_X m_{Y_1} m_{Y_2}^2 + 24 I_1 m_X m_{Y_1} m_{Y_2}^2\\
&  + 24 I_{01} m_X^3 m_{Y_1} m_{Y_2}^2 + 24 I_{01} m_X m_{Y_1}^3 m_{Y_2}^2 - 24 I_{01} m_X m_{Y_1} m_{Y_2}^4\\
\Omega_{11}^{LR} =& 24 I_{11} m_X m_{Y_1}^3 m_{Y_2}^2\\\nn
\Omega_{00}^{LL} =& -2 I m_X^2 - 6 I_0 m_X^4 - 4 I_{00} m_X^6 - 2 I m_{Y_1}^2 + 4 I_0 m_X^2 m_{Y_1}^2 + 8 I_{00} m_X^4 m_{Y_1}^2\\\nn
&  + 2 I_0 m_{Y_1}^4 - 4 I_{00} m_X^2 m_{Y_1}^4 + 4 I m_{Y_2}^2 - 2 I_0 m_X^2 m_{Y_2}^2 - 4 I_{00} m_X^4 m_{Y_2}^2\\
&  + 2 I_0 m_{Y_1}^2 m_{Y_2}^2 - 4 I_{00} m_X^2 m_{Y_1}^2 m_{Y_2}^2 - 4 I_0 m_{Y_2}^4 + 8 I_{00} m_X^2 m_{Y_2}^4\\\nn
\Omega_{01}^{LL} =& -2 I_0^1 m_X^2 - 2 I_1^0 m_X^2 + 2 I_0 m_X^4 - 6 I_1 m_X^4 - 4 I_{01} m_X^6 - 2 I_0^1 m_{Y_1}^2 - 2 I_1^0 m_{Y_1}^2\\\nn
&  + 4 I_0 m_X^2 m_{Y_1}^2 + 4 I_1 m_X^2 m_{Y_1}^2 + 4 I_{01} m_X^4 m_{Y_1}^2 - 6 I_0 m_{Y_1}^4 + 2 I_1 m_{Y_1}^4 + 4 I_{01} m_X^2 m_{Y_1}^4\\\nn
&  - 4 I_{01} m_{Y_1}^6 - 8 I m_{Y_2}^2 - 4 I_0^1 m_{Y_2}^2 - 4 I_1^0 m_{Y_2}^2 - 2 I_0 m_X^2 m_{Y_2}^2 - 6 I_1 m_X^2 m_{Y_2}^2\\\nn
&  - 6 I_0 m_{Y_1}^2 m_{Y_2}^2 - 2 I_1 m_{Y_1}^2 m_{Y_2}^2 - 16 I_{01} m_X^2 m_{Y_1}^2 m_{Y_2}^2 + 12 I_0 m_{Y_2}^4 + 12 I_1 m_{Y_2}^4\\
&  + 12 I_{01} m_X^2 m_{Y_2}^4 + 12 I_{01} m_{Y_1}^2 m_{Y_2}^4 - 8 I_{01} m_{Y_2}^6\\\nn
\Omega_{11}^{LL} =& -2 I m_X^2 + 2 I_1 m_X^4 - 2 I m_{Y_1}^2 + 4 I_1 m_X^2 m_{Y_1}^2 - 4 I_{11} m_X^4 m_{Y_1}^2 - 6 I_1 m_{Y_1}^4\\\nn
&  + 8 I_{11} m_X^2 m_{Y_1}^4 - 4 I_{11} m_{Y_1}^6 + 4 I m_{Y_2}^2 + 2 I_1 m_X^2 m_{Y_2}^2 - 2 I_1 m_{Y_1}^2 m_{Y_2}^2\\
&  - 4 I_{11} m_X^2 m_{Y_1}^2 m_{Y_2}^2 - 4 I_{11} m_{Y_1}^4 m_{Y_2}^2 - 4 I_1 m_{Y_2}^4 + 8 I_{11} m_{Y_1}^2 m_{Y_2}^4
\end{align}}

\subsection{$S\to VV$}

The real corrections for the decays $S\to VV$ are given by
\begin{align}
\Gamma_{S\to VV+\gamma}=\frac{e^2}{(4\pi)^3m_{Y_1}^2m_{Y_2}^2m_X}C'_Scc^*\Omega\,.
\end{align}
The coupling $c$ is the tree-level coupling of the vertex of $S\to VV$.
Given that both final state particles~$V$ are heavy gauge bosons only
photon emission is of relevance for this process. 
We identify $S=X$, $V=Y_1$ and $V=Y_2$ with the indices $0,1$ and $2$, respectively.
The individual contributions $\Omega_{ij}$ are given by:
{\allowdisplaybreaks\begin{align}\nn
\Omega_{00} =&  -6 I m_X^2 - 6 I_0 m_X^4 - 2 I_{00} m_X^6 + 4 I m_{Y_1}^2 + 8 I_0 m_X^2 m_{Y_1}^2 + 4 I_{00} m_X^4 m_{Y_1}^2 - 2 I_0 m_{Y_1}^4\\\nn
&  - 2 I_{00} m_X^2 m_{Y_1}^4 + 4 I m_{Y_2}^2 + 8 I_0 m_X^2 m_{Y_2}^2 + 4 I_{00} m_X^4 m_{Y_2}^2 - 20 I_0 m_{Y_1}^2 m_{Y_2}^2\\
&  - 20 I_{00} m_X^2 m_{Y_1}^2 m_{Y_2}^2 - 2 I_0 m_{Y_2}^4 - 2 I_{00} m_X^2 m_{Y_2}^4\\\nn
\Omega_{01} =& 8 I m_X^2 - 2 I_1^0 m_X^2 + 4 I_0 m_X^4 - 4 I_1 m_X^4 - 2 I_{01} m_X^6 - 4 I m_{Y_1}^2 + 2 I_1^0 m_{Y_1}^2 - 4 I_0 m_X^2 m_{Y_1}^2\\\nn
&  + 4 I_1 m_X^2 m_{Y_1}^2 + 2 I_{01} m_X^4 m_{Y_1}^2 + 2 I_{01} m_X^2 m_{Y_1}^4 - 2 I_{01} m_{Y_1}^6 - 4 I m_{Y_2}^2 + 2 I_1^0 m_{Y_2}^2\\\nn
&  - 4 I_0 m_X^2 m_{Y_2}^2 + 8 I_1 m_X^2 m_{Y_2}^2 + 6 I_{01} m_X^4 m_{Y_2}^2 - 20 I_1 m_{Y_1}^2 m_{Y_2}^2 - 20 I_{01} m_X^2 m_{Y_1}^2 m_{Y_2}^2\\
&  - 18 I_{01} m_{Y_1}^4 m_{Y_2}^2 - 4 I_1 m_{Y_2}^4 - 6 I_{01} m_X^2 m_{Y_2}^4 + 18 I_{01} m_{Y_1}^2 m_{Y_2}^4 + 2 I_{01} m_{Y_2}^6\\\nn
\Omega_{02} =& 8 I m_X^2 - 2 I_2^0 m_X^2 + 4 I_0 m_X^4 - 4 I_{2} m_X^4 - 2 I_{02} m_X^6 - 4 I m_{Y_1}^2 + 2 I_2^0 m_{Y_1}^2 - 4 I_0 m_X^2 m_{Y_1}^2\\\nn
&  + 8 I_{2} m_X^2 m_{Y_1}^2 + 6 I_{02} m_X^4 m_{Y_1}^2 - 4 I_{2} m_{Y_1}^4 - 6 I_{02} m_X^2 m_{Y_1}^4 + 2 I_{02} m_{Y_1}^6 - 4 I m_{Y_2}^2\\\nn
&  + 2 I_2^0 m_{Y_2}^2 - 4 I_0 m_X^2 m_{Y_2}^2 + 4 I_{2} m_X^2 m_{Y_2}^2 + 2 I_{02} m_X^4 m_{Y_2}^2 - 20 I_{2} m_{Y_1}^2 m_{Y_2}^2\\
&  - 20 I_{02} m_X^2 m_{Y_1}^2 m_{Y_2}^2 + 18 I_{02} m_{Y_1}^4 m_{Y_2}^2 + 2 I_{02} m_X^2 m_{Y_2}^4 - 18 I_{02} m_{Y_1}^2 m_{Y_2}^4 - 2 I_{02} m_{Y_2}^6\\\nn
\Omega_{11} =& -2 I m_X^2 + 2 I_1^0 m_X^2 + 2 I_1 m_X^4 + 4 I m_{Y_1}^2 + 6 I_1^0 m_{Y_1}^2 + 4 I_{11}^{00} m_{Y_1}^2 - 2 I_{11} m_X^4 m_{Y_1}^2\\\nn
&  - 2 I_1 m_{Y_1}^4 + 4 I_{11} m_X^2 m_{Y_1}^4 - 2 I_{11} m_{Y_1}^6 + 4 I m_{Y_2}^2 - 2 I_1^0 m_{Y_2}^2 - 4 I_1 m_X^2 m_{Y_2}^2\\
&  + 4 I_{11} m_X^2 m_{Y_1}^2 m_{Y_2}^2 - 20 I_{11} m_{Y_1}^4 m_{Y_2}^2 + 2 I_1 m_{Y_2}^4 - 2 I_{11} m_{Y_1}^2 m_{Y_2}^4\\\nn
\Omega_{12} =& -8 I m_X^2 - 2 I_1^2 m_X^2 - 2 I_2^1 m_X^2 + 4 I_1 m_X^4 + 4 I_{2} m_X^4 - 2 I_{12} m_X^6 + 4 I m_{Y_1}^2 - 6 I_1^2 m_{Y_1}^2\\\nn
&  + 2 I_2^1 m_{Y_1}^2 - 4 I_1 m_X^2 m_{Y_1}^2 - 8 I_{2} m_X^2 m_{Y_1}^2 + 6 I_{12} m_X^4 m_{Y_1}^2 + 4 I_{2} m_{Y_1}^4 - 6 I_{12} m_X^2 m_{Y_1}^4\\\nn
&  + 2 I_{12} m_{Y_1}^6 + 4 I m_{Y_2}^2 + 2 I_1^2 m_{Y_2}^2 - 6 I_2^1 m_{Y_2}^2 - 8 I_1 m_X^2 m_{Y_2}^2 - 4 I_{2} m_X^2 m_{Y_2}^2\\\nn
&  + 6 I_{12} m_X^4 m_{Y_2}^2 + 20 I_1 m_{Y_1}^2 m_{Y_2}^2 + 20 I_{2} m_{Y_1}^2 m_{Y_2}^2 - 28 I_{12} m_X^2 m_{Y_1}^2 m_{Y_2}^2\\
&  + 22 I_{12} m_{Y_1}^4 m_{Y_2}^2 + 4 I_1 m_{Y_2}^4 - 6 I_{12} m_X^2 m_{Y_2}^4 + 22 I_{12} m_{Y_1}^2 m_{Y_2}^4 + 2 I_{12} m_{Y_2}^6\\\nn
\Omega_{22} =& -4 I m_X^2 - 2 I_2^1 m_X^2 + 2 I_{2} m_X^4 + 6 I m_{Y_1}^2 + 2 I_2^1 m_{Y_1}^2 - 4 I_{2} m_X^2 m_{Y_1}^2 + 2 I_{2} m_{Y_1}^4\\\nn
&  + 2 I m_{Y_2}^2 + 2 I_2^1 m_{Y_2}^2 + 4 I_{22}^{11} m_{Y_2}^2 - 2 I_{22} m_X^4 m_{Y_2}^2 + 4 I_{22} m_X^2 m_{Y_1}^2 m_{Y_2}^2\\
&  - 2 I_{22} m_{Y_1}^4 m_{Y_2}^2 - 2 I_{2} m_{Y_2}^4 + 4 I_{22} m_X^2 m_{Y_2}^4 - 20 I_{22} m_{Y_1}^2 m_{Y_2}^4 - 2 I_{22} m_{Y_2}^6
\end{align}}

\subsection{$S\to SS$}

The real corrections for the decays $S\to SS$ are given by
\begin{align}
\Gamma_{S\to SS+\gamma/g}=\frac{c_g^2}{(4\pi)^3m_X}C'_Scc^*\Omega\,.
\end{align}
The coupling $c_g$ is the electromagnetic coupling $e$ or the strong coupling $g_s$ for photon and gluon emission, respectively.
$c$ is the coupling of the tree-level vertex of $S\to SS$. 
Therein we identify $S=X$, $S=Y_1$ and $S=Y_2$ with the indices $0,1$ and $2$, respectively.
The individual contributions $\Omega_{ij}$ are given by:
\begin{align}
\Omega_{00} =& -4 I_0 - 8 I_{00} m_X^2\\
\Omega_{01} =& -4 I_0 - 4 I_1 + I_{01} (-8 m_X^2 - 8 m_{Y_1}^2 + 8 m_{Y_2}^2)\\
\Omega_{02} =& -4 I_0 - 4 I_{2} + I_{02} (-8 m_X^2 + 8 m_{Y_1}^2 - 8 m_{Y_2}^2)\\
\Omega_{11} =& -4 I_1 - 8 I_{11} m_{Y_1}^2\\
\Omega_{12} =& 4 I_1 + 4 I_{2} + I_{12} (-8 m_X^2 + 8 m_{Y_1}^2 + 8 m_{Y_2}^2)\\
\Omega_{22} =& -4 I_{2} - 8 I_{22} m_{Y_2}^2
\end{align}

\subsection{$S\to SV$}

The real corrections for the decays $S\to SV$ are given by
\begin{align}
\Gamma_{S\to SV+\gamma/g}=\frac{c_g^2}{(4\pi)^3m_{Y_2}^2m_X}C'_Scc^*\Omega\,.
\end{align}
The coupling $c_g$ is the electromagnetic coupling $e$ or the strong coupling $g_s$ for photon and gluon emission, respectively.
In contrast to previous decays, $S\to SV+\gamma/g$ also has a contribution from a four-point interaction,
which accordingly does not contribute to the infra-red divergent part of the real corrections,
but yields a finite contribution. The sum over the gauge factors thus includes four indices $0,\ldots,3.$ Since we assume a non-extended gauge sector in this work, the four-point vertex is only present for photon emission, so we can simplify the calculation of the group theory factors to $C_{ij}=Q_iQ_j$ for $i,j\in \lbrace 0,1,2\rbrace$, where the charges of the incoming and outgoing scalars are $Q_0$ and $Q_1$ and
the charge of the outgoing gauge boson is $Q_2$; the factors involving the four-point interaction are then 
\begin{align}
 C_{03} &= Q_0(Q_0+Q_1)\,, 
 &C_{13} = Q_1(Q_0+Q_1)\\
 C_{23} &= Q_2(Q_0+Q_1)\,,
 &C_{33} = (Q_0+Q_1)^2\,.
\end{align}
For the emission of a gluon we can use $Q_2=0$ and $Q_0=Q_1=1$. However, we stress that the results below are true in general for any gauge groups; for extended gauge sectors we would just need to employ \eqn{EQ:GeneralCFactors}.
The individual contributions $\Omega_{ij}$ are given by:
{\allowdisplaybreaks
\begin{align}\nn
\Omega_{00} =& -4 I m_X^2 + I_0 (-8 m_X^4 + 8 m_X^2 m_{Y_1}^2 + 8 m_X^2 m_{Y_2}^2)\\
&  + I_{00} (-4 m_X^6 + 8 m_X^4 m_{Y_1}^2 - 4 m_X^2 m_{Y_1}^4 + 8 m_X^4 m_{Y_2}^2 + 8 m_X^2 m_{Y_1}^2 m_{Y_2}^2 - 4 m_X^2 m_{Y_2}^4)\\\nn
\Omega_{01} =& I_0^1 (4 m_X^2 - 4 m_{Y_1}^2 + 4 m_{Y_2}^2) + I_1^0 (-4 m_X^2 + 4 m_{Y_1}^2 + 4 m_{Y_2}^2) + I (4 m_X^2 + 4 m_{Y_1}^2 + 4 m_{Y_2}^2)\\\nn
&  + I_1 (-8 m_X^4 + 8 m_X^2 m_{Y_1}^2 + 16 m_X^2 m_{Y_2}^2 + 8 m_{Y_1}^2 m_{Y_2}^2 - 8 m_{Y_2}^4) + I_0 (8 m_X^2 m_{Y_1}^2 - 8 m_{Y_1}^4\\\nn
&  + 8 m_X^2 m_{Y_2}^2 + 16 m_{Y_1}^2 m_{Y_2}^2 - 8 m_{Y_2}^4) + I_{01} (-4 m_X^6 + 4 m_X^4 m_{Y_1}^2 + 4 m_X^2 m_{Y_1}^4 - 4 m_{Y_1}^6\\
&  + 12 m_X^4 m_{Y_2}^2 + 8 m_X^2 m_{Y_1}^2 m_{Y_2}^2 + 12 m_{Y_1}^4 m_{Y_2}^2 - 12 m_X^2 m_{Y_2}^4 - 12 m_{Y_1}^2 m_{Y_2}^4 + 4 m_{Y_2}^6)\\\nn
\Omega_{02} =& I_0^2 (-2 m_X^2 + 2 m_{Y_1}^2 - 2 m_{Y_2}^2) + I (2 m_X^2 + 6 m_{Y_1}^2 + 2 m_{Y_2}^2) + I_0 (2 m_X^4 + 4 m_X^2 m_{Y_1}^2\\\nn
&  - 6 m_{Y_1}^4 + 4 m_X^2 m_{Y_2}^2 + 12 m_{Y_1}^2 m_{Y_2}^2 - 6 m_{Y_2}^4) + I_{2} (-4 m_X^4 + 8 m_X^2 m_{Y_1}^2 - 4 m_{Y_1}^4\\\nn
&  + 4 m_{Y_2}^4) + I_{02} (-4 m_X^6 + 12 m_X^4 m_{Y_1}^2 - 12 m_X^2 m_{Y_1}^4 + 4 m_{Y_1}^6 + 4 m_X^4 m_{Y_2}^2 + 8 m_X^2 m_{Y_1}^2 m_{Y_2}^2\\
&  - 12 m_{Y_1}^4 m_{Y_2}^2 + 4 m_X^2 m_{Y_2}^4 + 12 m_{Y_1}^2 m_{Y_2}^4 - 4 m_{Y_2}^6)\\\nn
\Omega_{03} =& I (2m_X^2 - 2m_{Y_1}^2 + 2m_{Y_2}^2) + I_0^2 (2m_X^2 - 2m_{Y_1}^2 +  2m_{Y_2}^2) + I_0 ( 2m_X^4 - 4 m_X^2 m_{Y_1}^2\\
&  +  2m_{Y_1}^4 - 4 m_X^2 m_{Y_2}^2 - 4 m_{Y_1}^2 m_{Y_2}^2 + 2 m_{Y_2}^4)\\\nn
\Omega_{11} =& -4 I m_{Y_1}^2 + I_1 (8 m_X^2 m_{Y_1}^2 - 8 m_{Y_1}^4 + 8 m_{Y_1}^2 m_{Y_2}^2) + I_{11} (-4 m_X^4 m_{Y_1}^2 + 8 m_X^2 m_{Y_1}^4\\
&  - 4 m_{Y_1}^6 + 8 m_X^2 m_{Y_1}^2 m_{Y_2}^2 + 8 m_{Y_1}^4 m_{Y_2}^2 - 4 m_{Y_1}^2 m_{Y_2}^4)\\\nn
\Omega_{12} =& I (-6 m_X^2 - 2 m_{Y_1}^2 - 2 m_{Y_2}^2) + I_1^2 (-2 m_X^2 + 2 m_{Y_1}^2 + 2 m_{Y_2}^2) + I_{2} (4 m_X^4 - 8 m_X^2 m_{Y_1}^2\\\nn
&  + 4 m_{Y_1}^4 - 4 m_{Y_2}^4) + I_1 (6 m_X^4 - 4 m_X^2 m_{Y_1}^2 - 2 m_{Y_1}^4 - 12 m_X^2 m_{Y_2}^2 - 4 m_{Y_1}^2 m_{Y_2}^2 + 6 m_{Y_2}^4)\\\nn
&  + I_{12} (-4 m_X^6 + 12 m_X^4 m_{Y_1}^2 - 12 m_X^2 m_{Y_1}^4 + 4 m_{Y_1}^6 + 12 m_X^4 m_{Y_2}^2 - 8 m_X^2 m_{Y_1}^2 m_{Y_2}^2\\
&  - 4 m_{Y_1}^4 m_{Y_2}^2 - 12 m_X^2 m_{Y_2}^4 - 4 m_{Y_1}^2 m_{Y_2}^4 + 4 m_{Y_2}^6)\\\nn
\Omega_{13} =& I (-2 m_X^2 + 2 m_{Y_1}^2 + 2 m_{Y_2}^2) + I_1^2 (-2 m_X^2 + 2 m_{Y_1}^2 + 2 m_{Y_2}^2) + I_1 (2 m_X^4 - 4 m_X^2 m_{Y_1}^2\\
&  + 2 m_{Y_1}^4 - 4 m_X^2 m_{Y_2}^2 - 4 m_{Y_1}^2 m_{Y_2}^2 + 2 m_{Y_2}^4)\\\nn
\Omega_{22} =& I (-2 m_X^2 - 2 m_{Y_1}^2 - m_{Y_2}^2) + 8 I_2^1 m_{Y_2}^2 + 8 I_{22}^{11} m_{Y_2}^2 + I_{2} (8 m_X^2 m_{Y_2}^2 + 8 m_{Y_1}^2 m_{Y_2}^2 - 8 m_{Y_2}^4)\\
&   + I_{22} (-4 m_X^4 m_{Y_2}^2 + 8 m_X^2 m_{Y_1}^2 m_{Y_2}^2 - 4 m_{Y_1}^4 m_{Y_2}^2 + 8 m_X^2 m_{Y_2}^4 + 8 m_{Y_1}^2 m_{Y_2}^4 - 4 m_{Y_2}^6)\\
\Omega_{23} =& I (2 m_X^2 - 2 m_{Y_1}^2 - 4 m_{Y_2}^2) - 8 I_2^1 m_{Y_2}^2\\
\Omega_{33} =& I m_{Y_2}^2
\end{align}}

\section{Infra-red divergent parts of Passarino-Veltman integrals}
\label{sec:appendix_ir}

Subsequently we present the infra-red divergent parts of the Passarino-Veltman integrals relevant for our purposes regularized through a photon or gluon mass $m_\Lambda$:
\begin{align}
\dot{B}_0(p^2,m_\Lambda^2,p^2)&=\dot{B}_0(p^2,p^2,m_\Lambda^2)=-\dot{B}_1(p^2,p^2,m_\Lambda^2) = -\frac{1}{2p^2}\log\left(\frac{m_\Lambda^2}{p^2}\right)\\
C_0(p_0^2,p_1^2,p_2^2,m_\Lambda^2,p_0^2,p_2^2) &= \frac{1}{\lambda(p_0^2,p_1^2,p_2^2)}\log\left(\frac{p_0^2-p_1^2+p_2^2+\lambda(p_0^2,p_1^2,p_2^2)}{2p_0p_2}\right)\log\left(\frac{m_\Lambda^2}{p_0p_2}\right)\\
C_0(p_0^2,p_1^2,p_2^2,p_0^2,m_\Lambda^2,p_1^2) &= \frac{1}{\lambda(p_0^2,p_1^2,p_2^2)}\log\left(\frac{p_0^2+p_1^2-p_2^2+\lambda(p_0^2,p_1^2,p_2^2)}{2p_0p_1}\right)\log\left(\frac{m_\Lambda^2}{p_0p_1}\right)\\
C_0(p_0^2,p_1^2,p_2^2,p_2^2,p_1^2,m_\Lambda^2) &= \frac{1}{\lambda(p_0^2,p_1^2,p_2^2)}\log\left(\frac{-p_0^2+p_1^2+p_2^2+\lambda(p_0^2,p_1^2,p_2^2)}{2p_1p_2}\right)\log\left(\frac{m_\Lambda^2}{p_1p_2}\right)\\
C_1(p_0^2,p_1^2,p_2^2,p_0^2,m_\Lambda^2,p_1^2) &= -C_0(p_0,p_1,p_2,p_0,m_\Lambda,p_1)\\
C_2(p_0^2,p_1^2,p_2^2,p_2^2,p_1^2,m_\Lambda^2) &= -C_0(p_0,p_1,p_2,p_2,p_1,m_\Lambda)\\
C_{11}(p_0^2,p_1^2,p_2^2,p_0^2,m_\Lambda^2,p_1^2) &= C_0(p_0,p_1,p_2,p_0,m_\Lambda,p_1)\\
C_{22}(p_0^2,p_1^2,p_2^2,p_2^2,p_1^2,m_\Lambda^2) &= C_0(p_0,p_1,p_2,p_2,p_1,m_\Lambda)
\end{align}
Therein we use the K\"all\'en function given in \eqn{eq:kaellen}.
The result for $C_0$ is consistent with the infra-red divergent part of Eq. (B.5) of \citere{Dittmaier:2003bc}.
For the cases $p_1^2,p_2^2\ll p_0^2$ and $p_2^2\ll p_1^2,p_0^2$ which is of
particular relevance for scalar decays into light fermions, we also implemented formulas 
\begin{align}
C_0(p_0^2,p_1^2,p_2^2,m_\Lambda^2,p_0^2,p_2^2) &= \frac{1}{p_2^2}\log\left(\frac{m_\Lambda^2}{-p_2^2} \right) \log\left(\frac{p_0 p_1}{-p_2^2} \right) \\
C_0(p_0^2,p_1^2,p_2^2,m_\Lambda^2,p_0^2,p_2^2) &= \frac{1}{p_2^2 - p_1^2} \log\left(\frac{p_0(p_1^2-p_2^2)}{m_\Lambda^2 p_1} \right) \log\left(\frac{p_1^3-p_2^2}{p_0 p_1} \right)\,.
\end{align}
These are equivalent to Eqs. (B.8) and (B.9) of \citere{Dittmaier:2003bc}.
\section{Goldstone boson vertices}
\label{app:goldstones}

For decays into scalars and fermions, the cancellation of infra-red divergences is straightforward: they correspond
to summing the real emission of a massless gauge boson with the virtual process of the same massless gauge boson in the loop.
Because the current of the unbroken gauge symmetries are necessarily flavour diagonal, the only vertices involved
are the original two-body decay vertex and the gauge couplings of the external states. Therefore, when we want
to use loop-corrected external masses to cancel the infra-red divergences for this case it is straightforward to either
put all masses of internal and external states to the loop-corrected values, or (as we do here) to subtract off the infra-red divergent parts
separately from the Bremsstrahlung and virtual corrections before multiplying by a kinematic factor employing loop-corrected masses.

\begin{figure}
\begin{center}
\includegraphics[width=0.75\textwidth]{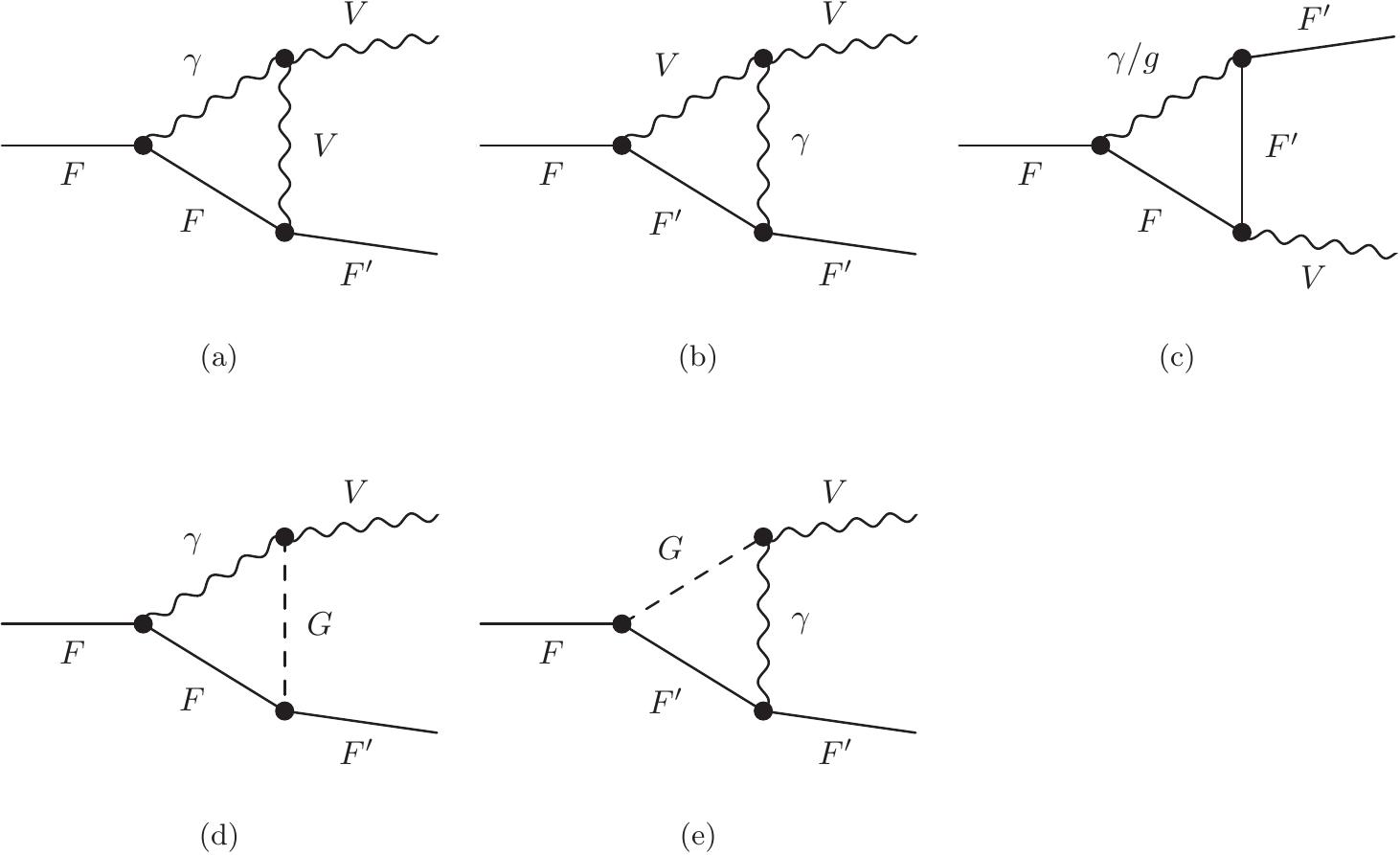}
\caption{Diagrams in the decay $F \rightarrow F'V$ that are potentially infra-red-divergent. Clearly (a) and (d) must
combine in a gauge-invariant way, so that when we put the masses of the external states to their loop-corrected values
we must also put the masses of the internal legs to those masses and adjust the couplings accordingly. The same is true of the pair (b) and (e). }
\label{fig:IRFermion}
\end{center}
\end{figure}

\begin{figure}
\begin{center}
\includegraphics[width=0.9\textwidth]{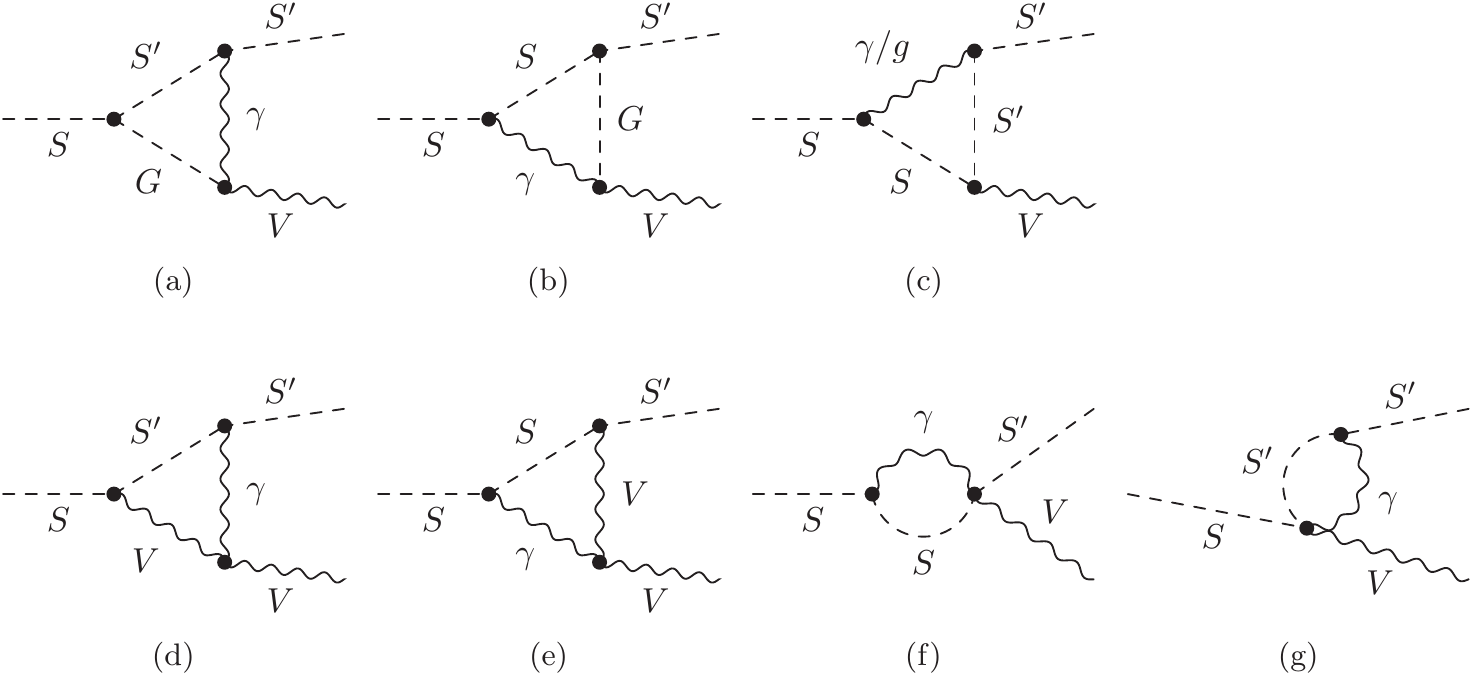}
\caption{Diagrams in the decay $S \rightarrow S'V$ containing a massless gauge boson propagator. The picture
is almost identical to the fermion case, except that we have more ``benign'' diagrams (f) and (g).}
\label{fig:IRScalar}
\end{center}
\end{figure}

\begin{figure}
\begin{center}
\includegraphics[width=0.9\textwidth]{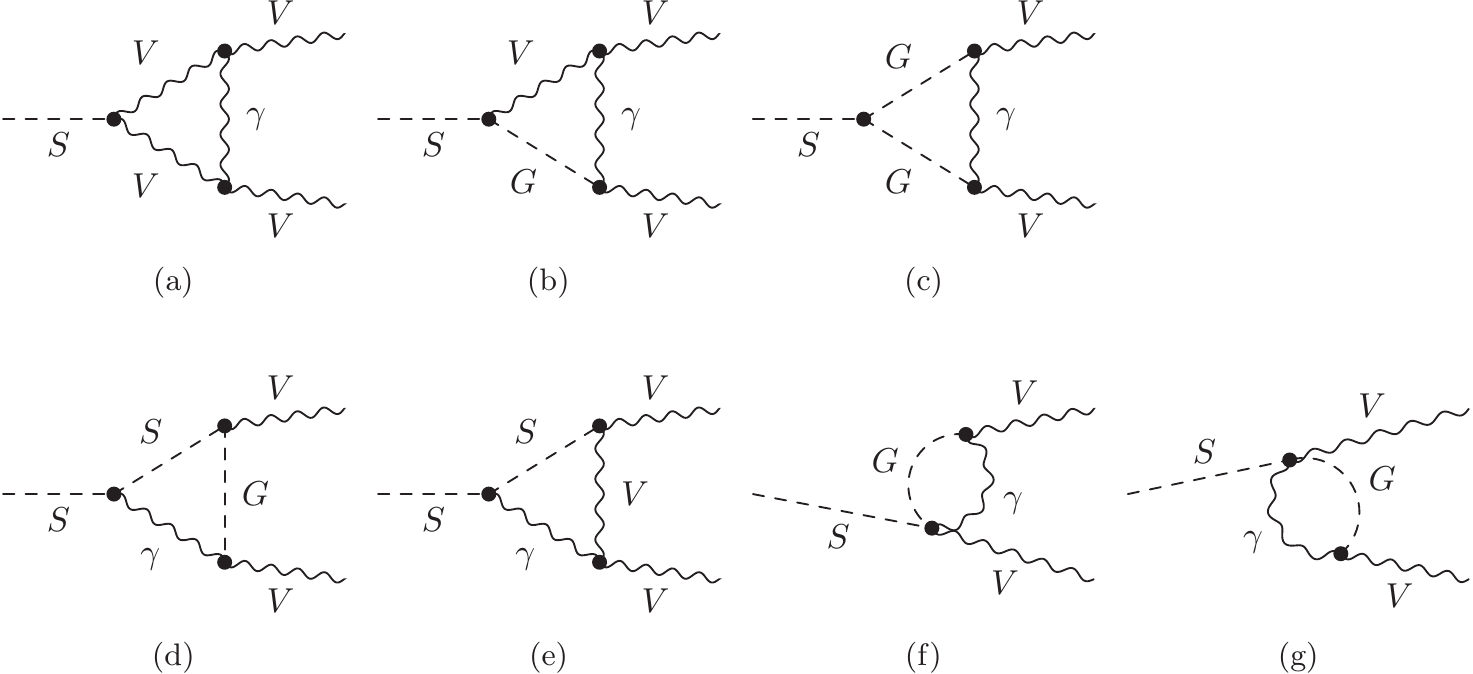}
\caption{Diagrams in the decay $S \rightarrow VV$ (for two heavy vectors) containing a massless gauge 
boson propagator. Diagrams (a), (b) and (c) must combine in a gauge-invariant way; as do (d) and (e); diagrams (f) and (g) are ``benign''. }
\label{fig:IRSVV}
\end{center}
\end{figure}

However, for decays with a massive gauge boson in the final state, the cancellation between the real and virtual
infra-red divergences is a little subtle. The reason is that a would-be Goldstone boson can propagate as an internal state.
While we perform the Bremsstrahlung calculation in the unitary gauge (so there are no Goldstone bosons), for the virtual
corrections for practical purposes the default choice is Feynman-'t Hooft gauge, and we must therefore sum the diagrams
with an internal massive gauge boson with those having a massive Goldstone boson. We show the relevant virtual corrections
for the processes $F\rightarrow FV$, $S\rightarrow SV$ and $S\rightarrow VV$ (for decays to massive vectors)
in Figs.\,\ref{fig:IRFermion}, \ref{fig:IRScalar} and \ref{fig:IRSVV} respectively. Denoting the heavy gauge boson
as ``$V$'', the massless one as ``$\gamma$,'' and the Goldstone boson as ``$G$,'' we see that we have the gauge coupling
of the unbroken group appearing in $FF\gamma, SS\gamma$ and $VV\gamma $ vertices, but we must maintain a relationship between
the $FFV$ and $FFG$ vertices, between the $SSV $ and $SSG$ vertices, the $VV\gamma$ and $VG\gamma$ vertices, in order
for these cancellations to occur. For the $S\rightarrow VV$ process, we require a relationship between the
\emph{three} sets of vertices $SVV$, $SGV$ and $SGG$. 

The required relations follow from Slavnov-Taylor identities, or alternatively  we could examine the infra-red divergent part of the loop amplitudes.
However, here we will more simply derive the conditions imposed by symmetry when inspecting the lagrangian at tree level. We first identify the would-be Goldstone boson by writing the symmetry transformations
of real scalars $S_i^0$ (before spontaneous symmetry breaking, and not necessarily in a mass-diagonal basis) as
\begin{align}
\delta S_i^0 \equiv& \epsilon^G \alpha_i^G = \epsilon^G a_{ij}^G S_j^0\,,
\end{align}
where $\epsilon^G$ is the gauge transformation parameter for each broken direction $G$ and $a_{ij}^G$
are numbers for a linearly realised gauge symmetry (as always assumed in \SARAH). Then 
\begin{align}
a_{ij}^G = t_{ij}^G
\end{align}
where $t_{ij}^G$ are the generators of the gauge symmetry that will be broken by appropriate vacuum expectation values;
since we are working with real scalars, the $t_{ij}^G$ are antisymmetric and real (to translate to complex fields
we require roughly $t_{ij}^G \rightarrow i T^G_{ij}$). Then we find that the Goldstones $G$ are defined by
\begin{align}
G =& N_G \alpha_j^G  S_j^0, \qquad S_i^0 = N_G \alpha_i^G  G + \ldots\,.
\end{align}
The vectors should be chosen to be orthogonal with normalisation $1$, so $N_G = 1/\sqrt{\sum_i (\alpha_i^G)^2}$.

Looking at the vector mass term we find for the scalars $S_i^0$ having vaccum expectation values~$v_i$
\begin{align}
-\frac{1}{2} D_\mu S_i^0 D^\mu S_i^0 \supset& g \partial_\mu S_i^0 V^{a\, \mu} t^a_{ij} v_j
- \frac{g^2}{2} t_{ij}^a v_j t_{ik}^b v_k V_\mu^a V^{b\, \mu} - g^2 t_{ij}^a S_j^0 t_{ik}^b v_k V_\mu^a V^{b\, \mu}\nn\\
&+\frac{g}{2} t_{ij}^a V^{a\, \mu} (S_j^0 \partial_\mu  S_i^0  -  S_i^0 \partial_\mu S_j^0) \nn\\
\supset& g \alpha_i^G \partial_\mu S_i^0 V^{G\, \mu}  - \frac{g^2}{2} \alpha_i^G \alpha_i^{G'}  v_k V_\mu^G V^{G'\, \mu} - g^2 \alpha_i^G t_{ij}^a S_j^0 V_\mu^a V^{G\, \mu} \nn\\
\supset& - \frac{g^2}{2} \sum_G \frac{1}{N_G^2} V_\mu^G V^{G\, \mu}\,,
\end{align}
we see that 
\begin{align}
N_G =& \frac{g}{m_V^G}\,,
\end{align}
where $m_V^G$ is the mass of the vector, 
and we define the $SSV$ coupling when we diagonalise the scalars to mass eigenstates through $S_i^0 \equiv R_{ik} S_k $ (where $R_{Gj} = N_G \alpha_j^G$):
\begin{align}
\lagr \supset& \frac{1}{2} c_{ij}^a V^{a\, \mu} (S_j \partial_\mu  S_i  -  S_i \partial_\mu S_j) \rightarrow c_{ij}^a = g R_{ik} R_{jl} t_{kl}^a\,.
\end{align}

For the $SGV$ coupling and $SVV$ couplings, we read off
\begin{align}
\lagr \supset& \frac{1}{2} c_{iG}^{G'} V^{G'\, \mu} (G \partial_\mu S_i - S_i \partial_\mu  G) + c_i^{GG'} V_{\mu}^G V^{G'\mu} \nn\\
\rightarrow c_{iG}^{G'} =& -\frac{g^2}{m_V^G} R_{ki} \alpha_j^G t_{jk}^{G'}\,, \qquad c_i^{GG'} = - g^2 R_{ki} \alpha_j^G  t^a_{jk} \nn\\
\rightarrow c_{iG}^{G'} =& \frac{1}{m_V^G} c_i^{GG'}\,.
\end{align}
This is the relationship that we enforce between the on-shell vertices to ensure that infra-red divergences are cancelled.

We can also read off the $GVV$ coupling
\begin{align}
 - g^2 \alpha_i^G T_{ij}^a S_j^0 V_\mu^a V^{G\, \mu} \supset& - g^2 N_G \alpha_i^{G'} t_{ij}^a \alpha_j^G G   V_\mu^a V^{G'\, \mu}\,.
\label{EQ:GVV'}
\end{align}

For any unbroken gauge groups, the Goldstones must transform as
\begin{align}
\lagr \supset& \frac{g}{2} N_G^2 \alpha_i^G t_{ij}^a \alpha_j^{G'} \gamma^{a\, \mu} (G' \partial_\mu  G  -  G \partial_\mu G') \nn\\
=& \frac{g}{2} t_{GG'}^a \gamma^{a\, \mu} (G' \partial_\mu  G  -  G \partial_\mu G')
\end{align}
and we therefore identify the $\alpha_i^{G'} t_{ij}^a \alpha_j^G$ factor in \eqn{EQ:GVV'} with $T_{GG'}^a$ to obtain
\begin{align}
\lagr \supset - g m_V^G t^a_{G'G} G'   \gamma_\mu^a V^{G\, \mu}\,.
\end{align}
For the photon, this just becomes the familiar vertex
\begin{align}
\lagr \supset - e m_W G^+ \gamma_\mu^a W^{-\,\mu} + \text{h.c.}\,.
\end{align}
Now the $VV\gamma$ coupling will be given just in terms of the gauge coupling of the unbroken gauge group, 
so the electromagnetic coupling $e$ here; indeed from decomposing the kinetic terms of the gauge bosons we will find
\begin{align}
\lagr \supset& c^{aGG'} \bigg(\partial^\mu \gamma_\nu^a V^{G}_\mu V^{G'\,\nu} -  V_\mu^G  \partial_\nu \gamma^{a\,\mu} V^{G'\,\nu}
-  V_\mu^G  \gamma^{a\,\mu} \partial_\nu  V^{G'\,\nu} + \gamma^{a}_\mu V^{G}_\nu\partial^\mu V^{G'\,\nu} \bigg)\,.
\end{align}
Of these, the first two terms vanish in the limit where the massless gauge boson is soft, while we
identify the last term as the conventional gauge current and 
\begin{align}
c^{aGG'} =& -g t^{a}_{GG'}\,.
\end{align}
This leads to the relation between the $\gamma GV$ and $\gamma VV$ vertices of simply a factor of $m_V^G$. 

To find the relationship between the Goldstone coupling to scalars and the gauge boson coupling, we can use Eq. (2.32)
of \citere{Braathen:2016cqe} to find the derivative of the (effective) potential $V$ with respect to the scalars:
\begin{align}
\frac{\partial^3V}{\partial G \partial S_i^0 \partial S_j^0} =& -N \frac{\partial\alpha_k^G}{\partial \phi_i}
\frac{\partial^2 V}{\partial \phi_k \partial \phi_j} -N\frac{\partial \alpha_k^G}{\partial \phi_j} \frac{\partial^2 V}{\partial \phi_k \partial \phi_i}  \nn\\
=& - \frac{1}{m_V} \bigg[ gt_{ki}^G \mathcal{M}_{kj}^2 +  \mathcal{M}_{ik}^2 gt_{kj}^G \bigg]\,.
\end{align}
When we diagonalise the masses this gives
\begin{align}
c_{ijG} \equiv - \frac{\partial^3V}{\partial G \partial S_i \partial S_j} =& \frac{1}{m_V^G} (m_i^2 - m_j^2)c_{ij}^G\,.
\end{align}
Again we enforce this relationship between the $GSS$ coupling and the $SSV$ coupling when we use loop-corrected masses in order to cancel infra-red divergences. 

The $GGS$ coupling is a special case of the above, with the understanding that we must use zero for the Goldstone boson mass to obtain
\begin{align}
c_{iGG'} =& \frac{m_i^2}{m_V^G m_V^{G'}} c_i^{GG'}\,.
\end{align}

We also find the a similar relationship for fermions, which can in that case also be derived from the fact that the masses
are Yukawa couplings; writing in terms of Weyl fermions we have: 
\begin{align}
\lagr \supset& -\frac{1}{2} Y_{i IJ} S_i^0  \psi_I^0 \psi_J^0 \nn\\
%=& - \frac{1}{2}Y_{iIJ} ( v_i + \frac{g}{m_V} \alpha_i^G G)  \psi_I^0 \psi_J^0 + ... \nn\\
=& -\frac{1}{2}\mathcal{M}_{IJ} \psi_I^0 \psi_J^0 - \frac{g}{2m_V} Y_{iIJ} t^G_{ij} v_j G\psi_I^0 \psi_J^0 + \ldots\,.
%\supset&  \ov{\psi}_I \ov{\sigma}^\mu \psi_J t^a_{IJ} V^a \rightarrow 
\end{align}
The gauge invariance of the Yukawa coupling gives
\begin{align}
Y_{i' IJ} t^a_{i'i} (i) + Y_{iI'J} t^a_{I'I} (I) + Y_{iIJ'} t^a_{J'J} (J) =& 0
\end{align}
and so 
\begin{align}
- \frac{g}{2m_V} Y_{iIJ} t^G_{ij} (i)  v_j =& \frac{g}{2m_V} Y_{iI'J} v_i t^G_{I'I} (I)  + \frac{g}{2m_V} Y_{iIJ'} v_i t^G_{J'J}(J) \nn\\
=& \frac{g}{2m_V} \bigg( \mathcal{M}_{I'J} t^G_{I'I} + \mathcal{M}_{IJ'} t^G_{J'J}(J) \bigg)\,.
\end{align}
Now we write down the gauge coupling
\begin{align}
\lagr \supset& g\ov{\psi}_I \ov{\sigma}^\mu \psi_J t^a_{IJ} V^a 
\end{align}
and we diagonalise into $\psi_I^0 = R_{IJ} \psi_J$ so we obtain 
\begin{align}
\lagr \supset&  c_{IJ}^G \ov{\psi}_I \ov{\sigma}^\mu \psi_J  V^{G\, \mu} + \frac{1}{2} c_{IJ G} \psi_I \psi_J G \nn\\
c_{IJG} =& \frac{m_I - m_J}{m_V^G} c_{IJ}^G\,.  
\end{align}
If we now split the fermions into left and right-handed states with separate rotation matrices $L$ and $R$, then we the gauge couplings are given in Dirac notation by
\begin{align}
\lagr \supset& g \ov{F}_I \gamma^\mu P_L F_J (L^\dagger t^a L)_{IJ} V^a + g \ov{F}_M \gamma^\mu P_R F_N (R^T t^a R^*)_{MN} V^a\,,
\end{align}
while the Goldstone couplings are
\begin{align}
\lagr \supset& \frac{g}{m_V} \bigg( \mathcal{M}_{I'J} t^G_{I'I} + \mathcal{M}_{IJ'} t^G_{J'J}(J) \bigg) G \psi_I^L \psi_J^R  + \text{h.c.} \nn\\
\rightarrow&  \frac{g}{m_V} \bigg( - m_J (R^T t^G R)_{IJ}   + (L^T t^G L)_{IJ} m_I \bigg) G \ov{F}_I P_L F_J \nn\\
&+ \frac{g}{m_V} \bigg( m_I^* (R^\dagger t^G R^*)_{IJ}   - (L^\dagger t^G L^*)_{IJ} m_J^* \bigg) G \ov{F}_I P_R F_J\, .
\end{align}
If we work in a basis where the fermion masses are complex and the rotation matrices $R,L$ are real,
then we find the relationship between the Goldstone boson couplings and the gauge couplings
\begin{align}
c_{IJG}^{L} =& \frac{1}{m_V^G} \big[  m_I c_{IJ}^{G, L} - m_J c_{IJ}^{G, R} \big] \nn\\
c_{IJG}^{R} =& -\frac{1}{m_V^G} \big[  m_J^* c_{IJ}^{G, L} - m_I^* c_{IJ}^{G, R} \big]\,. 
\end{align}

\newpage
\bibliographystyle{h-physrev5}
\bibliography{NLO_Decays}

\end{document}